\documentclass[a4paper,11pt]{article}

\usepackage{jinstpub}

\usepackage[lofdepth,lotdepth]{subfig}
\usepackage{rotating}
\usepackage{multirow}

\title{Development, Configuration and Performance Characterization of a Scalable BETA ASIC-Based Readout System for Multi-Channel SiPM Detectors}

\author[a,1]{Giulio Lucchetta,\note{Corresponding authors.}}
\author[a,1]{Keerthana Rajan Lathika,}
\author[a,1]{Benjamin Nobre Hauptmann,}
\author[b]{Philipp Azzarello,}
\author[a]{Oscar Blanch,}
\author[a]{Juan Boix,}
\author[a]{Laia Cardiel,}
\author[a,2]{Arturo Casta\~{n}o Gallardo,\note{Now at Delft University of Technology, Netherlands.}}
\author[c,d]{Albert Espinya,}
\author[e]{Jennifer Maria Frieden,}
\author[c,d]{David Gascon,}
\author[c]{Daniel Gubermann,}
\author[e]{Kimberly Sarah Keyser,}
\author[a]{Manel Martinez,}
\author[c]{Joan Mauricio,}
\author[c,d]{Marina Orta-Terr\'e,}
\author[e]{Chiara Perrina,}
\author[a]{Javier Rico,}
\author[c,d]{Andreu Sanuy}
\author[a,3]{and Monong Yu \note{Now at Purple Mountain Observatory, Chinese Academy of Science, China.}}

\affiliation[a]{Institut de F{\'i}sica d'Altes Energies (IFAE), The Barcelona Institute of Science and Technology (BIST),\\ E-08193 Bellaterra, Barcelona, Spain}
\affiliation[b]{D{\'e}partement de Physique Nucl{\'e}aire et Corpusculaire (DPNC), Universit{\'e} de Gen{\`e}ve,\\ CH-1211 Gen{\`e}ve 4, Switzerland}
\affiliation[c]{Department of F{\'i}sica Qu{\`a}ntica i Astrofísica, Institut de Ci{\`e}ncies Del Cosmos, Universidad de Barcelona, \\ 08007, Barcelona, Spain}
\affiliation[d]{Institut d’Estudis Espacials de Catalunya (IEEC), \\ 08860 Castelldefels, Spain}
\affiliation[e]{Institute of Physics, Ecole Polytechnique F{\'e}d{\'e}rale de Lausanne (EPFL),\\ CH-1015 Lausanne, Switzerland}

\emailAdd{glucchetta@ifae.es}
\emailAdd{klathika@ifae.es}
\emailAdd{bnobre@ifae.es}

\abstract{We present the design, configuration, and performance characterization of a scalable readout system based on the BETA application-specific integrated circuit (ASIC), developed to meet stringent requirements on noise, linearity, dynamic range, and power consumption for multi-channel silicon photomultiplier (SiPM) detectors in spaceborne instrumentation. The readout electronics consists of modular interface boards (FIBs) hosting multiple BETA ASICs and controlled by a field-programmable gate array (FPGA), which provides configuration, data acquisition, and global trigger generation. Multiple BETA FIBs were tested in a dedicated optical setup enabling simultaneous readout of a large number of channels. The system performance was evaluated using three S13552-10 SiPM arrays manufactured by Hamamatsu for the FIT detector, a scintillating-fiber tracker developed for charged cosmic-ray particle tracking and charge measurement in the HERD mission. We describe the configuration procedures and performance measurements of the readout system, including gain calibration, linearity characterization, and threshold response. In addition, we present the development of a global internal trigger logic for the identification of ionizing particles in the FIT detector. The results demonstrate the stability, scalability, and suitability of the developed BETA-based readout system for large-scale multi-channel SiPM detector applications in space experiments.}


\begin{document}
\maketitle
\flushbottom

\section{Introduction}
\label{sec:introduction}
In recent years, significant advancements have been achieved in scintillating fiber systems, with improvements in optical fiber uniformity, transparency, and cladding techniques. These developments have enhanced production and collection of the scintillation light, improving overall fiber efficiency. In parallel, the rapid development and commercialization of Silicon Photomultipliers (SiPMs) have allowed the implementation of compact and robust read-out systems immune to electromagnetic fields and featuring high granularity, easy scalability and relatively low operating voltages. As a result, tracking detectors based on scintillating fibers read out by SiPMs have emerged as a competitive alternative to silicon strip detectors in high-energy and astroparticle physics experiments. The main advantages of this technology include high position resolution, reduced cost, fast readout, and excellent timing performance. In high-energy physics, scintillating fiber trackers have been implemented, for example, in the SciFi sub-detector upgrade of the LHCb experiment \cite{SciFi,SciFi2}. In astroparticle physics, although the current technology readiness level is still lower than that of silicon strip detectors, the rapid progress in the field has motivated in recent years an extensive R\&D for future missions such as HERD \cite{HERD}, NUSES \cite{NUSES}, APT/ADAPT \cite{ADAPT}, and AMS-100 \cite{AMS-100}.

The Scintillating Fiber Tracker (FIT) \cite{FIT, FIT2}, initially proposed as the tracking detector for HERD, consists of several (seven in its baseline configuration) parallel tracking planes, each including two layers of scintillating fiber mats that measure the particles' interaction in the two transverse coordinates, respectively. In its baseline configuration, the FIT detector does not include high-density conversion foils, which limits the conversion probability for gamma rays. Instead, this design prioritizes higher precision in the reconstruction of the gamma-ray incoming direction, and also provides charge determination of cosmic rays. The scintillating fiber layers are segmented into modules, each comprising one fiber mat and three SiPM arrays. The mat is composed by stacking six layers of fibers, manufactured by \textit{Kuraray} (type SCSF-78MJ),\footnote{\url{https://www.kuraray.com/products/psf}.} with an average diameter of $250\,\mathrm{\mu m}$. The fiber mat is read out by three S13552-10 SiPM arrays, customized by \textit{Hamamatsu Photonics}.\footnote{\url{https://www.hamamatsu.com/jp/en/product/optical-sensors/mppc/mppc_mppc-array/S13552.html}.} These arrays are based on the S13552-HRQ model developed for the LHCb tracker upgrade, but feature an increased number of pixels, allowing the measurements of the charge of cosmic-ray nuclei. Each SiPM consists of $3749$ pixels of $10\,\mathrm{\mu m} \times 10\,\mathrm{\mu m}$ area, and each array contains $128$ SiPMs. The photo-detection efficiency (PDE) of the SiPMs peaks at $450\,\mathrm{nm}$, matching the maximum emission of SCSF-78MJ fibers.\\
A custom-made BETA application-specific integrated circuit (ASIC) \cite{BETAASIC}, was developed for the readout of the SiPM signals, in order to fulfill the stringent requirements in terms of noise, linearity, dynamic range, and power consumption for the FIT space adoption.

In this work we characterize the performance of three S13552-10 SiPM arrays read out by the BETA ASICs. The paper is organized as follows: Section~\ref{sec:setup} describes the optical setup used to test the SiPMs, as well as the readout electronics and data acquisition system developed to handle multiple BETA ASICs. Section~\ref{sec:optical_measurements} presents the optical characterization and calibration of the devices. Section~\ref{sec:thr_calibration} discusses the threshold calibration and Section~\ref{sec:internal_trigger} the generation of a global internal trigger for the identification of ionizing particles in the FIT detector. Finally, Section~\ref{sec:conclusions} summarizes the main findings and outlines future prospects.

\section{Hardware configuration}
\label{sec:setup}
This section describes the hardware configuration and experimental setup developed for the characterization and performance evaluation of the S13552-10 SiPM arrays and the BETA ASICs. Section~\ref{sec:BETA} introduces the architecture and main features of the BETA ASIC, while Section~\ref{sec:opticalsetup} describes the scalable readout system, front-end electronics, and optical setup employed in the measurements.

\subsection{The BETA ASIC architecture}
\label{sec:BETA}
\begin{figure}[h]
\centering
\includegraphics[width=0.9\columnwidth]{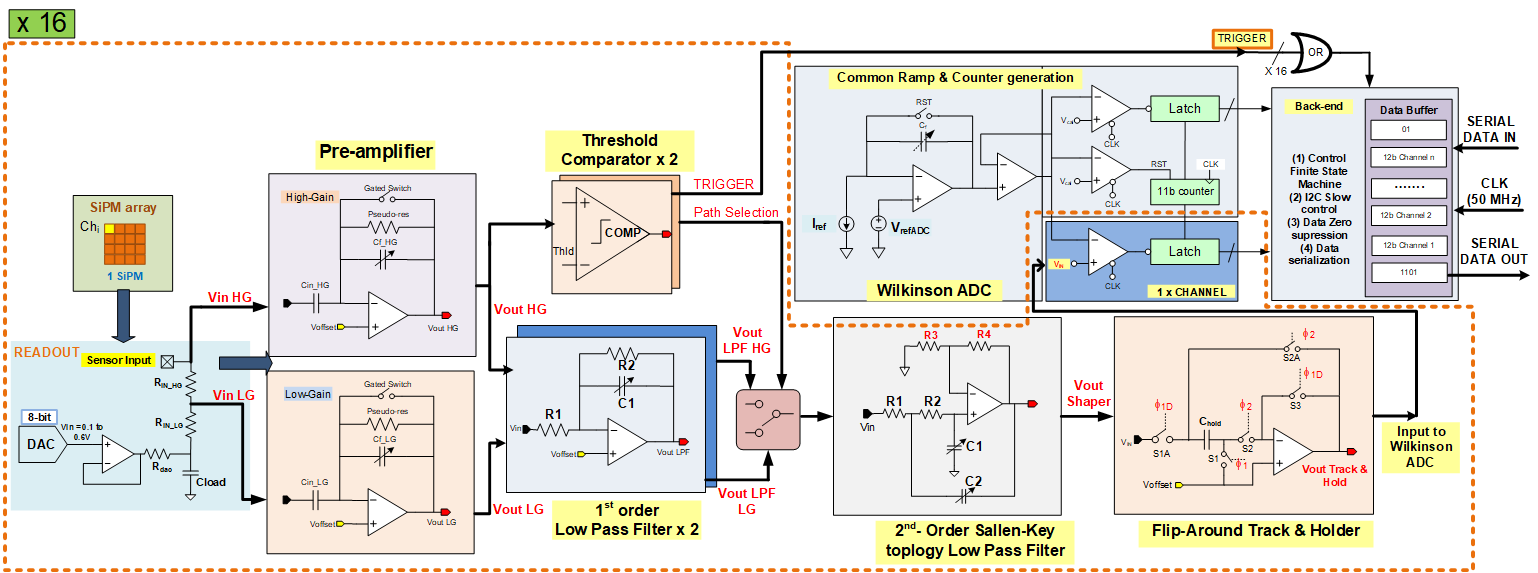}
\caption{\label{fig:BETAschematics} Functional block diagram of the BETA ASIC.}
\end{figure}

A functional block diagram of the BETA ASIC is shown in Figure~\ref{fig:BETAschematics}. The chip features a dual-gain charge-sensitive preamplifier system consisting of a high-gain (HG) and a low-gain (LG) preamplifier. For both the HG and LG paths, the gain can be independently configured among 16 selectable values. 

Two rail-to-rail comparators, connected to the HG preamplifier output, are implemented to generate a trigger signal and a path-selection signal, respectively. The path-selection signal is used to determine whether the HG or LG path is used in the read-out, preventing saturation and extending the dynamic range, if the authopath mode is selected for the readout. Alternatively, the readout can be forced in either HG or LG. This architecture enables single-photoelectron resolution while simultaneously providing a dynamic range sufficient for charge measurements up to $Z = 26$. The preamplifier is followed by a shaper composed of two stages: a first-order low-pass filter (LPF) followed by a second-order Sallen-Key filter. The shaper's output signal is captured using a flip-around track-and-hold circuit, and then digitized using a $12$-bit Wilkinson-type analog-to-digital converter (ADC). Eleven bits are used for the signal readout while one bit flags if either the HG or LG is selected. The power consumption of the ASIC is $\sim 1\,\mathrm{mW/channel}$. Further details about the BETA ASIC can be found in \cite{BETAASIC}. Different versions of the ASIC were produced: a preliminary 16-channel prototype version (BETA-16) and a full 64-channel version (BETA-64). Throughout this manuscript, the term "BETA ASIC" is used when statements apply to both versions of the device. The specific designations "BETA-16" and "BETA-64" are used only when differences between the two versions are relevant or when specifying the version employed for a particular set of measurements.

\subsection{Experimental setup}
\label{sec:opticalsetup}
The characterization and calibration of the BETA ASICs and S13552-10 SiPM arrays were conducted in an optical bench housed within a sealed dark box. A Picoquant PDL 800-D driver and a PLS-450 pulsed LED were used to generate sub-nanosecond pulses at $460\,\mathrm{nm}$\footnote{\url{https://www.picoquant.com/products/category/picosecond-pulsed-driver}} illuminating the silicon photomultipliers. The LED can either be triggered internally at various repetition frequencies, or through an external pulse. The LED intensity can be adjusted using a potentiometer, in a range between $0$ and $10$, and with a minimum step of $0.02$. The LED's wavelength corresponds to the peak PDE of the S13552-10 SiPM arrays.

\begin{figure}[ht]
	\centering
	\subfloat[S13552-10 SiPM array mounted in its readout PCB with the flat flexible cables.]{
	\includegraphics[width=0.99\columnwidth]{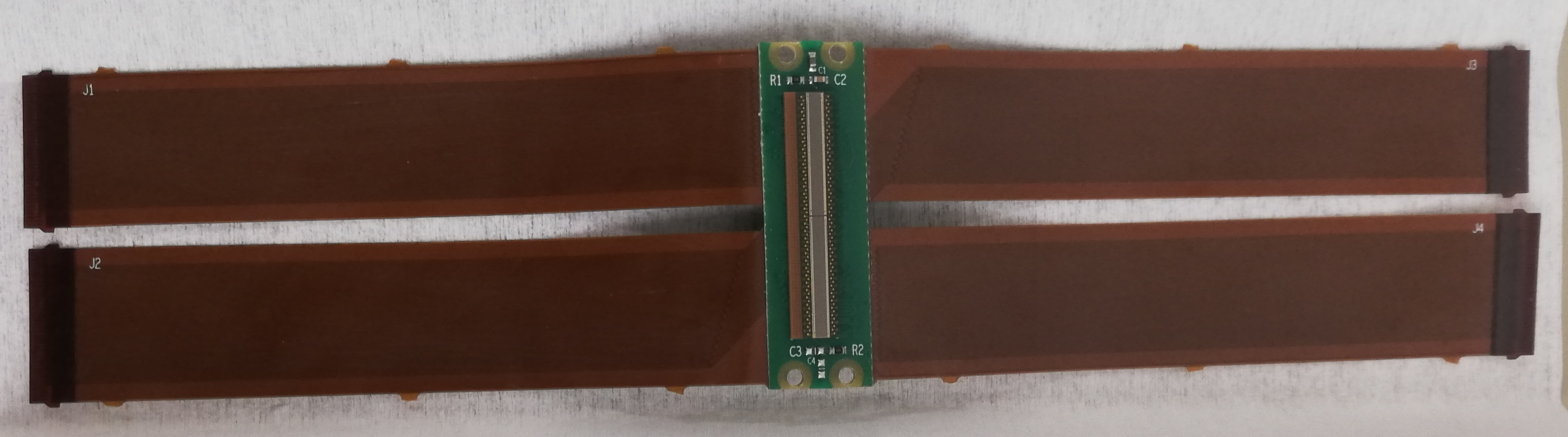}
	\label{fig:SiPM}}
	\vspace{0.4cm}
	\subfloat[Front-end electronics, SiPM arrays, and LED inside the dark box]{
	\includegraphics[width=0.48\columnwidth]{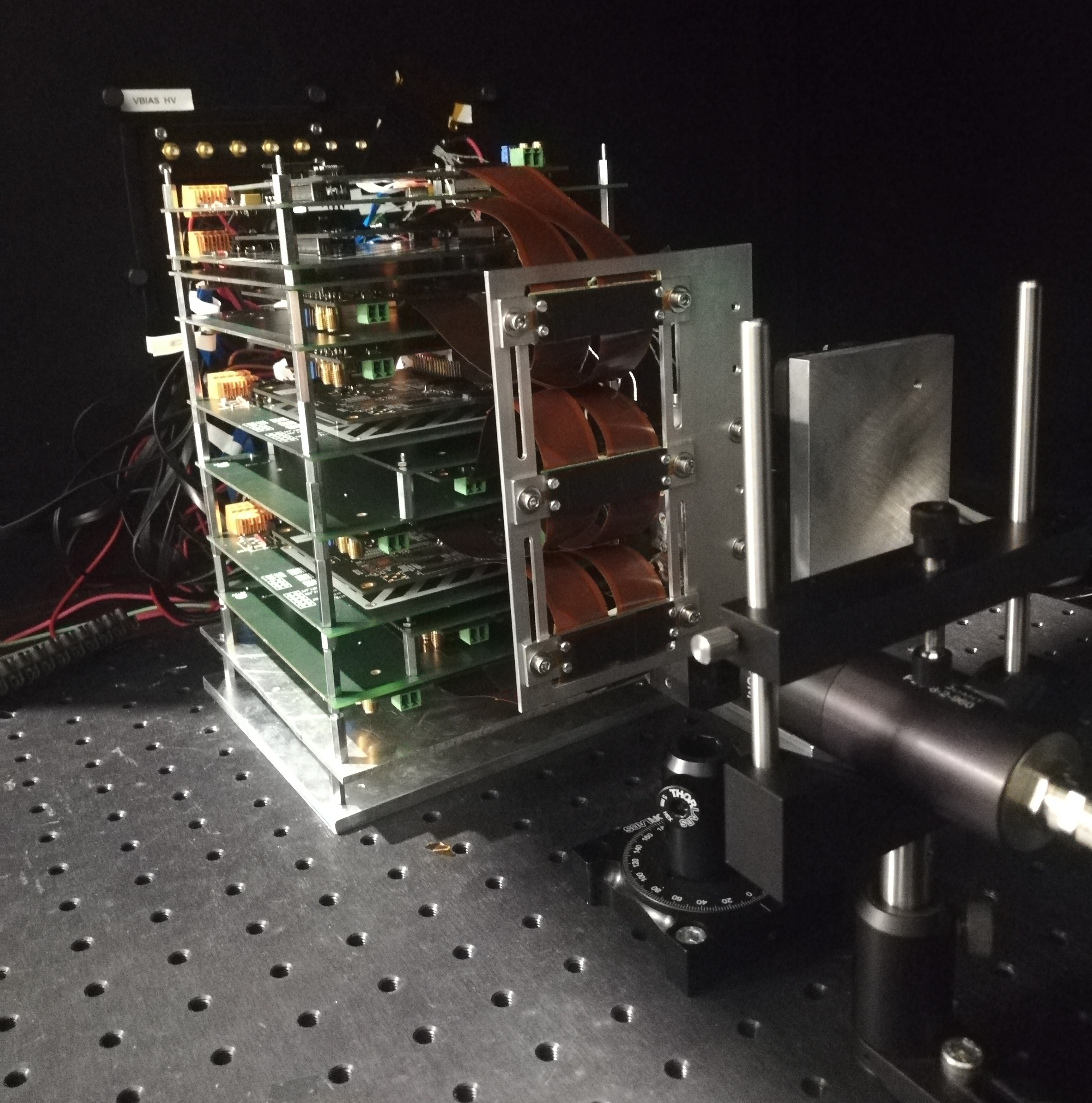}
	\label{fig:BlackBox}}
	\hfill
	\subfloat[Xilinx EK-Z7-ZC706-G FPGA (left) and interface motherboard (right)]{
	\includegraphics[width=0.48\columnwidth]{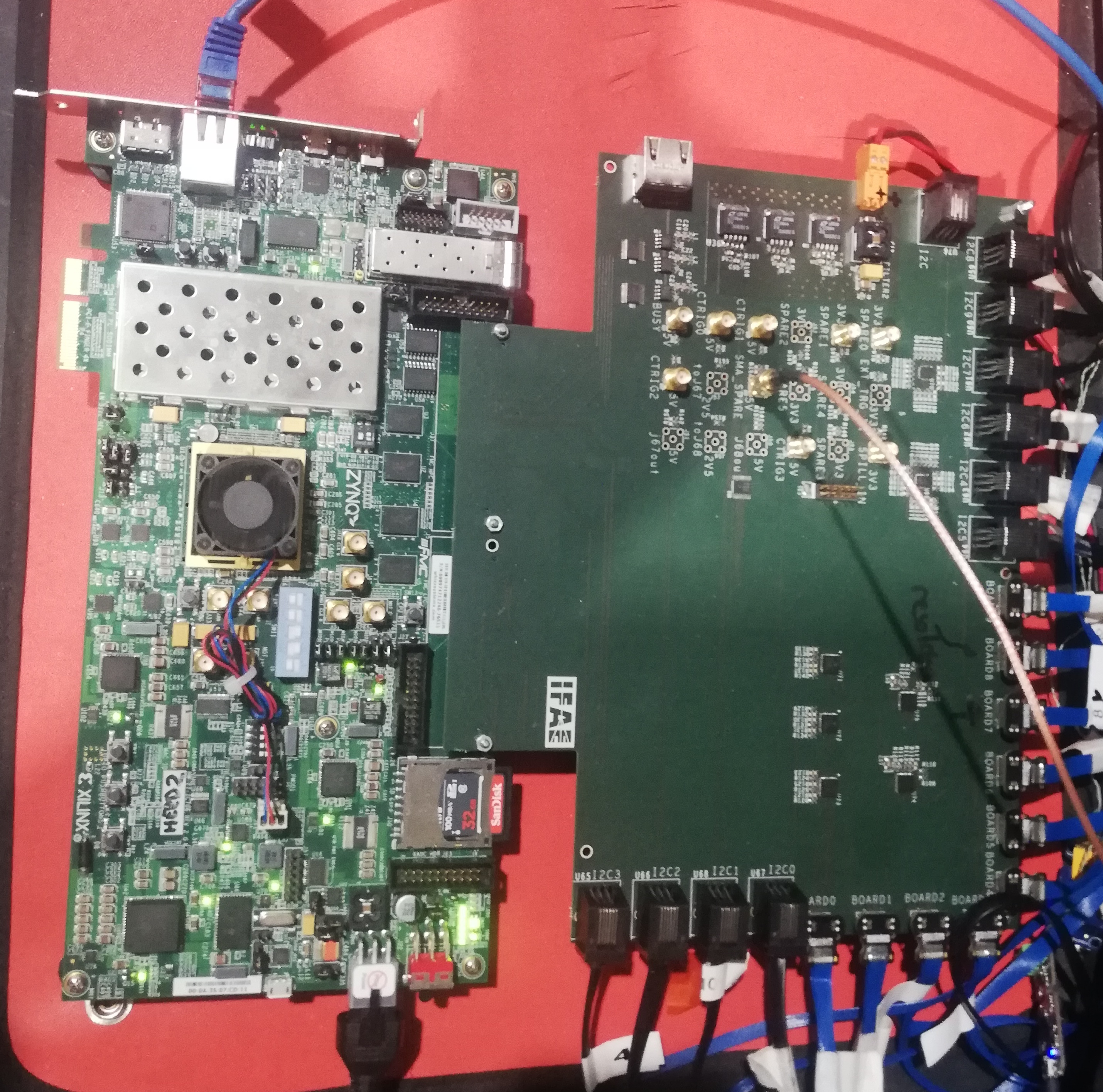}
	\label{fig:FPGA_Mother}}
	\caption{\label{fig:Set-up} Pictures of the components involved in the SiPM arrays characterisation setup.}
\end{figure}

Three S13552-10 SiPM arrays (one of them is shown in Figure~\ref{fig:SiPM}) were investigated in this study. Hereafter, the three SiPM arrays are referred to as SiPMarray-1, SiPMarray-2, and SiPMarray-3. Each SiPM array is mounted on a printed circuit board (PCB) and secured in a mechanical frame. A custom-designed mask with thin, precisely aligned and equally spaced slits, can be added to the frame to selectively illuminate specific channels of the SiPM arrays. The arrays are connected to the front-end electronics via flat flexible cables.

The readout electronics consist of multiple PCBs, referred to as FIT interface boards (FIBs), each housing two BETA ASICs, allowing for the simultaneous readout and characterization of a large number of channels. Between $2024$ and $2025$, $10$ BETA-16 FIBs (corresponding to a total number of $20$ ASICs) and $14$ BETA-64 FIBs (corresponding to $28$ ASICs in total) were configured and tested using the optical setup. The BETA-16 boards were extensively characterized, used to perform calibration studies and define the optimal operating range for the detection of ionizing particles in the FIT detector during beam-test campaigns. In contrast, for the BETA-64 boards, only validation tests and the configuration of the relevant ASIC registers were carried out prior to the beam tests to ensure proper operation, while the full calibration was performed directly using beam-test data. A photograph of the experimental setup inside the black box is shown in Figure~\ref{fig:BlackBox}.

Control, configuration, and signal processing of the BETA ASICs are handled by a Xilinx EK-Z7-ZC706-G field-programmable gate array (FPGA) board, interfaced with a motherboard (see Figure~\ref{fig:FPGA_Mother}). The interface motherboard sends the configuration to the ASICs via I2C buses and receives the datastreams from the BETA FIBs via Samtec cables. Furthermore, the FPGA firmware generates a global internal trigger signal, from the combinational logic of pre-trigger signals from individual BETA ASICs. Data acquisition can be initiated by four different triggering modes: the mentioned global internal trigger, a periodic trigger based on the FPGA clock, a $3.3\,\mathrm{V}$ TTL-level external trigger and an external trigger via an I2C interface. During data acquisition, additional information such as event ID, timestamp, pre-trigger maps, and individual ASIC pre-trigger counters are provided to facilitate monitoring and debugging of the readout system.

For laboratory calibration of the S13552-10 SiPM arrays and BETA ASICs, an Agilent 81160A pulse function generator is used to trigger both the PDL 800-D driver and the FPGA acquisition system. To maintain a stable gain for the SiPM arrays, their bias voltage is adjusted according to the temperature, on an active feedback loop. The relationship between the bias voltage, $V_{\text{bias}}$, and the temperature, $T$, is described by a linear parametrization,
\begin{equation}
V_{\text{bias}}/\text{V} = 0.035 \cdot T/^\circ\text{C} + 43.7
\end{equation}
where the coefficients were obtained from an experimental calibration of the temperature dependence of the SiPM arrays used in this work. The values result from averaging the temperature coefficients measured for all channels within each array and across multiple arrays. A detailed description of the measurement procedure will be presented in a dedicated publication.

All measurements were performed at room temperature. Temperature was measured using Yocto-Thermocouples\footnote{\url{http://www.yoctopuce.com/EN/products/usb-environmental-sensors/yocto-thermocouple}.} attached to the backside of the SiPM PCBs. The bias voltage is supplied through a Keithley 6514A electrometer, which adjusts the voltage every $10$ seconds based on the temperature reading from the thermocouples.

\section{Characterization and calibration of the S13552-10 SiPM arrays and BETA ASIC readout}
\label{sec:optical_measurements}
This section describes the experimental method adopted to calibrate the readout chain. A full calibration of the setup was performed for each gain configuration of the HG and LG preamplifiers, both in dark conditions and under pulsed LED illumination at various intensities. The gain of each preamplifier path is controlled by the $4$-bit registers \texttt{Dcap\textunderscore CH\textunderscore HG\textunderscore Cf} and \texttt{Dcap\textunderscore CH\textunderscore LG\textunderscore Cf}. The registers configure the feedback capacitance, $C_f$, of the HG and LG preamplifiers, respectively. The register values span the range from $0$ to $15$, corresponding to a linear increase in $C_f$. Since the gain of a charge-sensitive preamplifier scales as $1/C_f$, the register value of $0$ yields the maximum gain, while $15$ corresponds to the minimum.

The calibration measurements were performed for a subset of selected channels of the $20$ BETA-16 ASICs investigated in this work, due to the difficulty of uniformly illuminating all SiPM arrays with sufficient light intensity to probe the full dynamic range of the system. Despite the reduced sample, the measurements are sufficient to determine the optimal gain configuration for the detection of minimum ionizing particles (MIPs) and charged cosmic rays, and to provide a first assessment of the channel-to-channel calibration variations associated with different ASICs and SiPM arrays. A complete calibration of all channels of both BETA-16 and BETA-64 boards was performed using beam-test data, and will be addressed in a dedicated publication.

The procedure starts with a calibration of the ADC response in photoelectron (phe) units, for the maximum possible gain (HG00), where individual photoelectron peaks are clearly resolved. This calibration is then propagated to all other gain configurations through a relative calibration, extending the results across the full operational range, including gain settings where individual photopeaks are no longer resolvable. Prior to these calibration measurements, the BETA ASICs configuration was optimized following the procedure described in Appendix~\ref{sec:BETA_optimization}.

\subsection{Photoelectron signal calibration at maximum gain}
\label{sec:calibration}
\begin{figure}[htbp]
	\centering
	\subfloat[ADC distribution measured in dark conditions, recorded with the BETA ASIC at gain HG00. The observed multi-peak structure originates from random sampling of thermally induced dark pulses (see Section~\ref{sec:calibration} for details).]{
	\includegraphics[width=0.7\columnwidth]{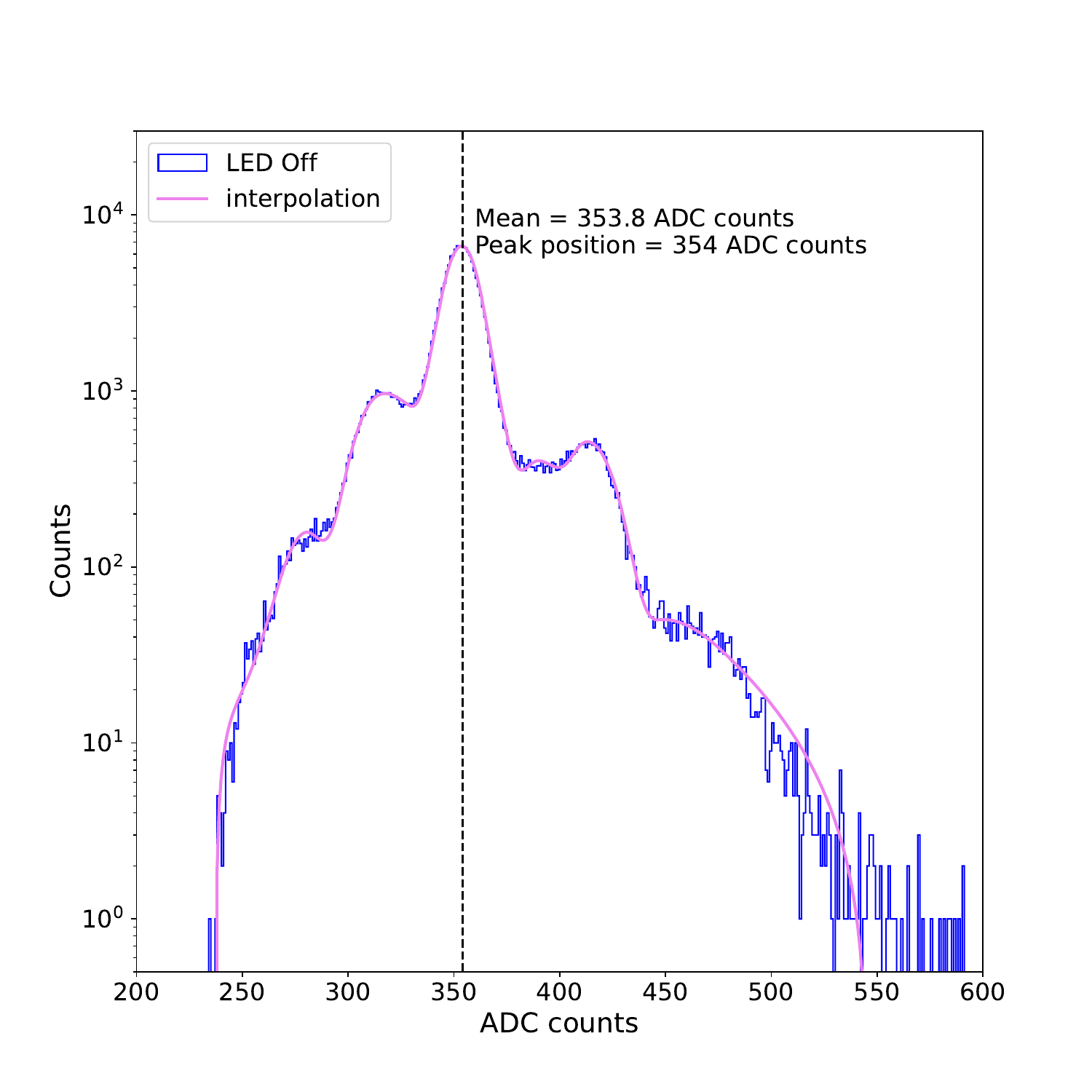}
	\label{fig:DarkCounts}}
	\vspace{-0.05cm}
	\subfloat[ADC distribution measured under pulsed LED illumination at several light intensities, recorded with the BETA ASIC at gain HG00.]{
	\includegraphics[width=0.85\columnwidth]{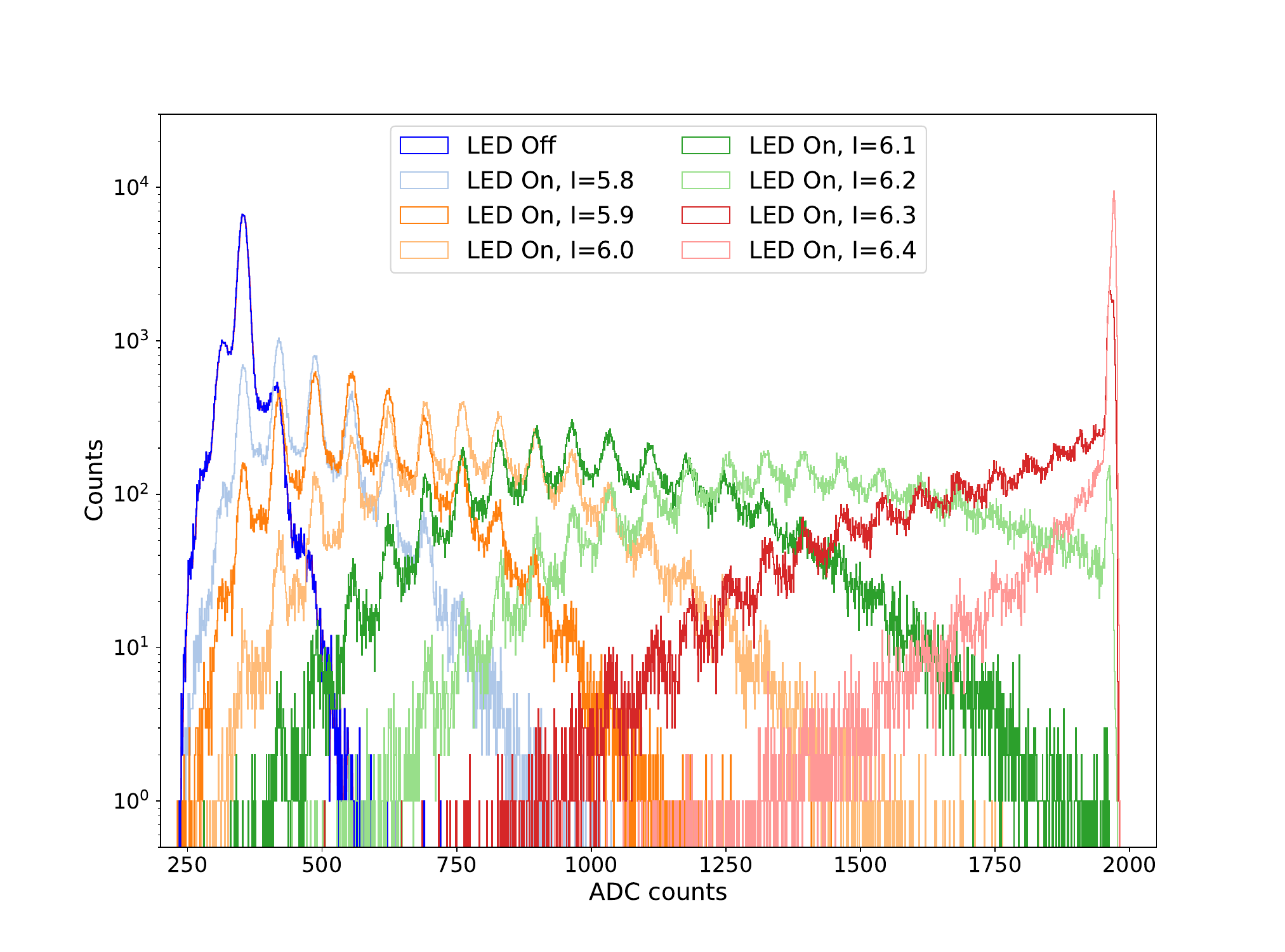}
	\label{fig:SignalDistribution}}
	\caption{\label{fig:PhotelectronDistributions} Dark-count and photoelectron spectra for several LED intensities produced by one SiPM of the S13552-10 array.}
\end{figure}

The photoelectron calibration was set using HG00. Data acquisition was triggered by an external pulse generated by the Agilent 81160A function generator. An example of the ADC distribution obtained in dark conditions (labelled as "LED Off") is shown in Figure~\ref{fig:DarkCounts}.

The observed multi-peak structure arises from random sampling of the BETA ASIC's bipolar shaper signal. Thermally generated charge carriers can trigger Geiger discharges in the SiPM, producing dark pulses that may coincide with the external trigger and be sampled on either the positive or negative lobe of the shaper. In Figure~\ref{fig:DarkCounts}, the central, most pronounced peak corresponds to the baseline (pedestal), while the peaks on the right and left sides correspond, respectively, to single-photoelectron events sampled on the positive and negative lobes. Optical crosstalk and afterpulses can occasionally generate signals at amplitudes higher than a single photoelectron. Since the positive and negative areas of a bipolar signal are equal, the mean of the measured ADC distribution coincides with the main peak position (i.e. the mode or most probable value), within the ADC precision. Thus, either estimator can be used to determine the pedestal position for calibration.

\begin{figure}[htbp]
\centering
\includegraphics[width=0.85\columnwidth]{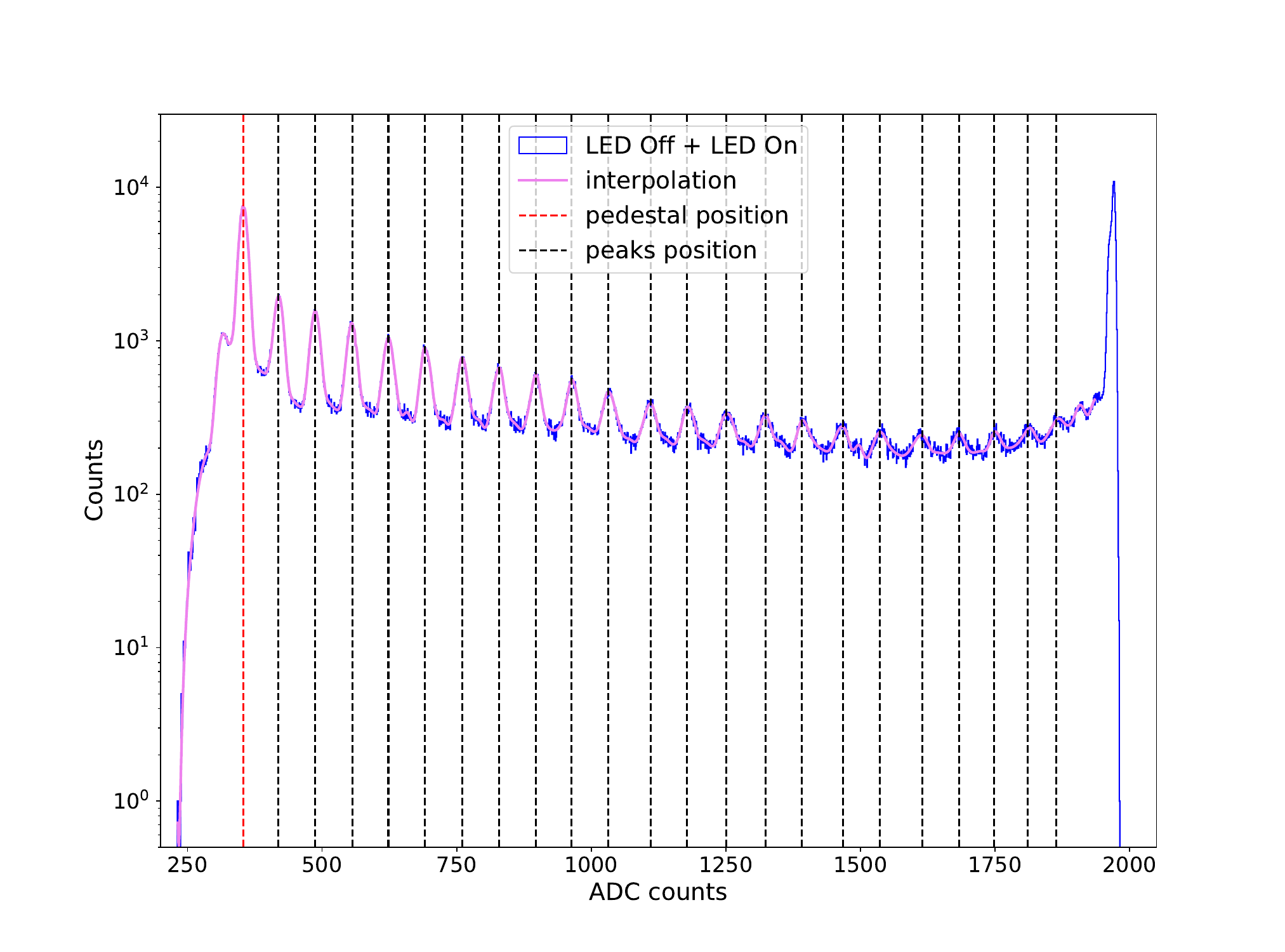}
\caption{\label{fig:JointDistribution} Combined ADC distribution obtained by summing spectra at different LED intensities, used for the photoelectron calibration of the BETA ASIC.}
\end{figure}

Figure~\ref{fig:SignalDistribution} shows the SiPM response to several low-light intensities (these measurements are labelled as "LED On, I=X", with X representing the intensity of the LED), tuned to produce spectra with high statistics over the full ADC range. For HG00, well-resolved phe peaks are identified up to $20$--$25\,\mathrm{phe}$, depending on the ASIC, before reaching saturation.\\

To calibrate the setup in terms of phe-to-ADC counts conversion ($C$), the spectra recorded at different LED intensities were summed together to obtain the combined distribution shown in Figure~\ref{fig:JointDistribution}. The resulting distribution was smoothed by an interpolation, and the positions of individual photoelectron peaks were automatically extracted using the peak-finding algorithm \texttt{find\textunderscore peaks} from the \texttt{SciPy} Python library.\footnote{\url{https://scipy.org/}} Statistical uncertainties on the photopeak positions were computed using a bootstrap method: starting from the distribution in Figure~\ref{fig:JointDistribution}, $300$ pseudo-datasets were generated by letting each histogram bin fluctuate according to a Poisson distribution of mean the measured value. Photopeak positions were recalculated for each pseudo-dataset, and the standard deviation of the resulting distribution is used as estimator of the peak position uncertainty.

The calibration factor $C$ was determined from the slope of a linear fit to the photopeak positions as a function of the photoelectron number, as illustrated in Figure~\ref{fig:Linearity}. The peak positions were fitted with a linear regression in which the intercept was fixed to the measured pedestal position, $P_0$, i.e. using the function:
\begin{equation}
\label{eq:fit_fixed_pedestal}
y = C \cdot x + P_0\, .
\end{equation}
The calibration factors obtained for HG$00$ in selected channels of the $20$ tested BETA-16 ASICs are summarized in Table~\ref{tab:CalibrationFactor}.

\begin{sidewaystable}[p]
\centering
\small
\caption{\label{tab:CalibrationFactor} Calibration factors $C$, obtained at HG$00$ for selected channels of $20$ tested BETA-16 ASICs. A mask was used to illuminate only the central channels of each ASIC. Missing entries correspond to channels with insufficient illumination or excessive noise, preventing a precise and reliable calibration. For each ASIC, the mean calibration factor $\bar{C}$ of the $N$ calibrated channels, the statistical uncertainty $\sigma_{\mathrm{stat}}$ and the dispersion uncertainty $\sigma_{\mathrm{disp}}$ are reported to assess the robustness of the calibration procedure and the channel-to-channel uniformity. The statistical uncertainty is defined as $\sigma_{\mathrm{stat}} = \frac{1}{N} \sqrt{\sum_{i=1}^{N} (\sigma_{C_i})^2}$, while the dispersion uncertainty is defined as the sample standard deviation $\sigma_{\mathrm{disp}} = \sqrt{\frac{1}{N - 1} \sum_{i=1}^{N} \left(C_i - \bar{C}\right)^2}$.}
\begin{center}
\resizebox{\columnwidth}{!}{%
\begin{tabular}{lcccccccc}
\hline
\hline
ASIC/SiPM array & \multicolumn{5}{c}{ $C$} & $\bar{C}$ & $\sigma_{\mathrm{disp}}$ & $\sigma_{\mathrm{stat}}$\\
 & \multicolumn{5}{c}{ $[\mathrm{ADC counts/phe}]$} & $[\mathrm{ADC counts/phe}]$ & $[\mathrm{ADC counts/phe}]$ & $[\mathrm{ADC counts/phe}]$\\
\hline
& Ch06 & Ch07 &	Ch08 & Ch09 & Ch10 & & & \\
ASIC0/SiPMarray-1 &	$68.17 \pm 0.03$	 &	$69.25 \pm 0.03$ &	$67.10 \pm 0.03$	 &	--- & --- & $68.173$ & $1.081$ & $0.016$ \\
ASIC1/SiPMarray-1 &	$73.93 \pm 0.03$	 &	$74.18 \pm 0.03$ &	$70.34 \pm 0.03$ &	--- & --- & $72.820$ & $2.149$ & $0.018$\\
ASIC2/SiPMarray-2 &	$75.79 \pm 0.03$	 &	$78.45 \pm 0.03$ &	--- 	& $77.25 \pm 0.03$ & --- & $77.163$ & $1.333$ & $0.018$ \\
ASIC3/SiPMarray-2 &	--- &	$73.384 \pm 0.021$ &	$72.92 \pm 0.03$ &	$73.65 \pm 0.03$	& --- & $73.319$ & $0.370$ & $0.016$ \\	
ASIC4/SiPMarray-3 &	---	&	$76.88 \pm 0.03$	 &	$76.60 \pm 0.04$ &	$77.11 \pm 0.05$	 & --- & $76.865$ & $0.254$ & $0.024$ \\
ASIC5/SiPMarray-3 &	---	&	$71.48 \pm 0.05$	&	$70.94 \pm 0.05$ &	$71.36 \pm 0.04$	 & --- & $71.258$ & $0.286$ & $0.025$ \\
ASIC6/SiPMarray-1 &	---	&	$74.65 \pm 0.04$	 &	$77.89 \pm 0.03$	 &	--- & --- & $76.270$ & $2.294$ & $0.023$ \\
ASIC7/SiPMarray-1 &	---	&	$75.65 \pm 0.04$	 &	$72.83 \pm 0.04$	 &	$72.65 \pm 0.03$ & --- &  $73.710$ & $1.680$ & $0.020$ \\
ASIC8/SiPMarray-3 &	$78.93 \pm 0.05$ &	$80.26 \pm 0.05$	 &	$73.78 \pm 0.03$	&	---	& --- & $79.19$ & $0.98$ & $0.03$ \\	
ASIC9/SiPMarray-3 &	$73.58 \pm 0.03$	 &	$72.55 \pm 0.04$	 &	$72.41 \pm 0.04$ 	&	--- & --- & $72.846$ & $0.636$ & $0.021$ \\	
ASIC10/SiPMarray-3 & ---	 &	--- &	$74.90 \pm 0.04$ &	$72.56 \pm 0.13$	 &	$78.89 \pm 0.03$ & $75.856$ & $2.685$ & $0.0182$ \\
ASIC11/SiPMarray-3 & ---	 &	---	&	$77.04 \pm 0.03$	 &	$73.91 \pm 0.03$	 &	$73.05 \pm 0.03$ & $74.667$ & $2.103$ & $0.018$ \\
ASIC12/SiPMarray-3 & ---	 &	$71.59 \pm 0.04$	 &	$70.64 \pm 0.03$	 &	$71.30 \pm 0.03$ & --- & $71.174$ & $0.485$ & $0.019$ \\
ASIC13/SiPMarray-3 & ---	 &	$71.27 \pm 0.04$	 &	$68.95 \pm 0.03$	 &	$68.35 \pm 0.03$ & --- & $69.523$ & $1.546$ & $0.021$ \\
ASIC14/SiPMarray-2 & ---	 &	---	&	$80.99 \pm 0.03$	 &	$82.29 \pm 0.04$	 & $82.52 \pm 0.04$ & $81.937$ & $0.824$ & $0.021$ \\
ASIC15/SiPMarray-2 & ---	 &	$72.43 \pm 0.03$	 &	$71.144 \pm 0.024$	 &	$73.149 \pm 0.024$ & --- & $72.241$ & $1.016$ & $0.014$ \\
ASIC16/SiPMarray-1 & ---	 &	$69.07 \pm 0.04$	 &	$68.78 \pm 0.05$	 & $69.20 \pm 0.03$ & --- & $69.016$ & $0.215$ & $0.023$ \\
ASIC17/SiPMarray-1 & $69.32 \pm 0.06$ & 	$70.61 \pm 0.10$	&	$70.15 \pm 0.03$ &	$71.94 \pm 0.03$ & --- & $70.51$ & $1.10$ & $0.05$ \\
ASIC18/SiPMarray-2 & $72.52 \pm 0.04$ &	$72.094 \pm 0.025$ &	$70.438 \pm 0.023$ & --- & --- & $71.684$ & $1.100$ & $0.016$ \\
ASIC19/SiPMarray-2 & ---	 &	---	&	$80.12 \pm 0.05$ &	 $80.10 \pm 0.06$ & 	--- & $80.110$ & $0.008$ & $0.038$ \\	
\hline
\hline
\end{tabular}
}
\end{center}
\end{sidewaystable}

\subsection{Linearity measurements and non-linearities}
\label{sec:linearity}
\begin{figure}[h]
\centering
\includegraphics[width=0.6\columnwidth]{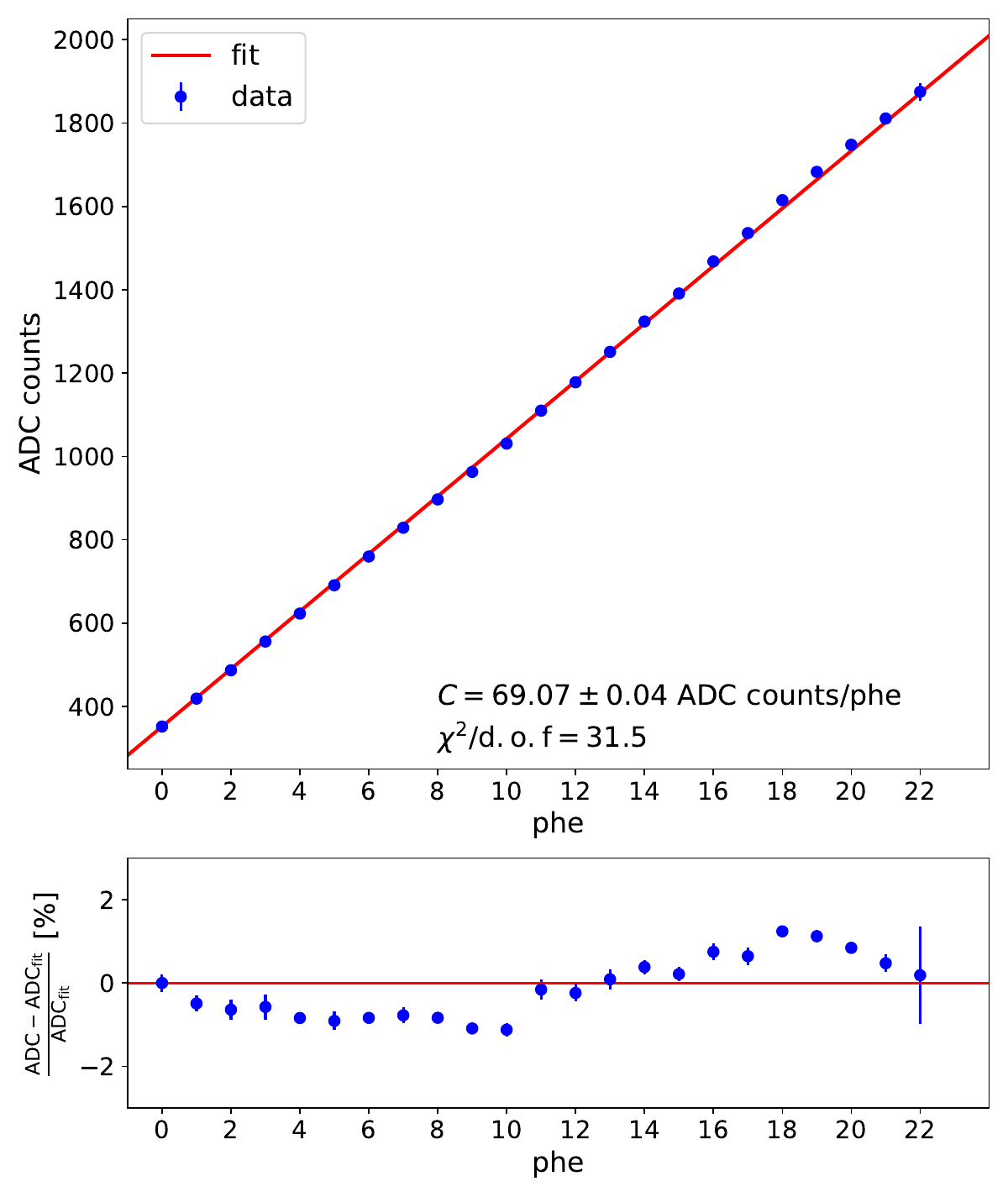}
\caption{\label{fig:Linearity} BETA-16 ASIC16, Ch07. Top: Linear fit of the photoelectron peak positions. The intercept was fixed to the measured pedestal position, $P_0$. Bottom: Non-linearity, quantified as the relative residual, as a function of the number of phe.}
\end{figure}

The identification of photoelectron peaks enables a direct assessment of the linearity of the readout chain, as outlined in Figure~\ref{fig:Linearity}. The non-linearity was then quantified as the relative deviation of the measured peak positions from the values predicted by the linear fit, as shown in the lower panel of Figure~\ref{fig:Linearity}.

\begin{figure}[h]
\centering
\includegraphics[width=0.95\columnwidth]{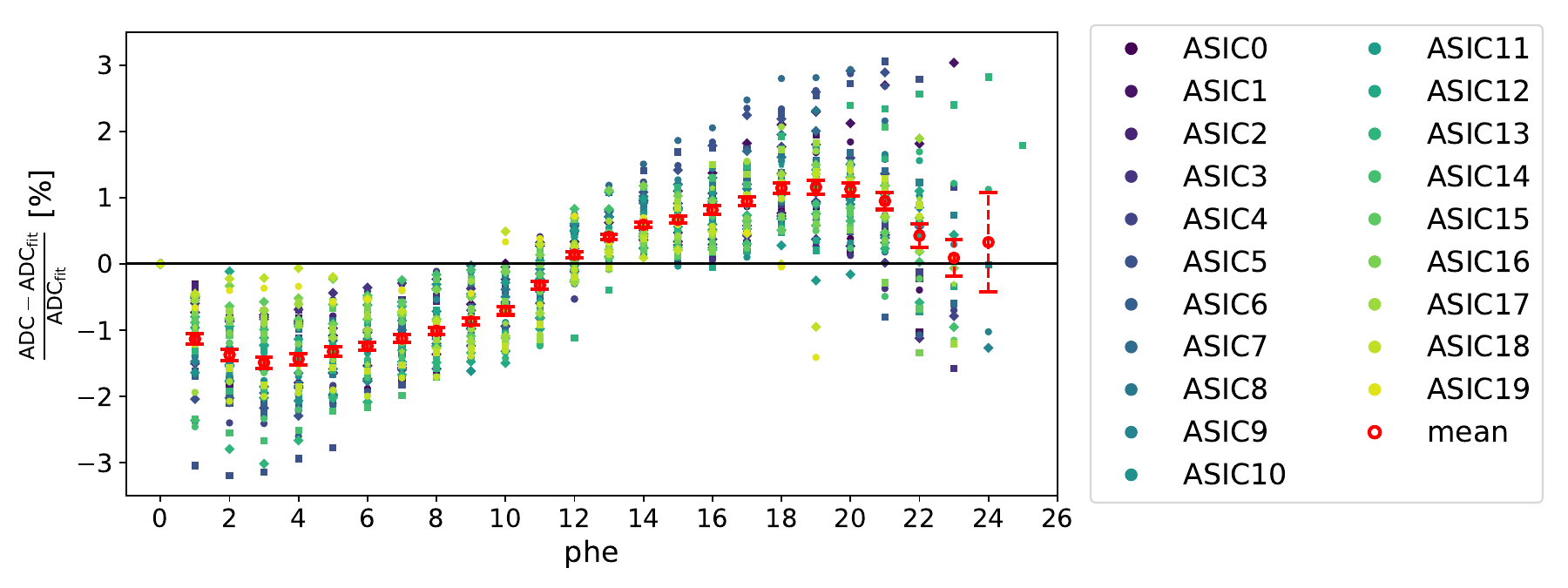}
\caption{\label{fig:LinearityError_all} Non-linearity as a function of number of phe for the 20 BETA-16 ASICs characterized in the laboratory. The error bars on individual data points are omitted for better clarity of the plot. The red dots show the mean non-linearity as a function of the phe number. The red error bars display the standard errors of the sample means.}
\end{figure}

A consistent dependence of the non-linearity was observed across all tested ASICs, as summarized in Figure~\ref{fig:LinearityError_all}. To disentangle non-linearities of the read-out electronics from potential distortions introduced by the SiPMs response or by the LED source, independent measurements were performed by injecting electrical pulses directly into the BETA ASIC. A complete description of the setup, methodology and results for these measurements is provided in Section~\ref{sec:charge_injection}.

\subsection{Relative calibration for lower gain values}
\label{sec:relative_calibration}

\begin{figure}[htbp]
\centering
\includegraphics[width=0.55\columnwidth]{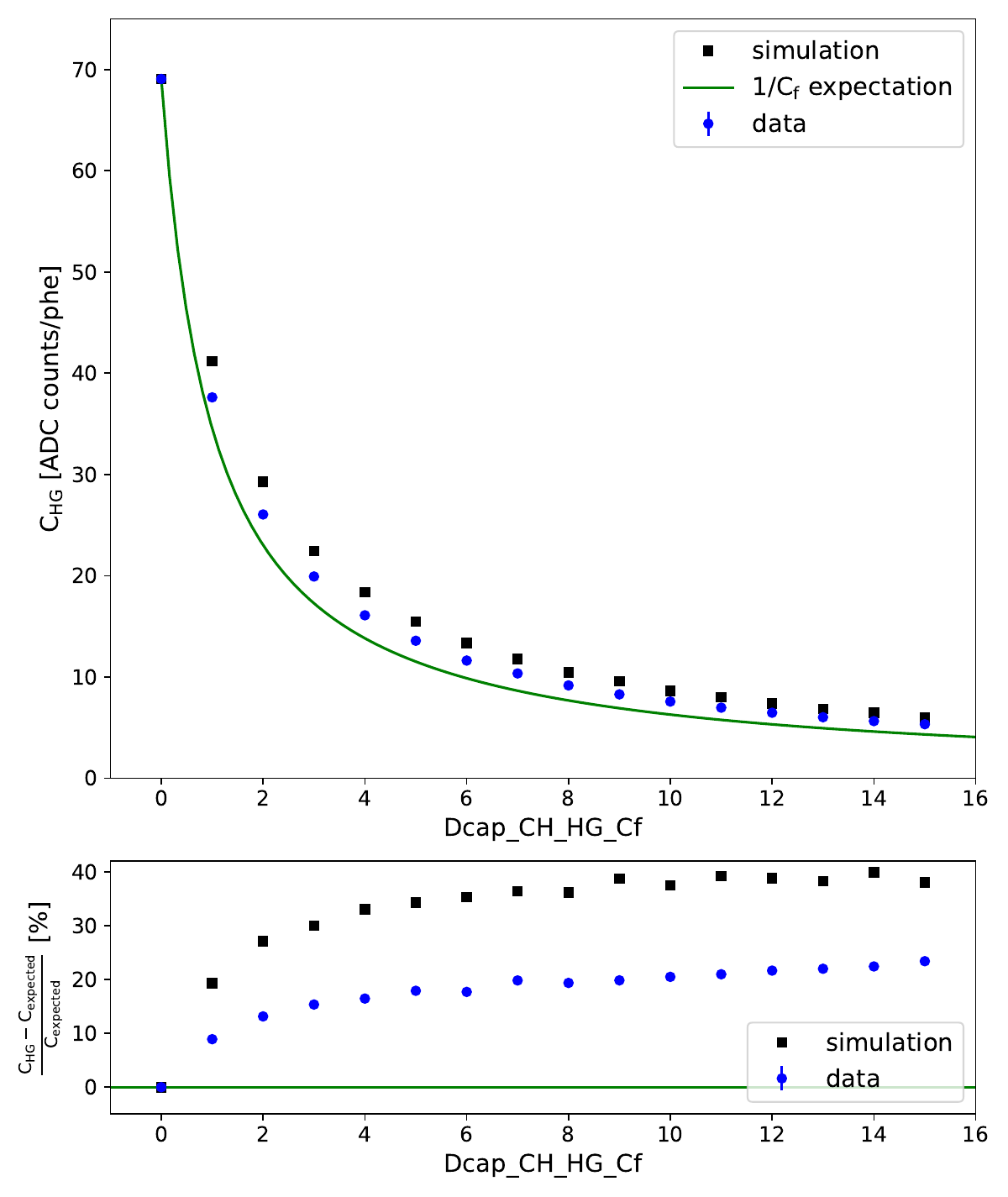}
\caption{\label{fig:Calibrations_HG} BETA-16 ASIC16, Ch07. Top: calibration factors (blue data points) measured for one representative SiPM of the S13552-10 array as a function of the $16$ HG-gain settings of the BETA ASIC. Statistical uncertainties are not visible because of the plot's scale. The green curve indicates the expected $1/C_f$ scaling, normalized to the calibration factor measured for HG00, while the black data points the calibration factors derived from simulations. Bottom: relative deviation of the measured (and simulated) calibration factors from the expected scaling.}
\end{figure}

\begin{figure}[htbp]
\centering
\includegraphics[width=0.55\columnwidth]{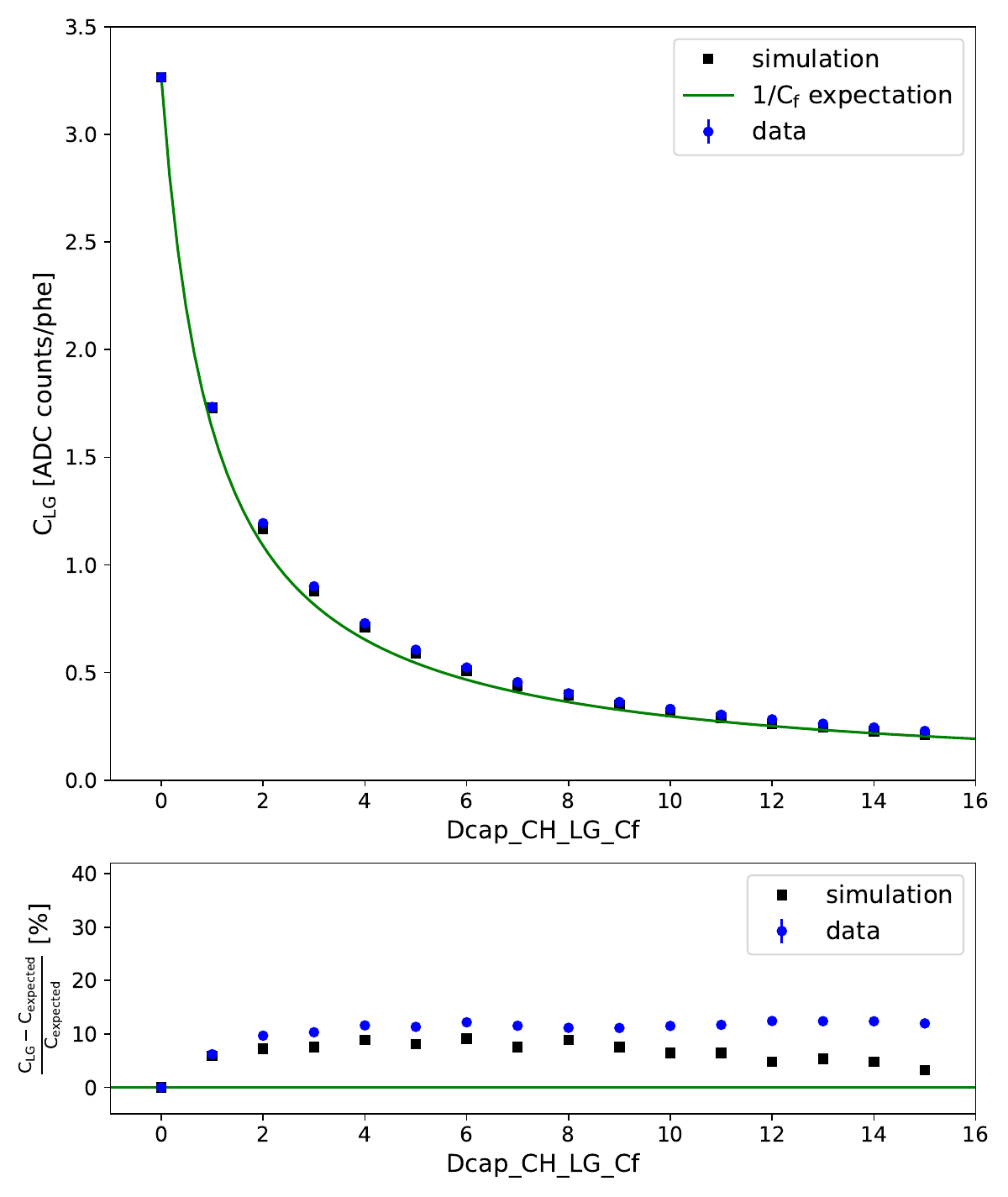}
\caption{\label{fig:Calibrations_LG} Same as in Figure~\ref{fig:Calibrations_HG}, for the 16 LG gain settings of the same BETA ASIC.}
\end{figure}

The BETA ASIC provides a total of $32$ gain configurations: $16$ settings for the HG path and $16$ for the LG path. To extend the calibration across all gain settings, a relative calibration procedure was implemented: for each consecutive gain configuration, the "LED-on" signals (after pedestal subtraction) recorded at a fixed LED intensity were compared. This approach allows for cross-calibration between different gain stages even when individual photopeaks are no longer resolved. For each given LED intensity, the mean of the ADC distribution was extracted to compute the signal strength. The calibration factor for a given gain $i$, $C_i$, is then computed as:
\begin{equation}
C_i = C_{i-1} \cdot \frac{S_i - P_i}{S_{i-1} - P_{i-1}}\, ,
\end{equation}
where $S$ and $P$ denote the signal and pedestal mean ADC values, respectively.

\begin{figure}[h]
\centering
\includegraphics[width=0.75\columnwidth]{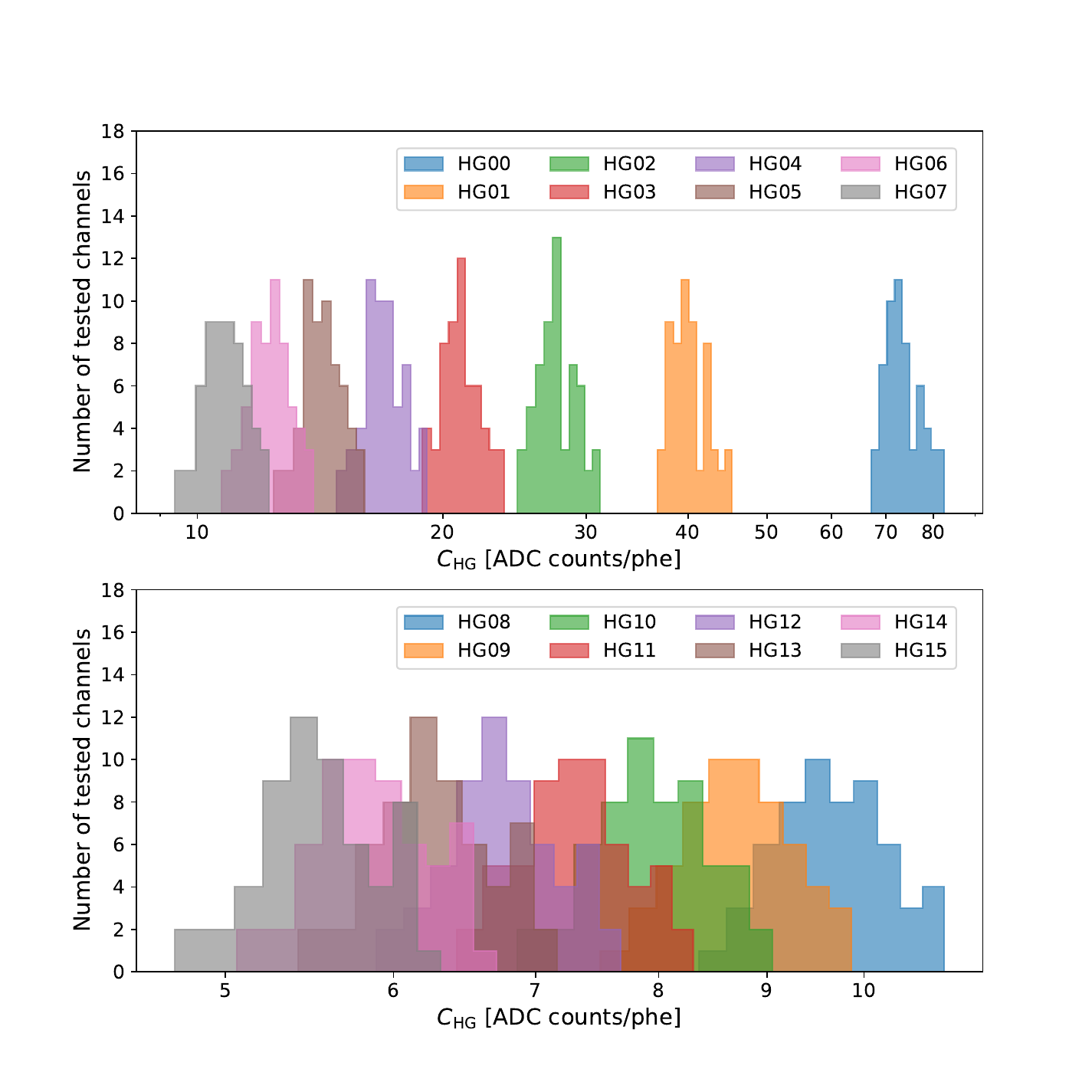}
\caption{\label{fig:CalibrationHisto_HG} Distribution of the calibration factor $C$ for the HG path across the tested BETA ASICs.}
\end{figure}

\begin{figure}[h]
\centering
\includegraphics[width=0.75\columnwidth]{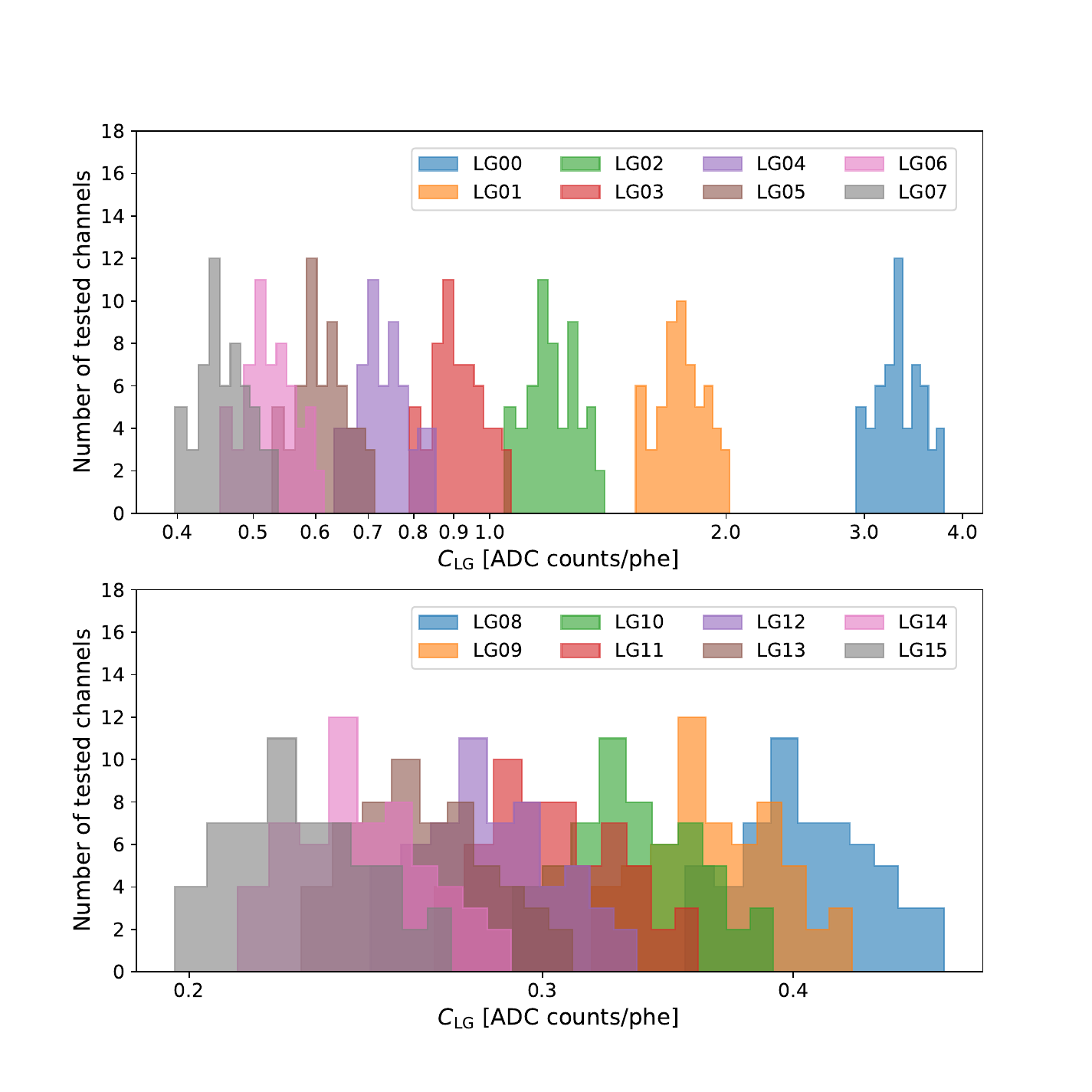}
\caption{\label{fig:CalibrationHisto_LG} Same as Figure~\ref{fig:CalibrationHisto_HG} for the LG path.}
\end{figure}

The calibration factors obtained for all $16$ HG and LG settings are shown in Figure~\ref{fig:Calibrations_HG} and Figure~\ref{fig:Calibrations_LG}, respectively, for one representative SiPM of the S13552-10 array. In both gain paths, deviations from the ideal $1/C_f$ scaling are observed. For HG, the deviations exceed $20\%$, while for the LG path they typically range between $10$ and $15\%$. The measured data are also compared to the gain values obtained from simulations and reported in the BETA datasheet. The observed mismatches are consistent with the expected process variations in ASIC fabrication and with parasitic effects associated with the switched-capacitor implementation of the preamplifier network. In particular, the gain selection relies on an array of switched capacitor elements whose effective capacitance is affected by manufacturing mismatch and parasitic contributions from the pass-gate switches used to connect the individual capacitors. Since the relative deviation is measured with respect to HG00 (and LG00), where only the minimum feedback capacitance is connected, the impact of parasitic switch capacitances becomes increasingly significant for the following gain settings, leading to the larger relative deviations observed in the measurements. Comparable results were obtained in independent measurements performed with direct charge injection, as discussed in Section~\ref{sec:charge_injection}. Figure~\ref{fig:CalibrationHisto_HG} and Figure~\ref{fig:CalibrationHisto_LG} show the distributions of the calibration factors measured across all characterized ASICs.

At the lowest LG setting, the system's dynamic range extends to approximately $8000\,\mathrm{phe}$. In the FIT, charged particles traversing the fibers produce scintillation light that is collected and detected by the coupled SiPMs. A MIP is expected to generate a signal of approximately $10\,\mathrm{phe}$. According to the Bethe–Bloch formula, and neglecting any saturation effect, the deposited energy - and hence the scintillation light output - scales with the square of the particle's charge. Under this assumption, an iron nucleus (Z = $26$) would produce a signal of order $10 \cdot 26^2 = 6760\,\mathrm{phe}$. The measured dynamic range of the BETA ASIC therefore enables reliable detection and charge discrimination of high-Z particles in the FIT tracker.

\subsection{BETA ASIC calibration and linearity measurements with charge injection}
\label{sec:charge_injection}

\begin{figure}[htbp]
\centering
\includegraphics[width=0.7\columnwidth]{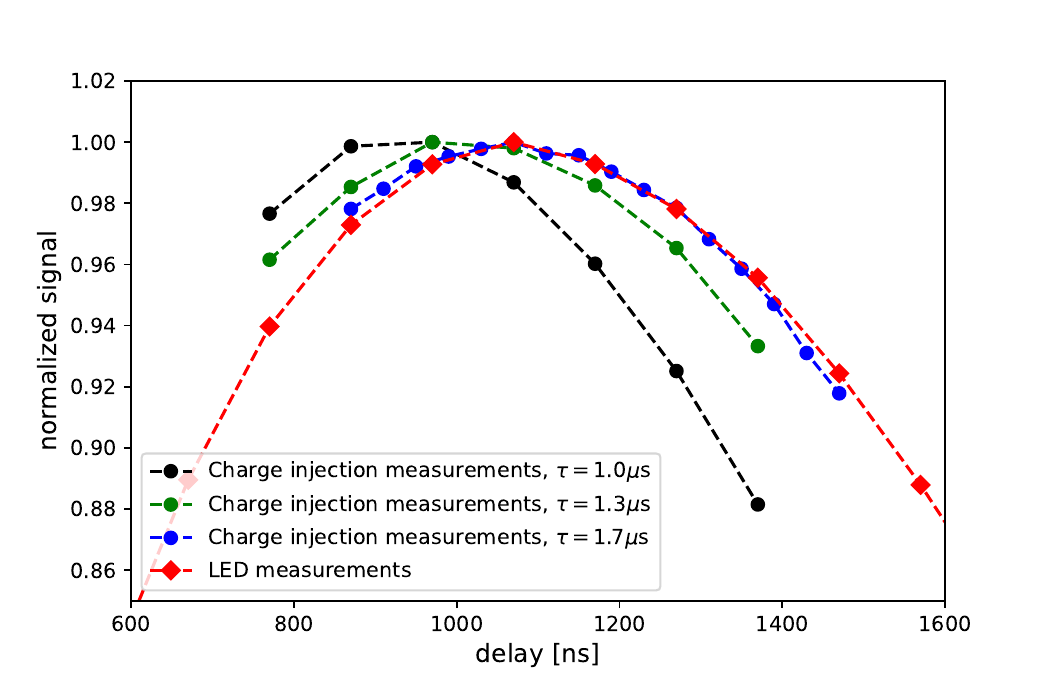}
\caption{\label{fig:TimingComparison} Comparison of the mean signal (normalized to its peak amplitude) obtained by injecting an exponential pulse directly into the BETA ASIC with that measured under LED illumination of the S13552-10 array, as a function of the sampling time. The different curves corresponding to the charge injection measurements are for different values of the decay time $\tau$. A decay time $\tau = 1.7\,\mathrm{\mu s}$ approximately reproduces the temporal response observed under LED illumination of the S13552-10 array.}
\end{figure}

The linearity of the BETA ASIC readout was independently evaluated by injecting electrical pulses directly into the chip. An arbitrary waveform generator, AWG5062,\footnote{\url{https://www.activetechnologies.it/products/signal-generators/arbitrary-waveform-generators/arb-rider-awg-5000/}.} was used to produce pulses with an exponential decay resembling the typical response of SiPM signals. The decay time constant was set to $\tau = 1.7\,\mathrm{\mu s}$, since this value approximately reproduces the temporal response observed under LED illumination of the S13552-10 array. Figure~\ref{fig:TimingComparison} compares the signal response as a function of the internal delay setting (see Appendix~\ref{sec:SamplingTime} for more detials about the internal delay configuration) obtained with LED illumination and with charge injection measurements using exponential input pulses with different decay time constants.

\begin{figure}[htbp]
\centering
\includegraphics[width=0.7\columnwidth]{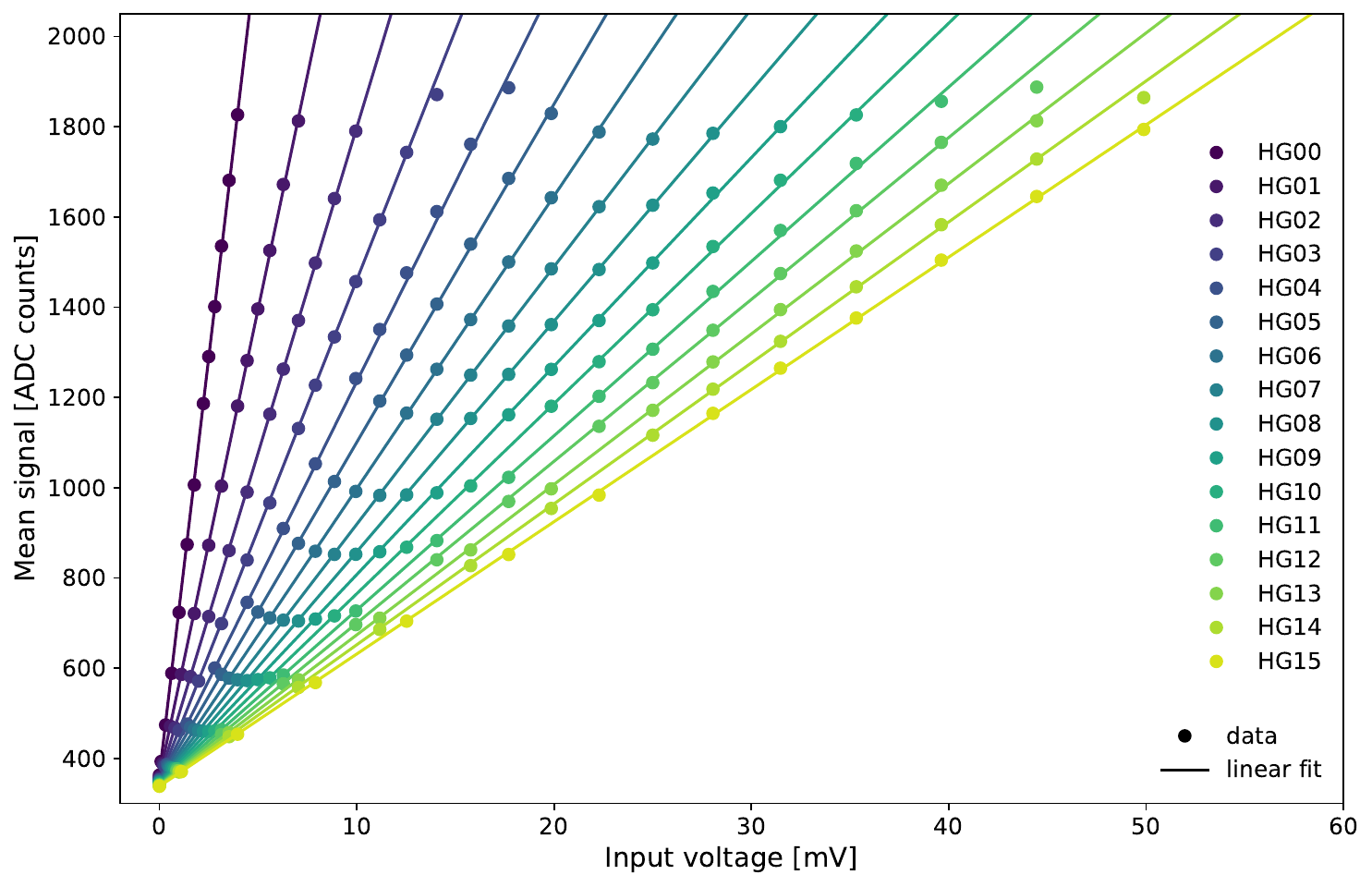}
\caption{\label{fig:Injector_Measurement_HG} BETA-16 ASIC16, Ch07. Relationship between the peak amplitude of the injected signal and mean ADC signal recorded in the BETA ASIC. The measurements span the entire range of HG path, from HG00 to HG15.}
\end{figure}

\begin{figure}[htbp]
\centering
\includegraphics[width=0.7\columnwidth]{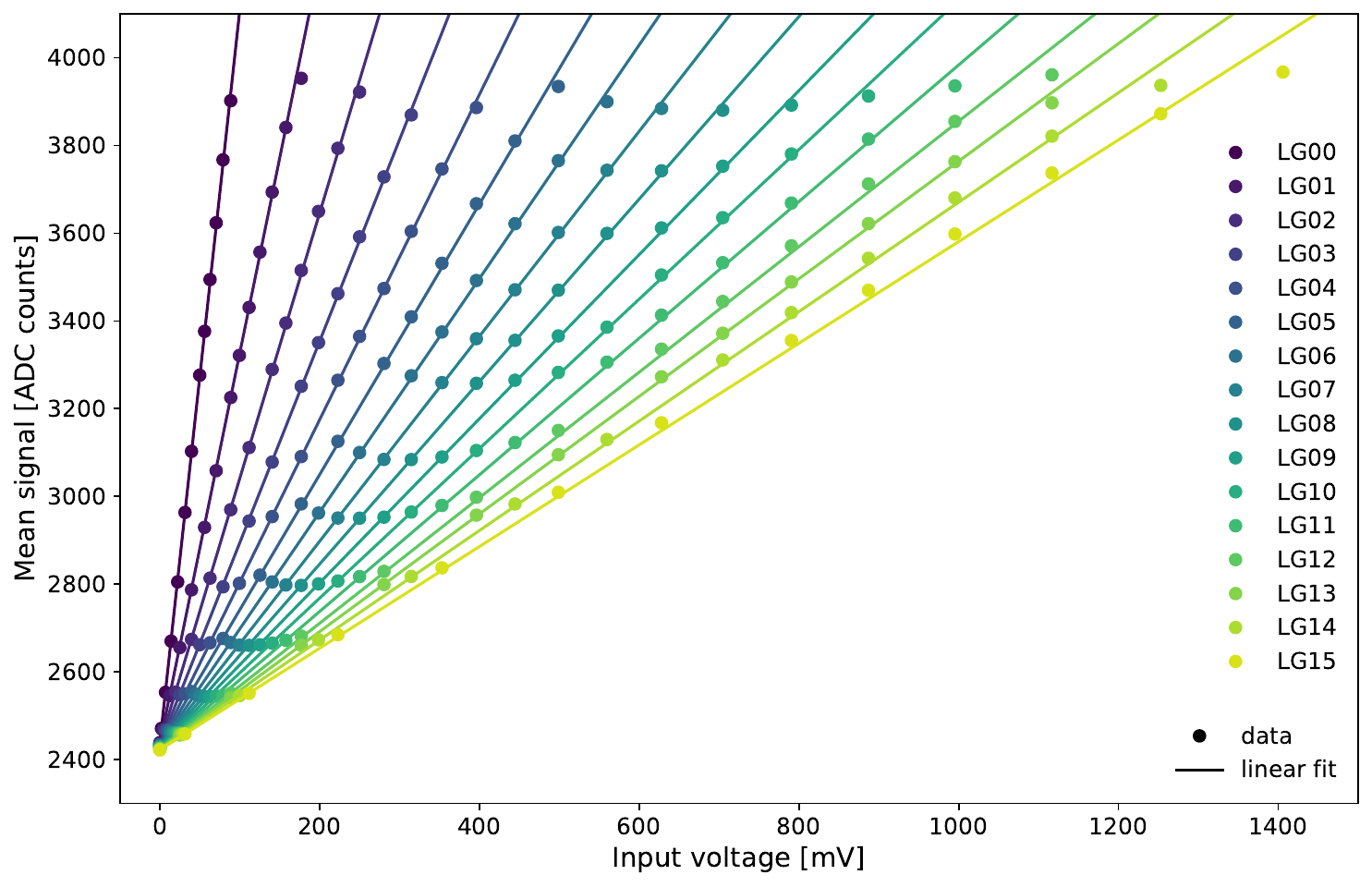}
\caption{\label{fig:Injector_Measurement_LG} Same as Figure~\ref{fig:Injector_Measurement_HG}, for LG path, from LG00 to LG15.}
\end{figure}

The generated waveforms were sent to a Keysight 11713D attenuator driver,\footnote{\url{https://www.keysight.com/us/en/product/11713D/11713d-attenuator-switch-driver.html}.} controlling two cascaded attenuators. The attenuators allow adjustment of the signal amplitude over a wide dynamic range, a key requirement to probe all gain configurations available in the BETA ASIC. Specifically, a Keysight 8496H attenuator provided coarse attenuation from $0\,\mathrm{dB}$ to $110\,\mathrm{dB}$ in $10\,\mathrm{dB}$ steps, while a second attenuator, 8494H, allowed fine adjustments in $1\,\mathrm{dB}$ steps, spanning a range from $0\,\mathrm{dB}$ to $11\,\mathrm{dB}$. The signal is injected into the BETA ASIC through a $220\,\Omega$ input resistor.\\
Compared to the measurements performed under LED illumination, this method extends the linearity study to all gain settings of the BETA ASIC, including those for which individual photoelectrons are no longer resolvable in the optical setup. Moreover, this approach isolates intrinsic electronic non-linearity of the BETA ASIC, from potential distortions introduced by the silicon photomultipliers response or by the light source. This setup can also be tailored for an automatic scan to quickly test each readout channel without requiring optical excitation from an LED or laser source.

\begin{figure}[htbp]
\centering
\includegraphics[width=0.85\columnwidth]{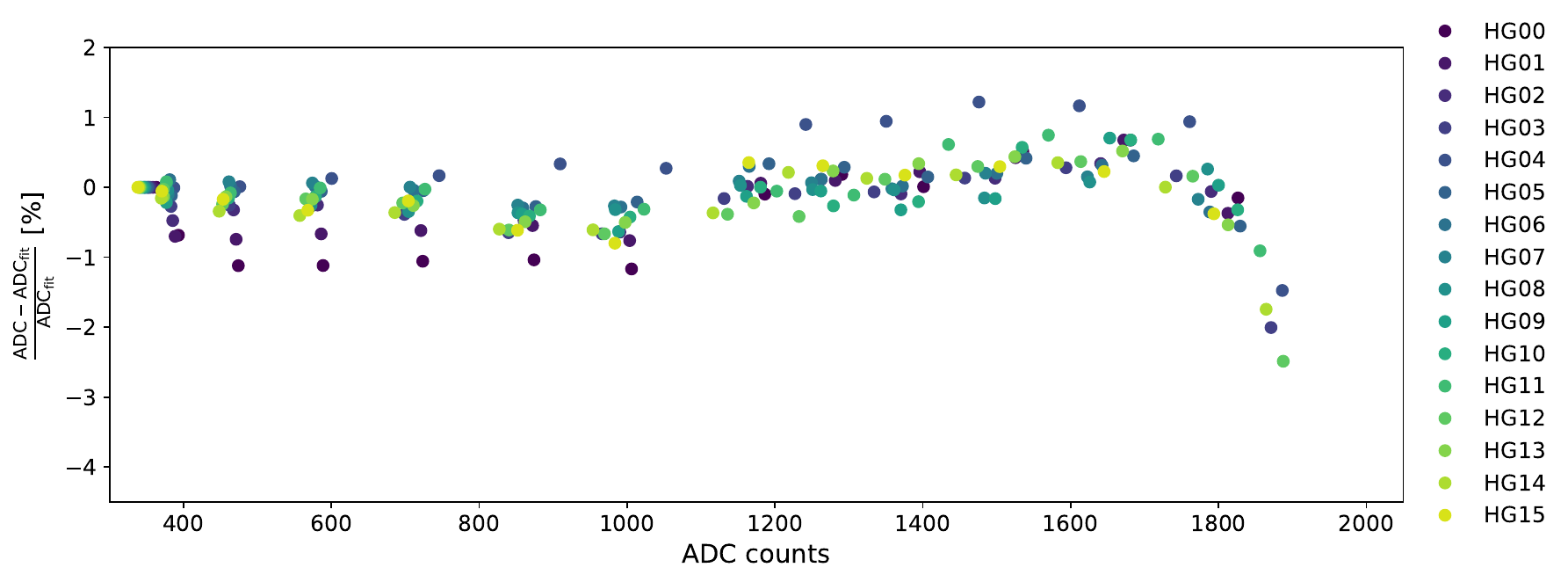}
\caption{\label{fig:LinearityError_ChargeInjection_HG} BETA-16 ASIC16, Ch07. Non-linearity measured for all HG settings, from HG00 to HG15.}
\end{figure}

\begin{figure}[htbp]
\centering
\includegraphics[width=0.85\columnwidth]{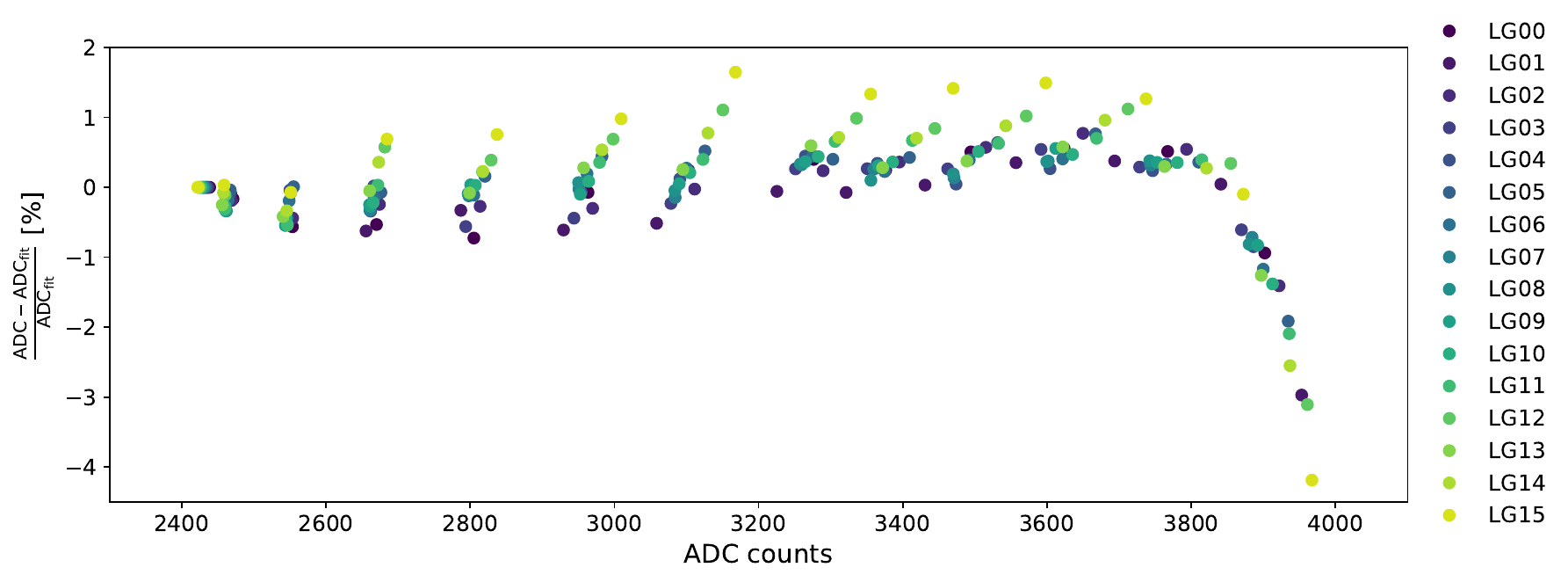}
\caption{\label{fig:LinearityError_ChargeInjection_LG} Same as Figure~\ref{fig:LinearityError_ChargeInjection_HG}, for all LG settings, from LG00 to LG15.}
\end{figure}

\begin{figure}[htbp]
\centering
\includegraphics[width=0.75\columnwidth]{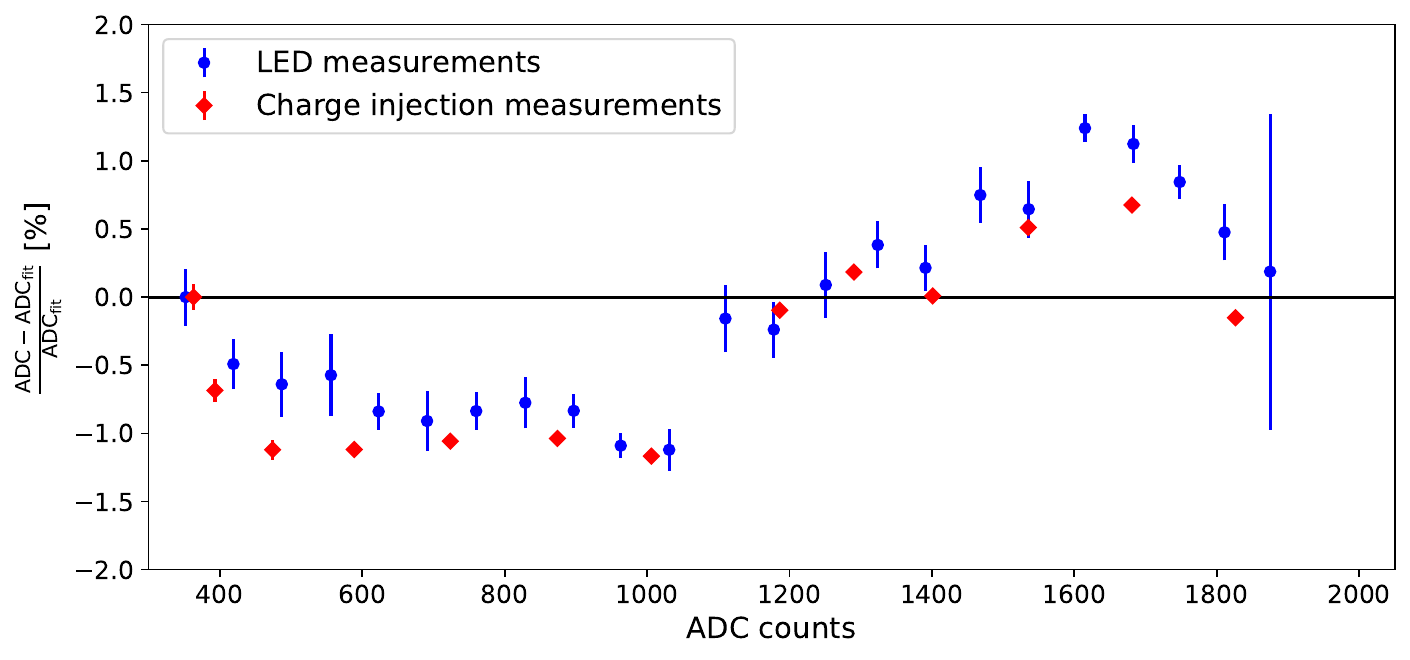}
\caption{\label{fig:LinearityError_comparison} BETA-16 ASIC16, Ch07. Comparison of the non-linearity measured independently using LED illumination and direct charge injection into the BETA ASIC for HG00.}
\end{figure}

Figure~\ref{fig:Injector_Measurement_HG} and Figure~\ref{fig:Injector_Measurement_LG} show the mean ADC recorded signal as a function of the injected pulse's peak amplitude for all HG and LG gain configurations of the BETA ASIC. Data were fitted with the same procedure described in Section~\ref{sec:calibration}, using Equation~\ref{eq:fit_fixed_pedestal}. The corresponding non-linearity remains below $2\%$ for all gain settings, except for the highest injected amplitudes approaching ADC saturation, as shown in Figure~\ref{fig:LinearityError_ChargeInjection_HG} and  Figure~\ref{fig:LinearityError_ChargeInjection_LG}, in agreement with the results reported in \cite{BETAASIC}. A comparison between the non-linearities obtained with LED illumination and with charge injection is provided in Figure~\ref{fig:LinearityError_comparison} for HG00, from which we conclude that the bulk of the non-linearity is caused by the read-out electronics.

\begin{figure}[htbp]
\centering
\includegraphics[width=0.5\columnwidth]{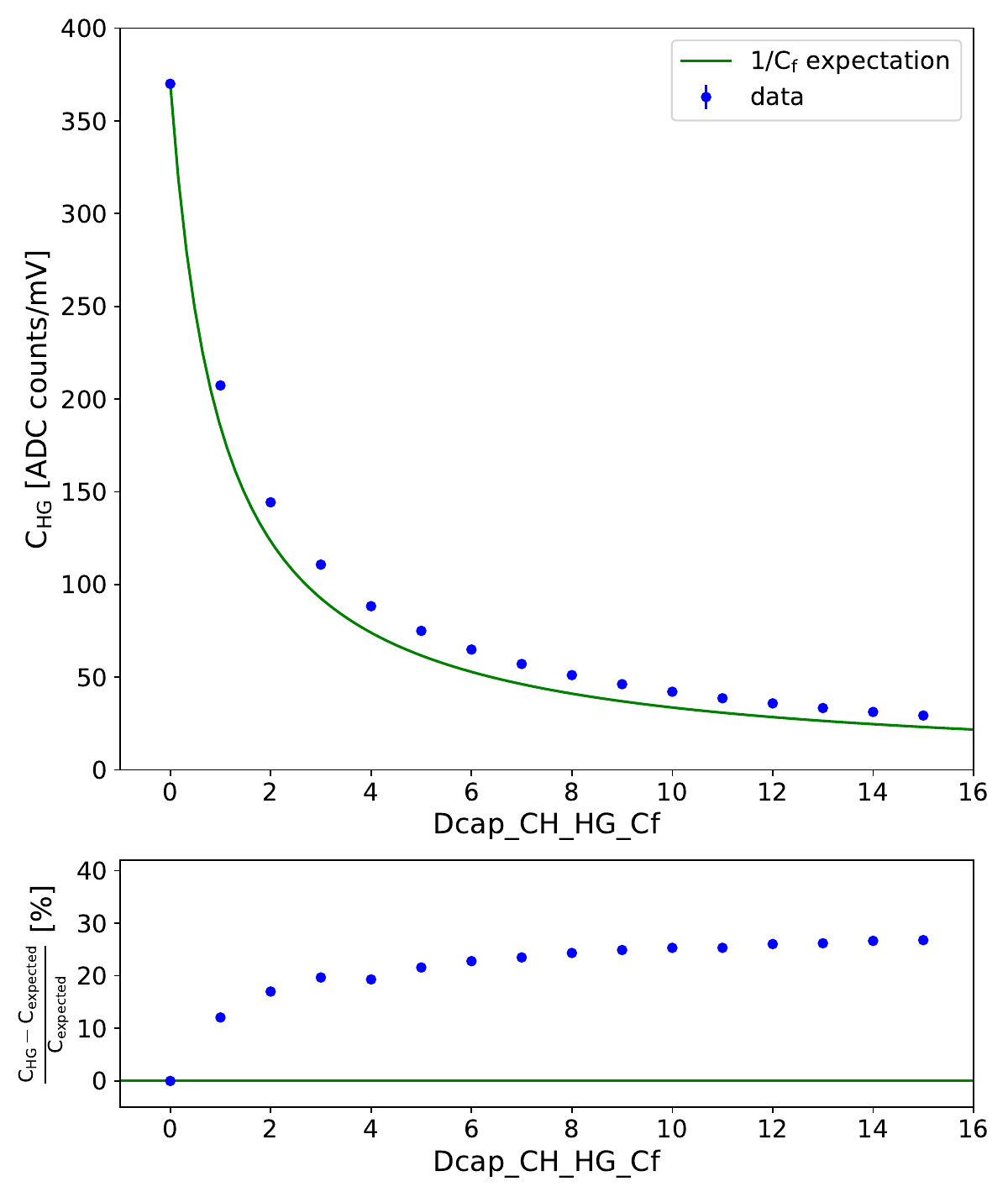}
\caption{\label{fig:ChargeInjection_Calibrations_HG} BETA-16 ASIC16, Ch07. Top: input-signal to ADC-response calibration factors measured via direct charge injection into the BETA ASIC, as a function of the $16$ HG-gain settings. Statistical uncertainties are not visible because of the plot's scale. The green curve indicates the expected $1/C_f$ scaling, normalized to the calibration factor measured for HG00. Bottom: relative deviation of the measured calibration factors from the expected scaling.}
\end{figure}

\begin{figure}[htbp]
\centering
\includegraphics[width=0.5\columnwidth]{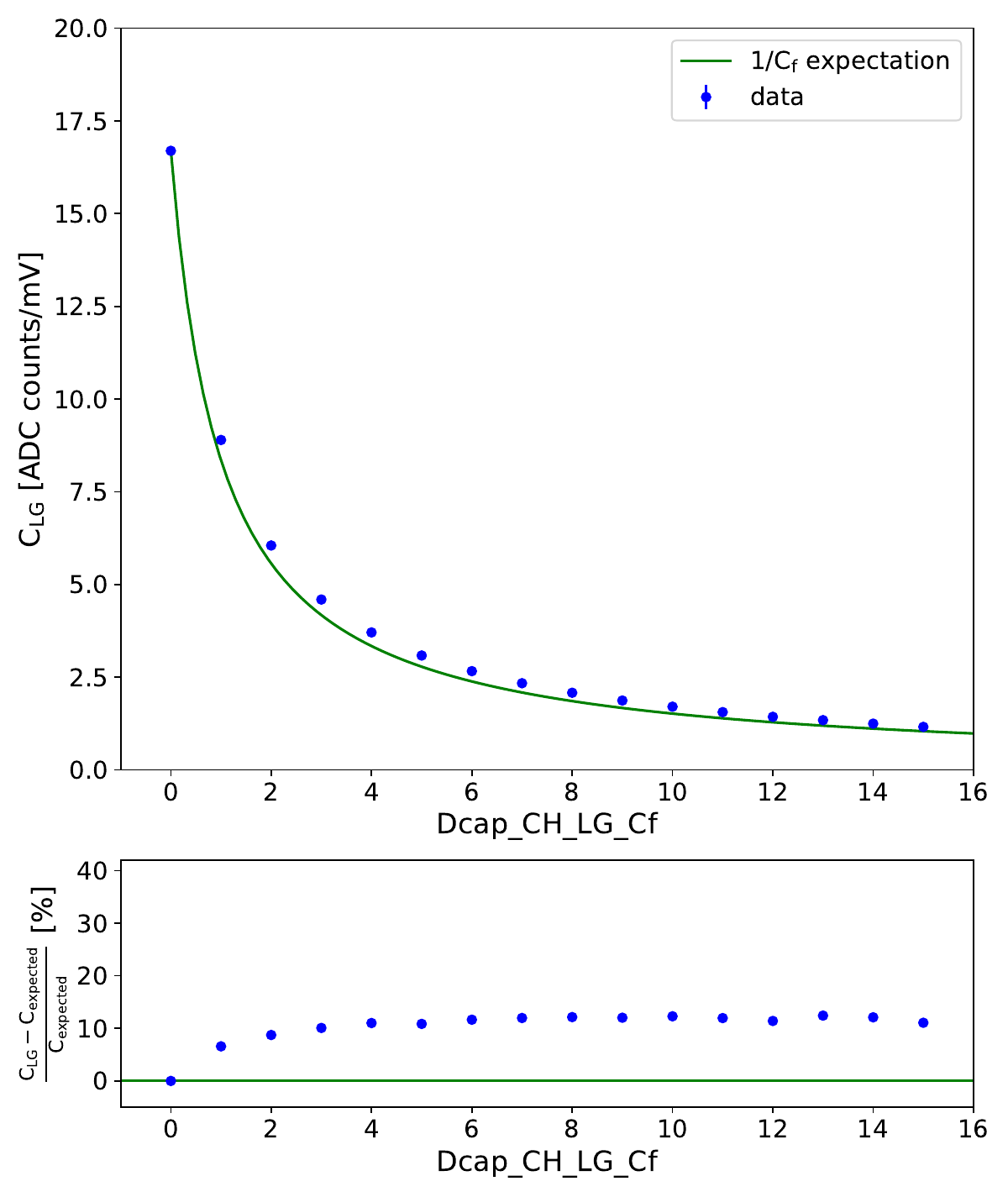}
\caption{\label{fig:ChargeInjection_Calibrations_LG} Same as Figure~\ref{fig:ChargeInjection_Calibrations_HG} for the $16$ LG-gain settings.}
\end{figure}

As already observed in the optical measurements, the calibration factors exhibit significant deviations from the ideal $1/C_f$ scaling expected for the charge-sensitive preamplifier, as shown in Figure~\ref{fig:ChargeInjection_Calibrations_HG} and Figure~\ref{fig:ChargeInjection_Calibrations_LG}. However, the ratios between calibration factors of consecutive gain settings obtained from charge injection and LED measurements agree within the experimental uncertainties, demonstrating that both methods provide an accurate relative gain calibration of the BETA ASIC. Therefore, charge injection measurements, combined with a single direct ADC-to-phe calibration for the gain setting HG00, provide a full calibration of the readout chain in phe units, as an alternative to the method described in Section~\ref{sec:relative_calibration}.

\section{Trigger threshold response}
\label{sec:thr_calibration}
\subsection{Threshold calibration}
Each channel of the BETA ASIC includes a comparator that generates a trigger whenever the HG preamplifier output exceeds a user-defined threshold. A dedicated calibration of the discriminator thresholds was performed for each BETA ASIC channel in order to establish a robust procedure for setting trigger levels corresponding to a well-defined number of photoelectrons. The calibration was performed under dark conditions, allowing for a rapid calibration of the setup without the requirement of illuminating each channel of the BETA ASICs individually. The HG preamplifier was set at its maximum gain setting, HG00, which provides the highest signal resolution. During the measurements, the count rates for each individual BETA ASIC pre-trigger were recorded while varying the \texttt{D\textunderscore SELFTRIGV\textunderscore CHx} register values, which are converted by an internal digital-to-analog converter (DAC) into the threshold voltage applied to the trigger comparator of channel x. Since the internal pre-trigger signal of the BETA ASIC is generated as the OR logical combination of all individual channel discriminated output, only one channel per time was enabled during the measurements, to characterize each channel independently and minimize channel-to-channel variations.

\begin{figure}[ht]
\centering
\includegraphics[width=0.7\columnwidth]{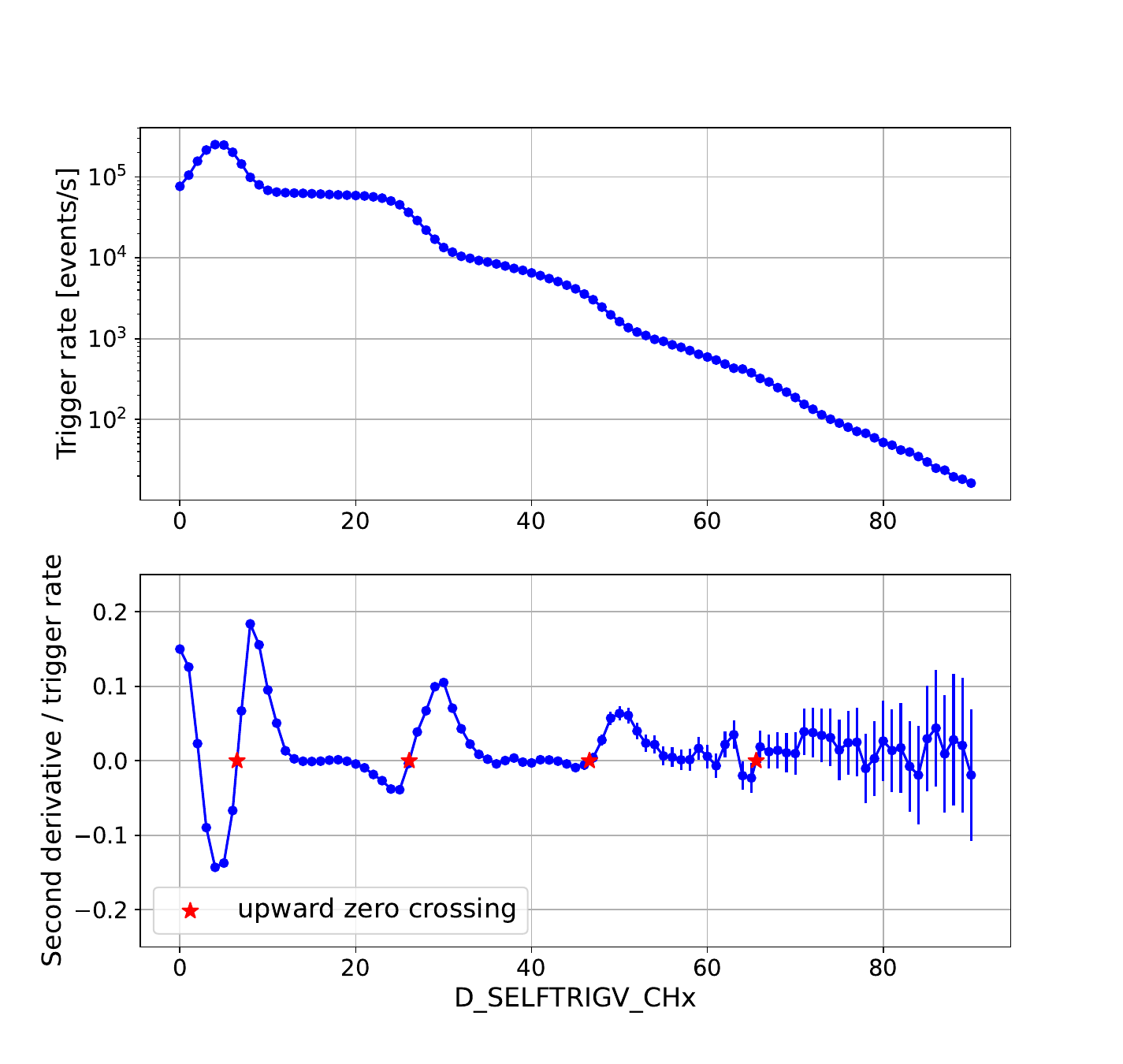}
\caption{\label{fig:Thr_staircase} Top: trigger count rate as a function of the discriminator threshold (\texttt{D\textunderscore SELFTRIGV\textunderscore CHx}) for a representative BETA ASIC channel. Bottom: normalized second derivative of the trigger count rate, as a function of the discriminator threshold \texttt{D\textunderscore SELFTRIGV\textunderscore CHx}. Upward zero crossings correspond to integer number of photoelectron threshold levels.}
\end{figure}

\begin{figure}[ht]
	\centering
	\subfloat{
	\includegraphics[width=0.48\columnwidth]{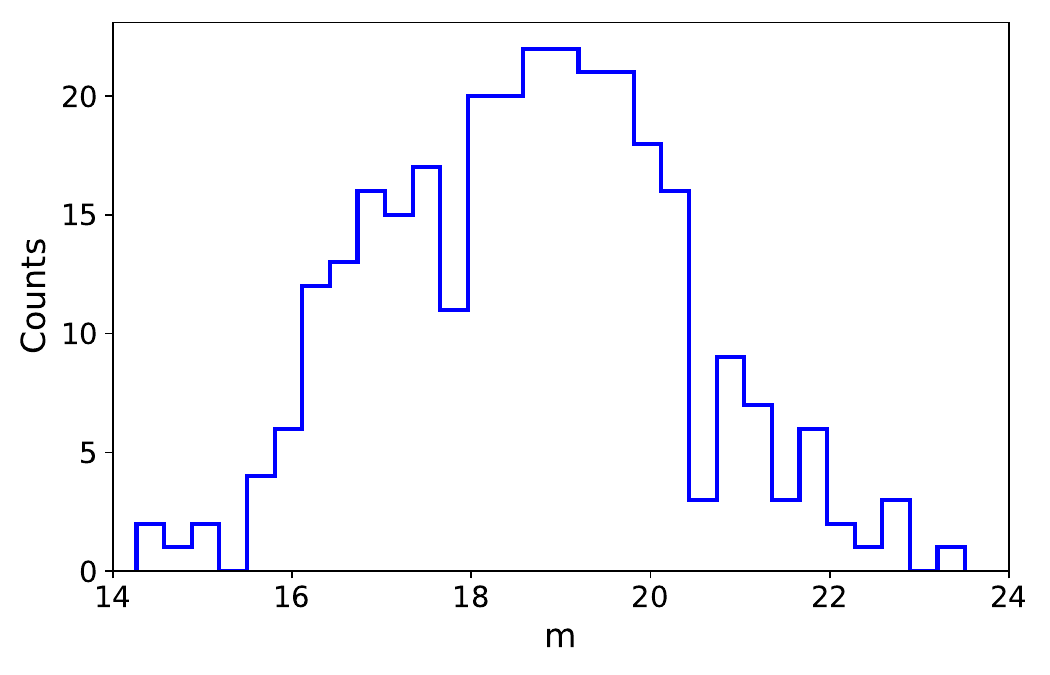}
	\label{fig:mvalue}}
	\hspace{-0.05cm}
	\subfloat{
	\includegraphics[width=0.48\columnwidth]{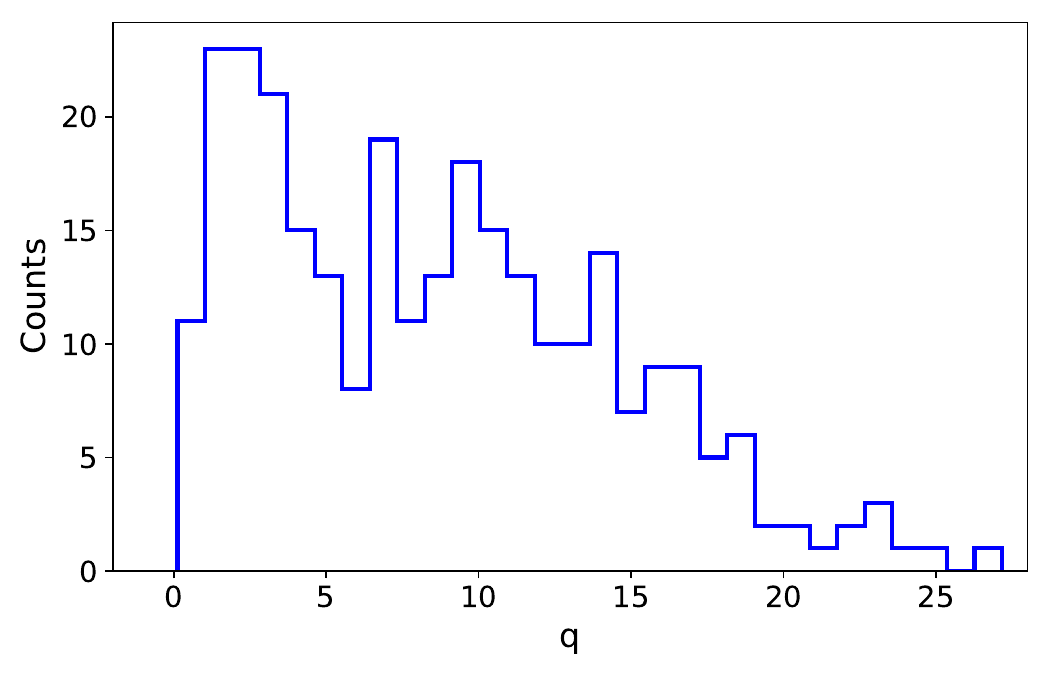}
	\label{fig:qvalue}}
	\caption{\label{fig:Thr_parameters} Fitted parameters m (left) and q (right) obtained in the threshold calibration performed for all channels of all BETA-16 ASICs.}
\end{figure}

An example of the trigger rate evolution as a function of the discrimination threshold, commonly referred to as a "staircase plot" \cite{Eckert_StairCase, Balagura_StairCase}, is shown in Figure~\ref{fig:Thr_staircase}. As the threshold increases, the cumulative count rate decreases in discrete steps (smeared out by electronic noise and afterpulses) corresponding to integer numbers of photoelectrons. Well-defined steps are identified for the first few photoelectrons, while at higher thresholds the step-like structure gradually disappears, and the rate decreases approximately exponentially with the threshold level. Discrete photoelectron thresholds were identified by analyzing the derivatives of the trigger rate curve. Specifically, the local minima of the first derivative or, equivalently, the upward zero crossings of the second derivative are commonly used for calibration of discrete phe levels (see the lower panel of Figure~\ref{fig:Thr_staircase}). The linearity of the threshold comparator register was evaluated for a few representative channels in Figure~\ref{fig:Thr_fit}. The associated statistical uncertainties in the threshold levels are computed as described in Appendix~\ref{sec:error_thr}. Data are fitted by a linear regression, parameterized by a slope $m$ and an intercept $q$; the fitted parameters provide the conversion between the \texttt{D\textunderscore SELFTRIGV\textunderscore CHx} discriminator register value and the equivalent phe number. The fitted parameters $m$ and $q$ obtained for all channels of all tested ASICs are shown in the histograms of Figure~\ref{fig:Thr_parameters}. The dispersions in m values arise from small channel-to-channel variations in gain, while the dispersion in q value is due to differences in baseline levels between channels.

\begin{figure}[ht]
	\centering
	\subfloat[BETA-16 ASIC02, Ch01.]{
	\includegraphics[width=0.48\columnwidth]{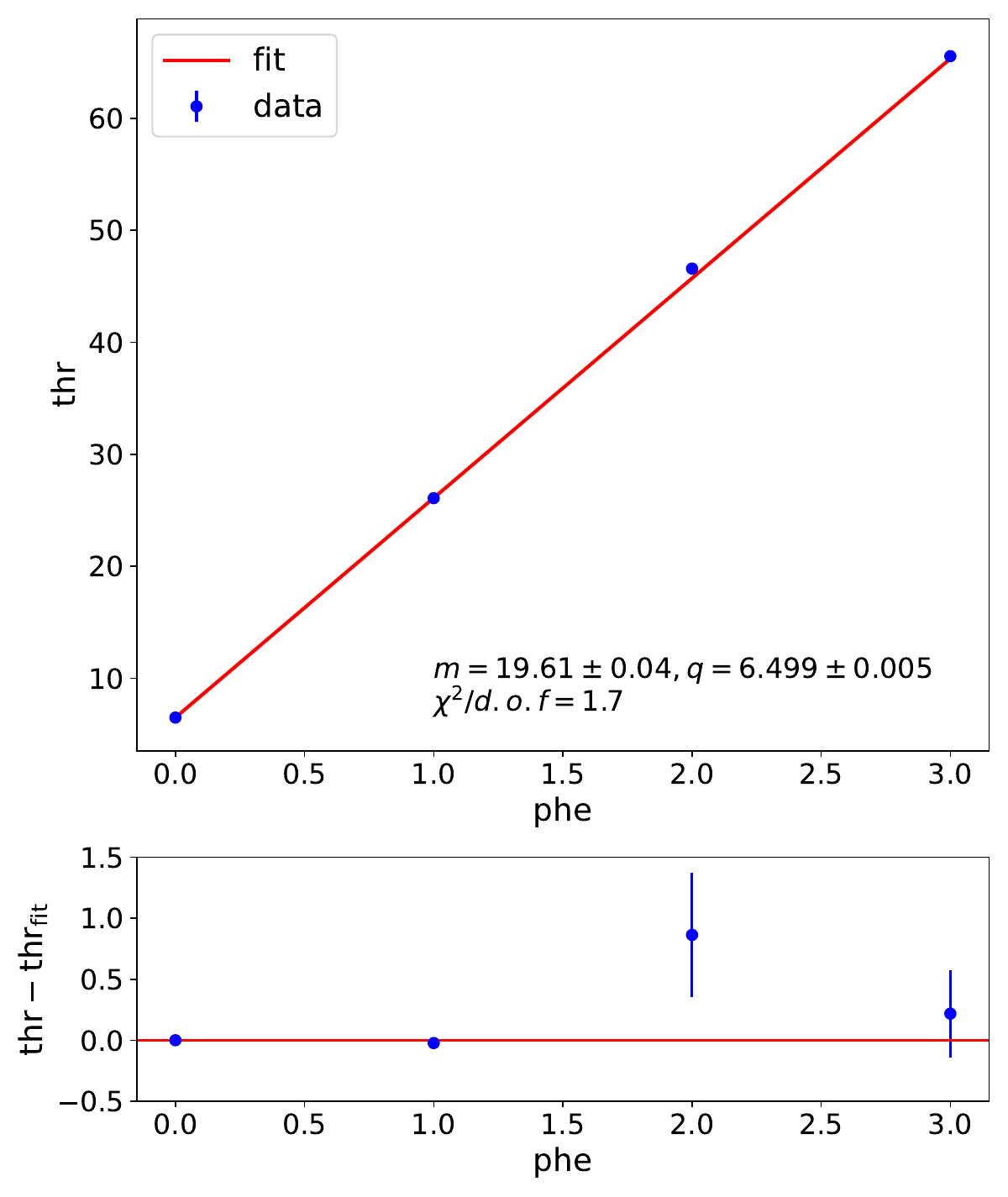}
	\label{fig:Thr_Ch01_ASI02}}
	\hfill
	\subfloat[BETA-16 ASIC15, Ch01.]{
	\includegraphics[width=0.48\columnwidth]{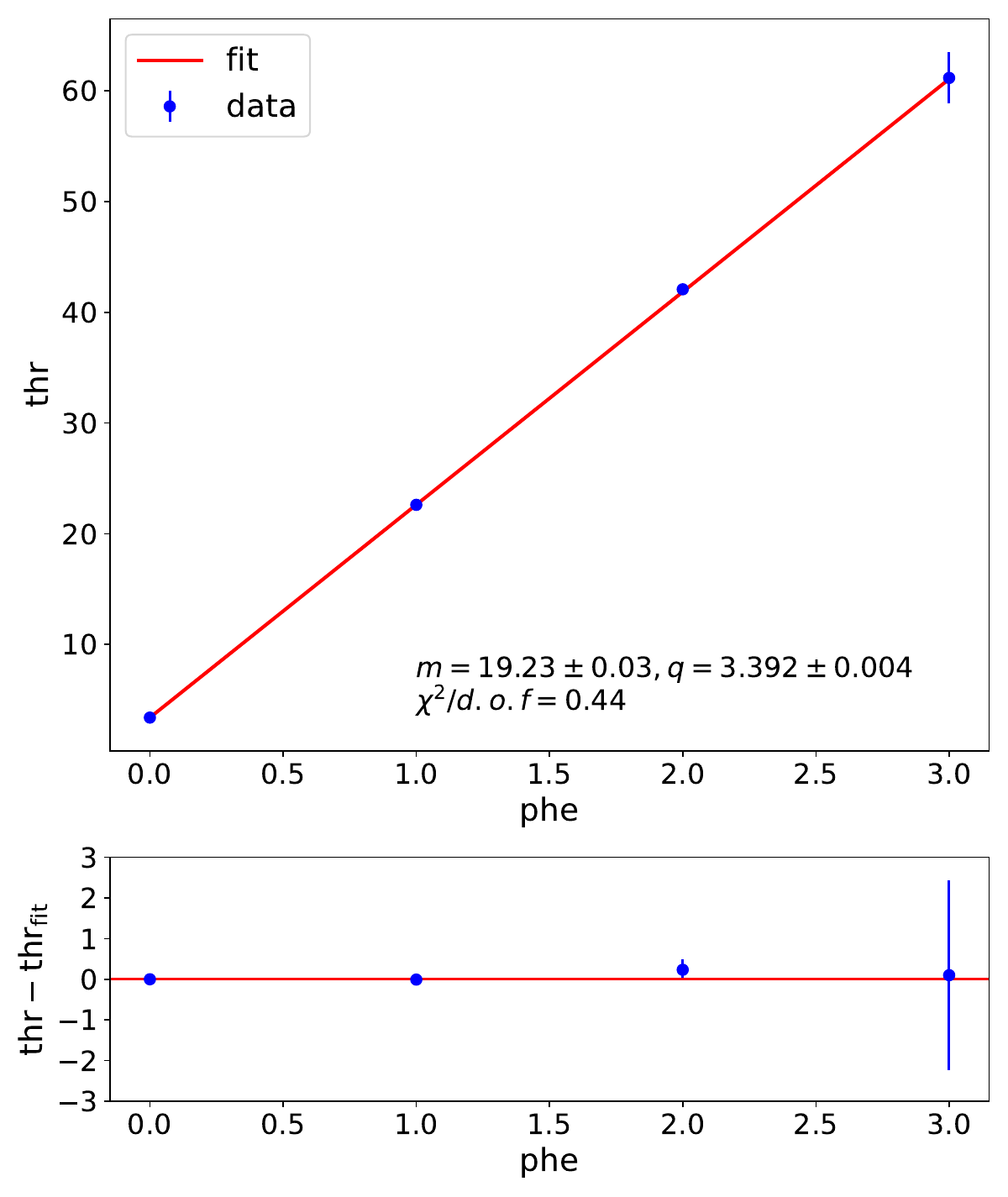}
	\label{fig:Thr_Ch01_ASIC15}}
	\vspace{0.5pt}
	\centering
	\subfloat[BETA-16 ASIC08, Ch00.]{
	\includegraphics[width=0.48\columnwidth]{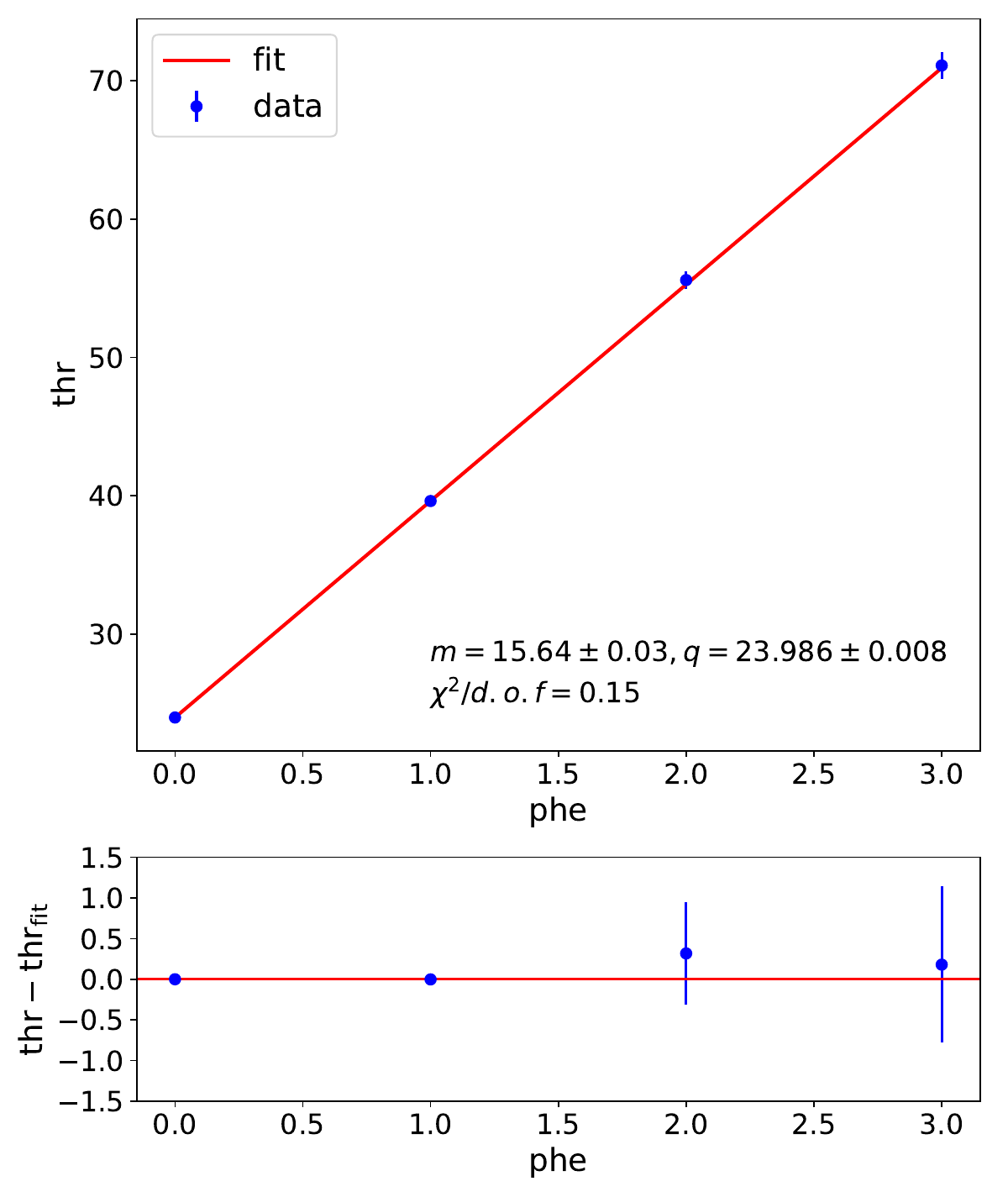}
	\label{fig:Thr_Ch00_ASIC08}}
	\hfill
	\subfloat[BETA-16 ASIC17, Ch00.]{
	\includegraphics[width=0.48\columnwidth]{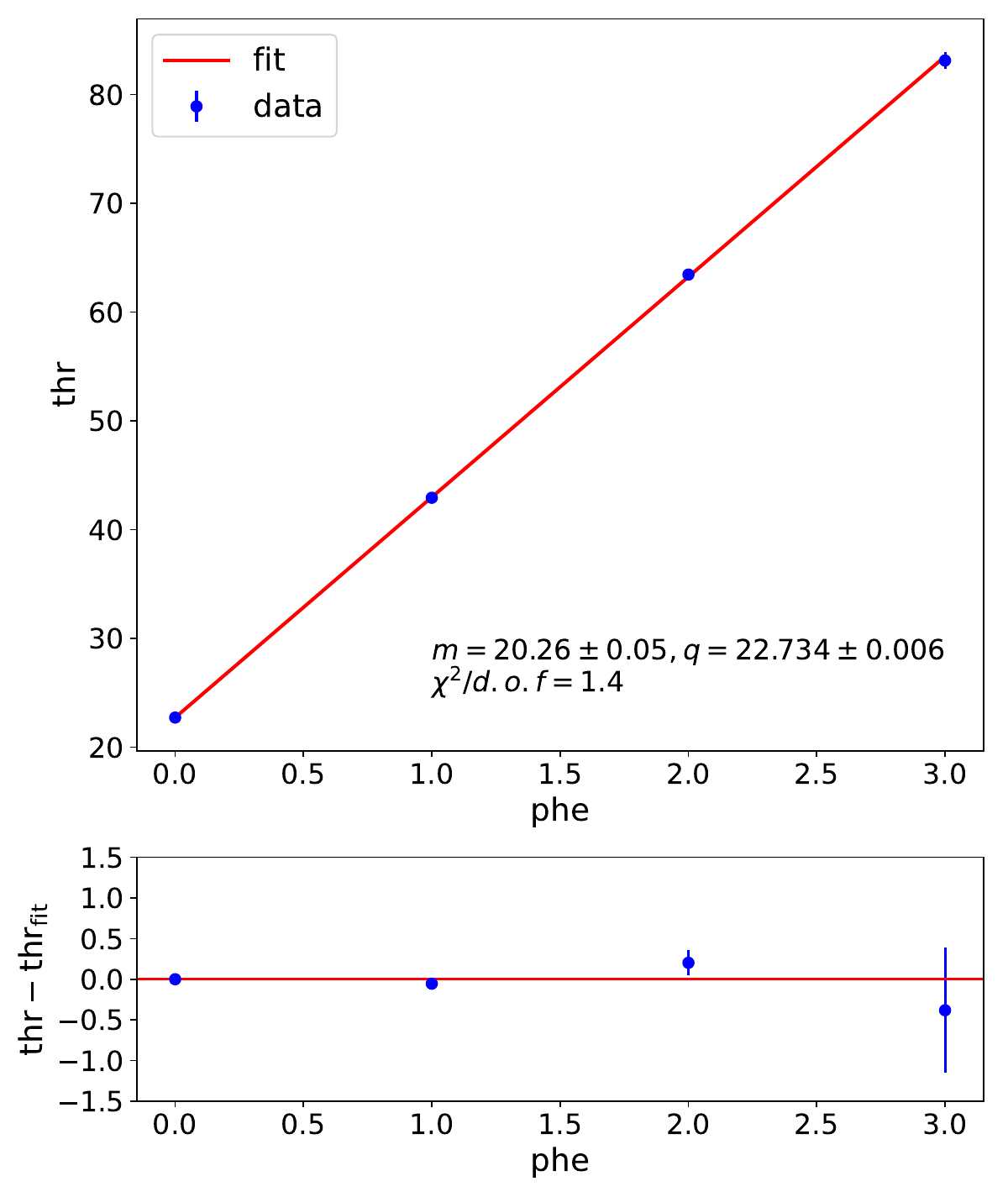}
	\label{fig:Thr_Ch00_ASIC17}}
	\caption{\label{fig:Thr_fit} Linearity of the threshold comparator register \texttt{D\textunderscore SELFTRIGV\textunderscore CHx} for a few representative channels.}
\end{figure}

For the S13552-10 arrays, this method allows a direct and reliable calibration generally up to approximately $3\,\mathrm{phe}$. At higher photoelectron numbers, the staircase structure becomes increasingly smeared, preventing an accurate identification of higher number of photoelectrons. Thresholds corresponding to higher phe numbers in HG00 can be obtained by linear extrapolation of these low-photoelectron calibration points. The achievable precision is limited by the statistical uncertainties of the fitting parameters and does not account for possible systemic effects. In particular the procedure assumes a perfectly linear response of the BETA charge-sensitive preamplifier. At lower gain settings, the reduced signal resolution further degrades the staircase structure, eventually preventing a reliable identification of single-photoelectron thresholds. However, assuming a linear response of the preamplifier and discriminator, the threshold corresponding to a given phe number is expected to scale proportionally with the gain setting, similarly to the calibration factors discussed in Section~\ref{sec:optical_measurements}, allowing us to set threshold levels in terms of phe numbers also for lower gain configurations.

\subsection{Crosstalk probability}
The staircase analysis enables an estimate of the SiPMs optical crosstalk probability. Since the probability of multiple independent thermal excitations occurring simultaneously is negligible, dark events with amplitudes exceeding one photoelectron are predominantly due to crosstalk-induced avalanches. The crosstalk probability is determined by comparing the dark count rate above a threshold level corresponding to one photoelectron with the total dark rate measured. Following the method described in~\cite{Eckert_StairCase}, the crosstalk probability is defined as:
\begin{equation}
P_\mathrm{CT} = \frac{r_\mathrm{1.5phe}}{r_\mathrm{0.5phe}} ,
\end{equation}
where $r_\mathrm{1.5phe}$ and $r_\mathrm{0.5phe}$ are the trigger count rates measured at thresholds corresponding to $1.5$ and $0.5$ photoelectrons, respectively. The corresponding threshold values were derived from the discriminator calibration shown in Figure~\ref{fig:Thr_fit}.

\begin{figure}[htbp]
\centering
\includegraphics[width=0.8\columnwidth]{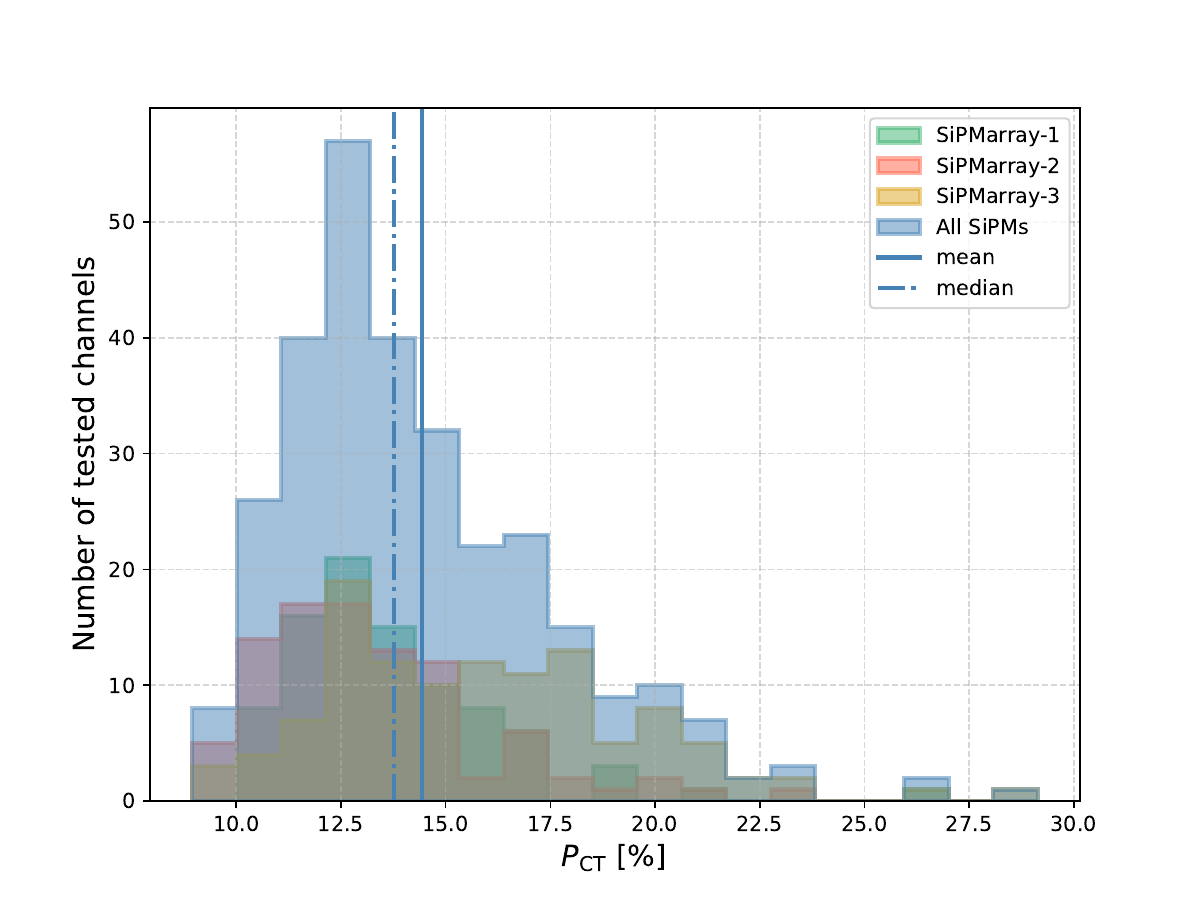}
\caption{\label{fig:Crosstalk_probability} Distribution of the optical crosstalk probability measured across  all tested channels of the three S13552-10 SiPM arrays available during the optical-bench tests.}
\end{figure}

The measured crosstalk probabilities for all tested SiPMs are shown in Figure~\ref{fig:Crosstalk_probability}. The mean crosstalk probability is $\sim 14\%$, with SiPMarray-3 exhibiting $\sim 20\%$ lower performance compared to the other two devices.

\section{Internal trigger logic and timing performance}
\label{sec:internal_trigger}
In the previous sections, the performance of individual S13552-10 SiPM arrays coupled to the BETA ASICs was characterized in terms of gain, linearity, and trigger response. In the following, we discuss the development and performance of a global internal trigger system integrated in the FIT detector. The internal trigger combines the information from multiple SiPM arrays and BETA ASICs pre-triggers to efficiently detect ionizing particles, while keeping the noise-induced trigger rate under control. In this context, the studies presented in this section are essential to evaluate the feasibility of a trigger architecture suitable for stable operation in tracking detectors.

\subsection{Internal trigger logic}
\begin{figure}[htbp]
\centering
\includegraphics[width=0.8\columnwidth]{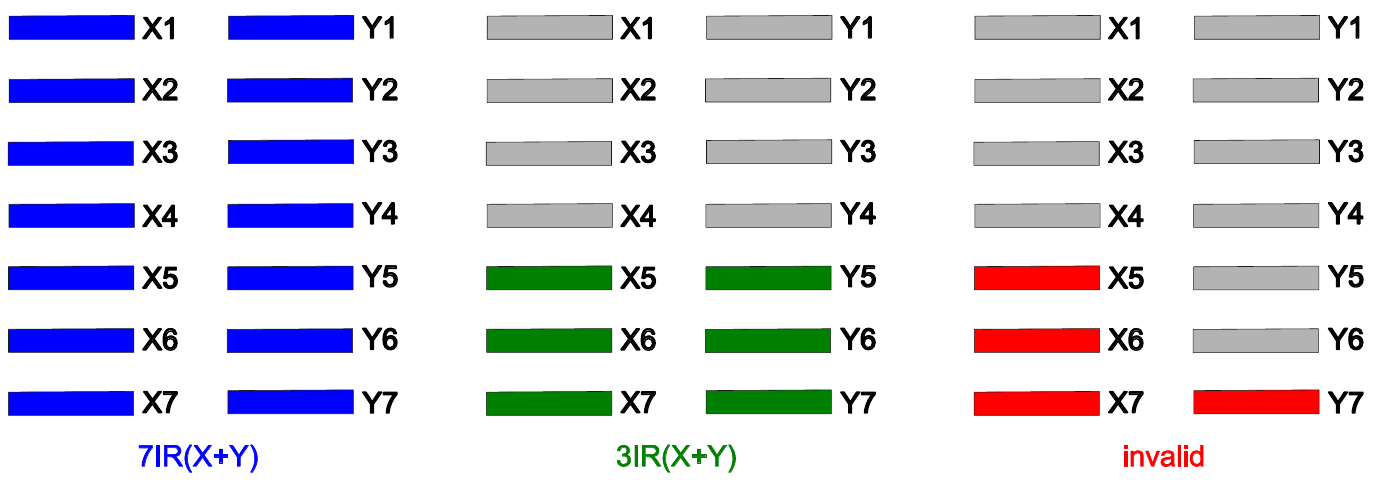}
\caption{\label{fig:Internal_trigger_schematics} Schematic representation of the main internal trigger logics implemented in the 2025 miniFIT test beam setup with seven instrumented layers per tracking direction (X and Y). Colored bars indicate layers contributing to the trigger decision. Blue: example of a valid 7IR(X+Y) trigger, requiring coincidence in all seven layers in both tracking directions. Green: example of a valid 3IR(X+Y) trigger, requiring coincidence in at least three consecutive layers in both directions with the same starting layer. Red: example of an invalid trigger that does not satisfy any implemented logic.}
\end{figure}

In tracking detectors, internal trigger logics based on coincidences between signals from consecutive layers are commonly employed to suppress noise-induced triggers while preserving a high detection efficiency for ionizing particles. The implementation of an effective trigger strategy is particularly critical for spaceborne instruments, where stringent constraints on data rate and telemetry require an efficient rejection of spurious signals at the hardware level.\\
In the FIT, an efficient internal trigger can be established by requiring pre-trigger signals in at least three consecutive layers in both orthogonal directions (the so-called three-in-a-row, or 3IR, trigger logic) \cite{Farina_3IRtrigger}. In the FPGA, combinational logic processes the pre-triggers generated by individual BETA ASICs, identifies patterns corresponding to hits in consecutive tracking layers, and generates the global internal trigger. For testing and debugging purposes, the global internal trigger information is encoded using $4$ bits, allowing the exploration of different combinations and triggering logic (e.g. different starting layers or a variable number of required layers). The generated internal trigger can be used both to initiate FIT data acquisition itself and, if required, for synchronization with external detectors via dedicated test points on the motherboard.\\
The definition of the global internal trigger logic has undergone several iterations, driven by successive prototype detector versions evaluated during test beam campaigns at CERN. During the $2024$ campaigns, a prototype "miniFIT" detector instrumented with four scintillating fiber layers in each tracking direction (4X+4Y) was operated \cite{miniFIT}. Each layer was read out by one S13552-10 SiPM array and a configuration of $20$ BETA-16 ASICs was implemented for the readout of a limited section of the SiPM sensitive area. In this preliminary configuration triggers could be generated independently in each tracking direction or in coincidence between them; the implemented trigger logic was therefore 3IR(X), 3IR(Y), and 3IR(X+Y). In $2025$ the setup was upgraded by instrumenting seven layers per tracking direction and deploying a total of $28$ BETA-64 ASICs. In this configuration each S13552-10 array was fully readout by two BETA ASICs per layer. The implemented trigger logic included:
\begin{enumerate}
\item 7IR(X), 7IR(Y) and 7IR(X+Y): combination of triggers in all seven instrumented layers, either independently in a single tracking direction or in coincidence between both directions.
\item 7IR-1(X), 7IR-1(Y) and 7IR-1(X+Y): same as above, but allowing one single missing trigger "cell".\footnote{We define the trigger cell as the elementary building block used in the trigger implementation. For the miniFIT2024 setup (based on the BETA-16 chip), the trigger cell was defined as the individual ASIC itself, whereas for the miniFIT2025 setup (based on the BETA-64 chip), the trigger cell was formed by the logical OR of the two ASICs connected to the same FIB.}
\item 3IR(X+Y): combination of triggers in three consecutive layers in both tracking directions; with the same starting layer required for X and Y.
\item 3IR-1(X+Y): same as above, but allowing one single missing trigger cell.
\end{enumerate}
A schematic representation of the main trigger configurations implemented for the 2025 test beam setup is shown in Fig.~\ref{fig:Internal_trigger_schematics}. Activated layers are highlighted in different colors: blue corresponds to a valid 7IR(X+Y) trigger, green to a valid 3IR(X+Y) trigger, and red to an invalid trigger.

\subsection{Noise trigger rate}
\label{sec:noise_rate}
\begin{figure}[htbp]
\centering
\includegraphics[width=0.8\columnwidth]{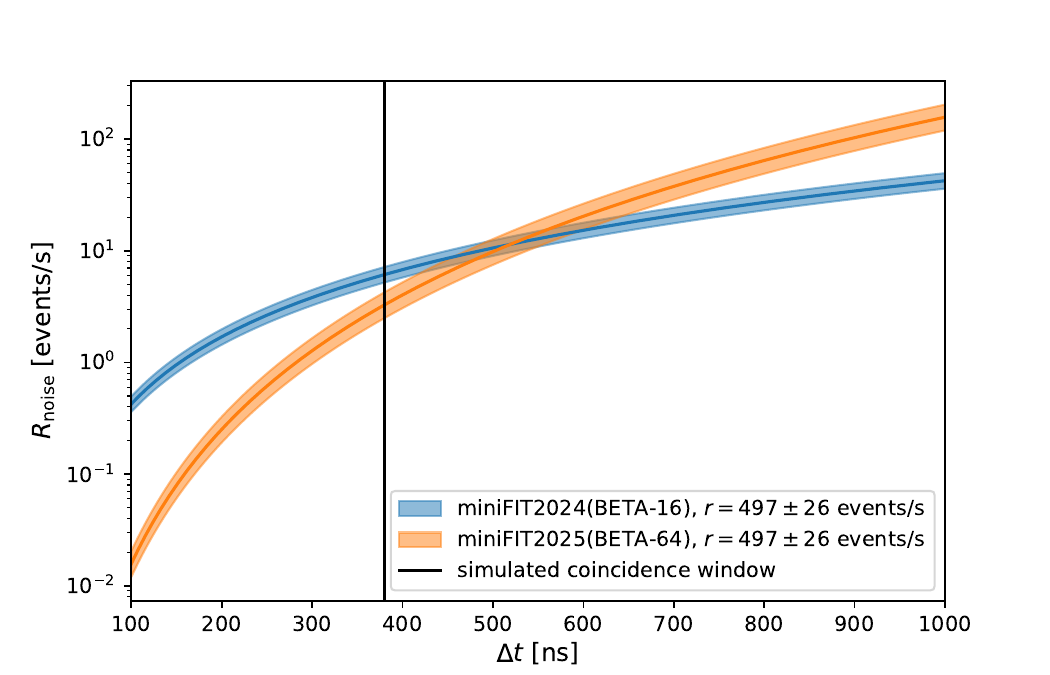}
\caption{\label{fig:Noise_rate_internal_trigger} Expected $R_{\mathrm{noise}}$ for the two tested miniFIT setups as a function of $\Delta t$. The trigger threshold was set to 3 phe for all channels, and the associated trigger rate and uncertainty measured for the three tested SiPM arrays in dark conditions. The vertical line indicates the $\Delta t$ value obtained from a toy Monte Carlo simulation of the time-over-threshold of the preamplifier signals (see Section~\ref{sec:noise_rate} for more details).}
\end{figure}

\begin{figure}[htbp]
\centering
\includegraphics[width=0.8\columnwidth]{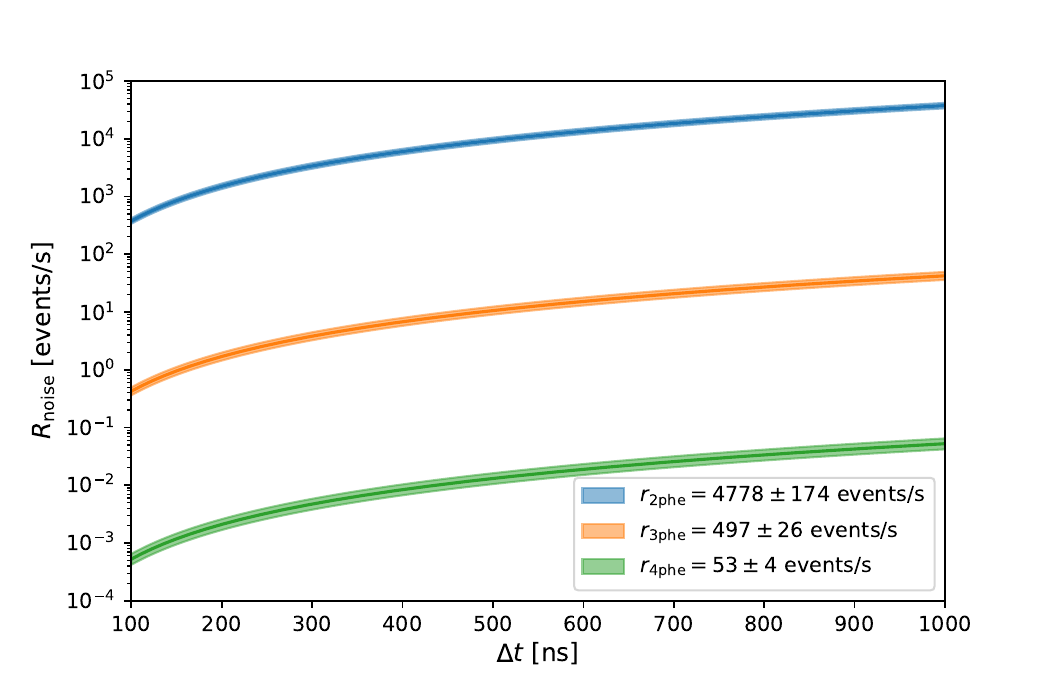}
\caption{\label{fig:Noise_rate_BETA16} Expected internal trigger noise rate for the miniFIT2024(BETA-16) setup as a function of $\Delta t$ for single-channel thresholds of $2$, $3$ and $4\,\mathrm{phe}$.}
\end{figure}

\begin{figure}[htbp]
\centering
\includegraphics[width=0.8\columnwidth]{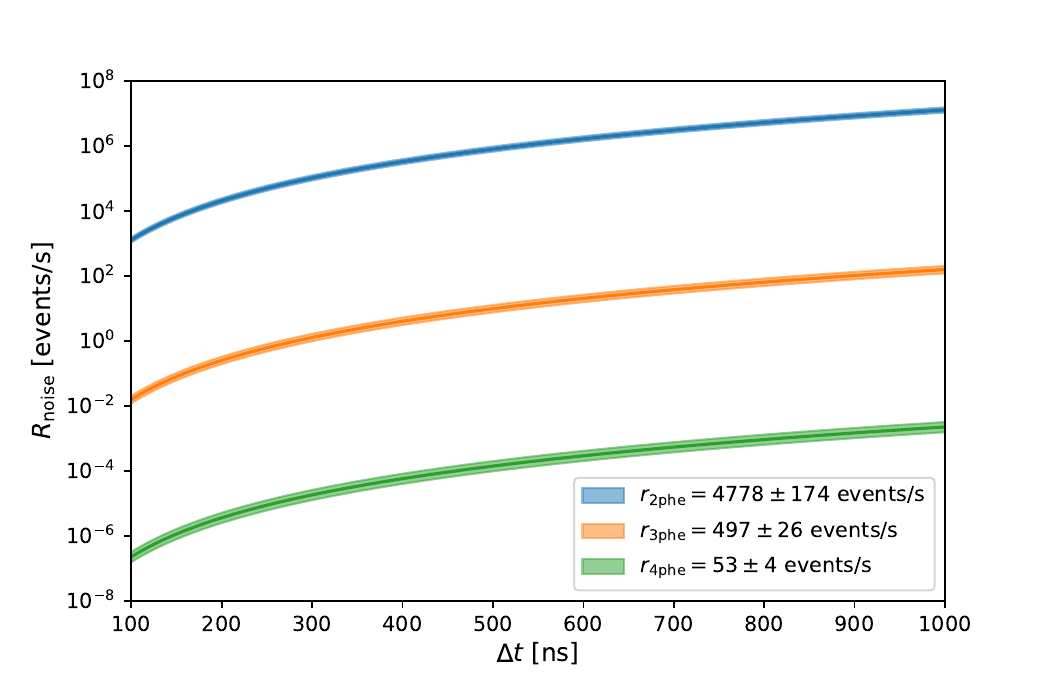}
\caption{\label{fig:Noise_rate_BETA64} Same as Figure~\ref{fig:Noise_rate_BETA16} for the miniFIT2025(BETA-64) setup.}
\end{figure}

The rate of random coincidences among independent random pre-trigger signals from $n$ trigger cells within a coincidence window $\Delta t$ is given by:
\begin{equation}
R_{\mathrm{coinc}} = n \cdot (r \cdot N_{\mathrm{ch}})^n \cdot \Delta t^{n -1}\, ,
\end{equation}
where $N_{\mathrm{ch}}$ is the number of channels per trigger cell and $r$ is the mean trigger rate per channel. The FIT internal trigger comprises multiple trigger logics characterized by different values of $n$.
Let $N_{\mathrm{comb}}$ denote the number of combinations satisfying a given trigger logic, the total noise-induced trigger rate can be computed as:
\begin{equation}
R_{\mathrm{noise}} = \sum_{i} N_{\mathrm{comb}, i} \cdot R_{\mathrm{coinc}, i}
\end{equation}
From the threshold calibration procedure described in Section~\ref{sec:thr_calibration}, $r$ can be determined for any given photoelectron threshold, while the parameters $N_{\mathrm{comb}}$, $n$ and $N_{\mathrm{ch}}$ are fixed by the detector configuration. Therefore it is possible to estimate the expected noise rate due to randomly coincident dark-count events in the FIT detector. Figure~\ref{fig:Noise_rate_internal_trigger} shows the expected trigger noise rate for the miniFIT2024(BETA-16) setup and the miniFIT2025(BETA-64) setup, as a function of $\Delta t$. The single-channel trigger rate was fixed for a threshold of $3\,\mathrm{phe}$. The dependence of the noise rate on the photoelectron threshold is shown separately in Figures~\ref{fig:Noise_rate_BETA16} and~\ref{fig:Noise_rate_BETA64} for the two miniFIT setups.

\begin{figure}[htbp]
\centering
\includegraphics[width=0.65\columnwidth]{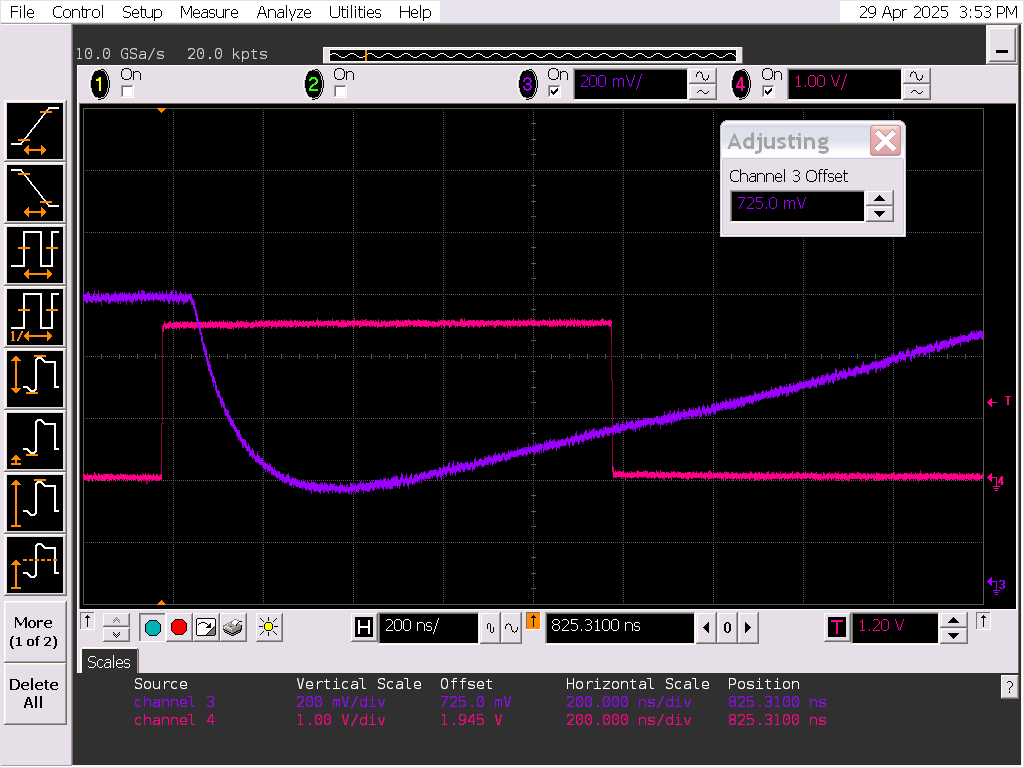}
\caption{\label{fig:Preamp_oscilloscope} Trigger signal used to pulse the LED (pink) and corresponding BETA preamplifier output waveform recorded with a DSO80204B Infiniium oscilloscope.}
\end{figure}

\begin{figure}[htbp]
\centering
\includegraphics[width=0.8\columnwidth]{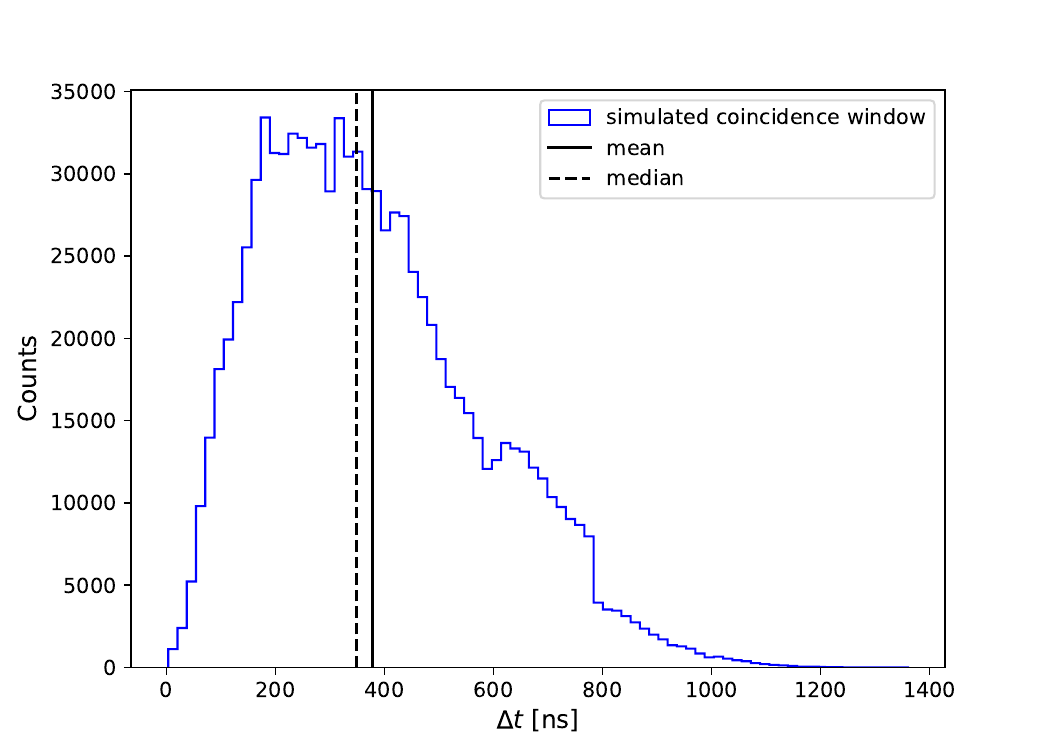}
\caption{\label{fig:CW_simulation} Distribution of the coincidence window width obtained from a toy Monte Carlo simulation. The mean value is approximately $380\,\mathrm{ns}$.}
\end{figure}

A coarse estimate of the $\Delta t$ can be obtained by analyzing the pulse shape of the BETA preamplifier output (see Figure~\ref{fig:Preamp_oscilloscope}). The coincidence window was derived from the time-over-threshold of the preamplifier signals using a toy Monte Carlo simulation. A reference waveform was recorded with a DSO80204B Infiniium oscilloscope.\footnote{\url{https://www.keysight.com/us/en/product/DSO80204B/infiniium-high-performance-oscilloscope-2ghz.html}.} The waveform amplitudes were scaled by a factor $N$ corresponding to the desired phe number. The phe number was generated according to the staircase probability distribution shown in Figure~\ref{fig:Thr_staircase}, and the discriminator threshold was fixed at $3\,\mathrm{phe}$. The resulting distribution of the coincidence window width is shown in Figure~\ref{fig:CW_simulation}, the mean value is $\Delta t \simeq 380\,\mathrm{ns}$.

Table~\ref{tab:NoiseRates} reports the expected noise trigger rates for the 2024 and 2025 miniFIT setups, assuming the mean coincidence window width, $380\,\mathrm{ns}$, and a mean single-channel trigger rate of $r = 497\,\mathrm{events/s}$. Under these conditions, the expected noise rate is of the order of a few events per second.

\begin{figure}[htbp]
\centering
\includegraphics[width=0.65\columnwidth]{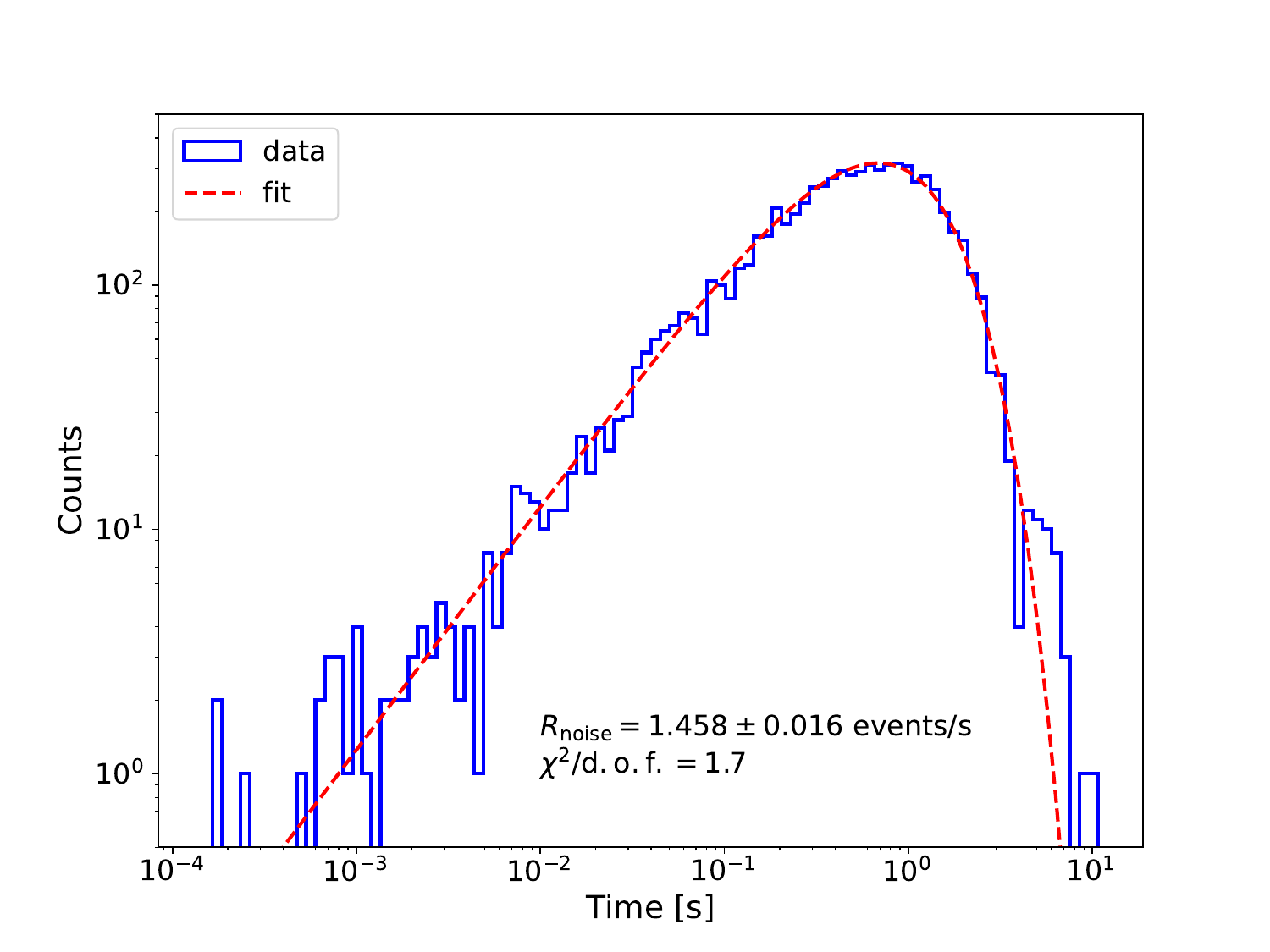}
\caption{\label{fig:InternalTrigger_lab} Distribution of time differences between consecutive triggers generated by random coincidences of dark count events in the miniFIT2024 setup.}
\end{figure}

\begin{table}[htbp]
\centering
\caption{\label{tab:NoiseRates} Expected trigger noise rates for different miniFIT setups and trigger logic. A coincidence window width $\Delta t = 380\,\mathrm{ns}$ and a mean single-channel trigger rate $r = 497\,\mathrm{events/s}$ are assumed. In the miniFIT2024(BETA-16) setup, triggers are generated at the individual ASIC level, while in the miniFIT2025(BETA-64) setup the logical OR of the two ASICs connected to the same FIB is used, leading to $N_{\mathrm{ch}} = 128$.}
\begin{center}
\begin{tabular}{llcccccc}
\hline
\hline
Setup & trigger logic & $N_{\mathrm{comb}}$ & $n$ & $r\, \mathrm{[ev/s]}$ & $N_{\mathrm{ch}}$ & $\Delta t [ns]$ & $R_{\mathrm{noise}}\, \mathrm{[ev/s]} $\\
\hline
\multirow{4}{*}{miniFIT2024(BETA-16)} & 3IR(X) & $14$ & $3$ & $497$ & $16$ & $380$ & $3.05$ \\
& 3IR(Y) & $14$ & $3$ & $497$ & $16$ & $380$ & $3.05$  \\
& 3IR(X+Y) & $200$ & $6$ & $497$ & $16$ & $380$ & $2 \cdot 10^{-6}$ \\
& total & &	 &  &  & & $6.1$ \\
\hline
\multirow{9}{*}{miniFIT2025(BETA-64)} & 3IR-1(X+Y) & $30$ &	$5$	& $497$ & $128$ & $380$ & $3.3$\\
& 3IR(X+Y) & $5$ &	$6$	& $497$ & $128$ & $380$ & $1.6 \cdot 10^{-2}$\\
& 7IR-1(X) & $7$ &	$6$	& $497$ & $128$ & $380$ & $2.2 \cdot 10^{-2}$\\
& 7IR-1(Y) & $7$ &	$6$	& $497$ & $128$ & $380$ & $2.2 \cdot 10^{-2}$\\
& 7IR(X) & $1$ &	$7$	& $497$ & $128$ & $380$ & $9 \cdot 10^{-5}$\\
& 7IR(Y) & $1$ &	$7$	& $497$ & $128$ & $380$ & $9 \cdot 10^{-5}$\\
& 7IR-1(X+Y) & $1$ &	$13$ 	& $497$ & $128$ & $380$ & $3 \cdot 10^{-14}$\\
& 7IR(X+Y) & $1$ &	$14$ 	& $497$ & $128$ & $380$ & $9 \cdot 10^{-16}$\\
& total &  & 	&  &  &  & $3.4$ \\
\hline
\hline
\end{tabular}
\end{center}
\end{table}
For the miniFIT2024(BETA-16) setup, it was possible to connect all readout channels simultaneously to the three SiPM arrays available during the optical-bench tests, enabling an experimental validation of the internal trigger behaviour. Figure~\ref{fig:InternalTrigger_lab} shows the distribution of time differences between consecutive triggers generated by random coincidences of dark-count events. The distribution follows an exponential law, as expected for a Poisson process. The expected count density is given by:
\begin{equation}
f(t) = N_{\mathrm{tot}} R_{\mathrm{noise}} e^{-R_{\mathrm{noise}} (t - t_0)} \Theta (t - t_0)\, ,
\label{eq:noise_rate_function}
\end{equation}
where $N_{\mathrm{tot}}$ is the total number of recorded events, $t_0=92.5\, \mathrm{\mu s}$ is the deadtime of the readout electronics and $\Theta (t - t_0)$ is the Heaviside step function. The expected number of counts in a time bin $[t_i,t_{i+1}]$ is obtained by integrating Equation~\ref{eq:noise_rate_function} over that interval. From a fit to the experimental data, a noise rate of $R_{\mathrm{noise}} = 1.458 \pm 0.016\,\mathrm{events/s}$ is obtained. This value is lower than the expected rate from Figure~\ref{fig:Noise_rate_internal_trigger} of approximately $6\,\mathrm{events/s}$, likely due to the approximate nature of the coincidence window width estimation and the consideration of $r$ as constant value for all channels.

The performance of the global internal trigger was further validated during CERN test beam campaigns and will be presented in future dedicated publications.

\subsection{Latency of the BETA ASIC and global internal triggers}
\begin{figure}[htbp]
\centering
\includegraphics[width=0.65\columnwidth]{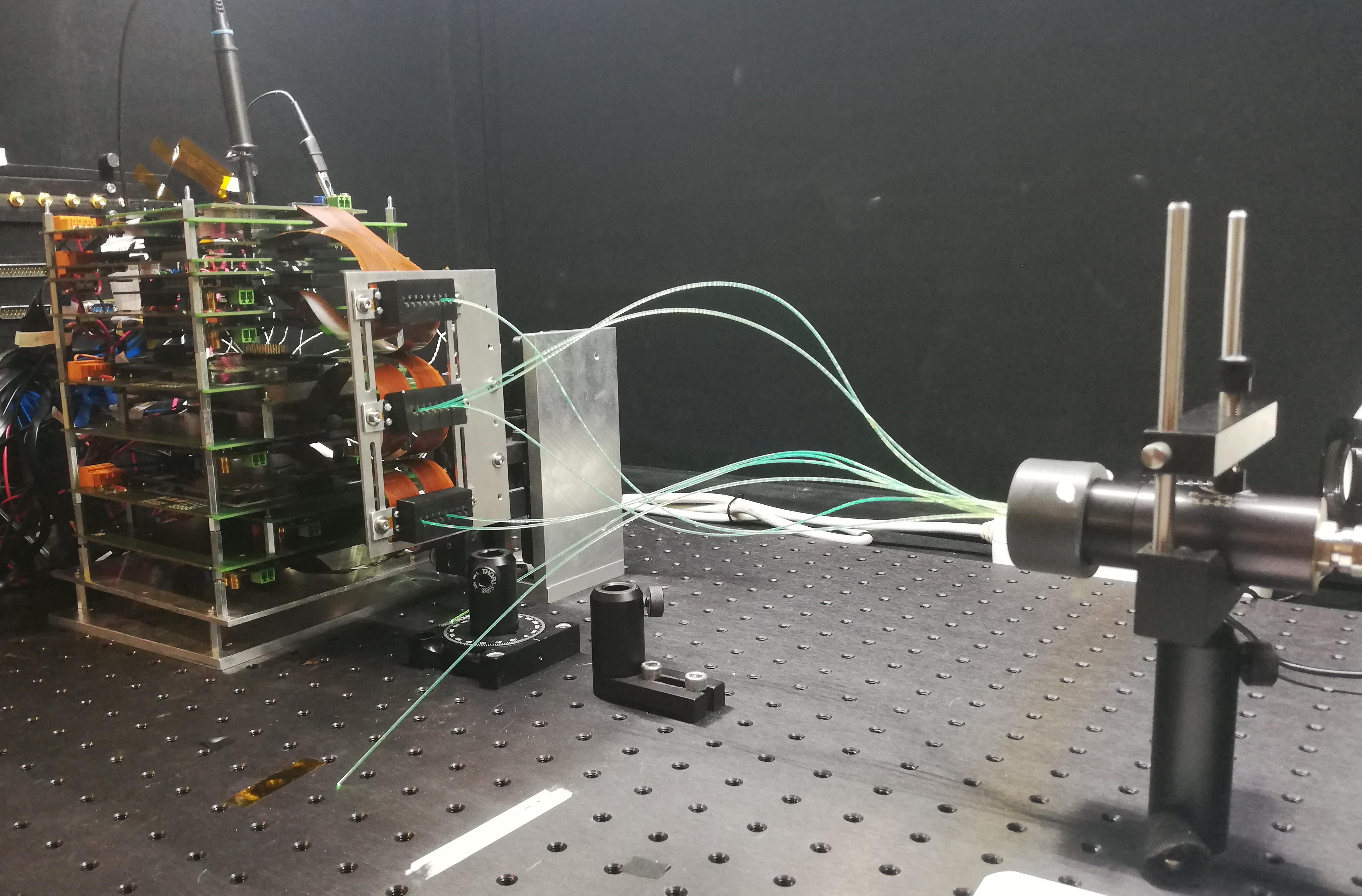}
\caption{\label{fig:Setup_OpticalFibers} Picture of the experimental setup used to compute the latency in the BETA ASIC and global internal triggers generation. Several optical fibers were used to produce signals in at least six different ASICs simultaneously.}
\end{figure}

Although exceptional timing performance was not a primary design goal of the BETA ASIC, latency measurements were conducted to characterize the trigger response and overall timing behaviour. To validate the global internal trigger generation and measure its latency, a light-distribution system made of several optical fibers was implemented in the experimental setup, to illuminate simultaneously SiPMs readout by at least six different ASICs  (see Figure~\ref{fig:Setup_OpticalFibers}), which build up a valid 3IR trigger combination.

Pre-trigger signals generated by the BETA ASICs (the logical OR of triggers from all its channels) and the global internal trigger produced by the FPGA combinational logic were recorded and analyzed using a DSO80204B Infiniium oscilloscope.

The trigger signal latency is defined as the time difference between the rising edge of the BETA ASIC pre-trigger or the global internal trigger signals and the rising edge of the Agilent 81160A signal, which is used to drive the LED.

\begin{figure}[htbp]
\centering
\includegraphics[width=0.7\columnwidth]{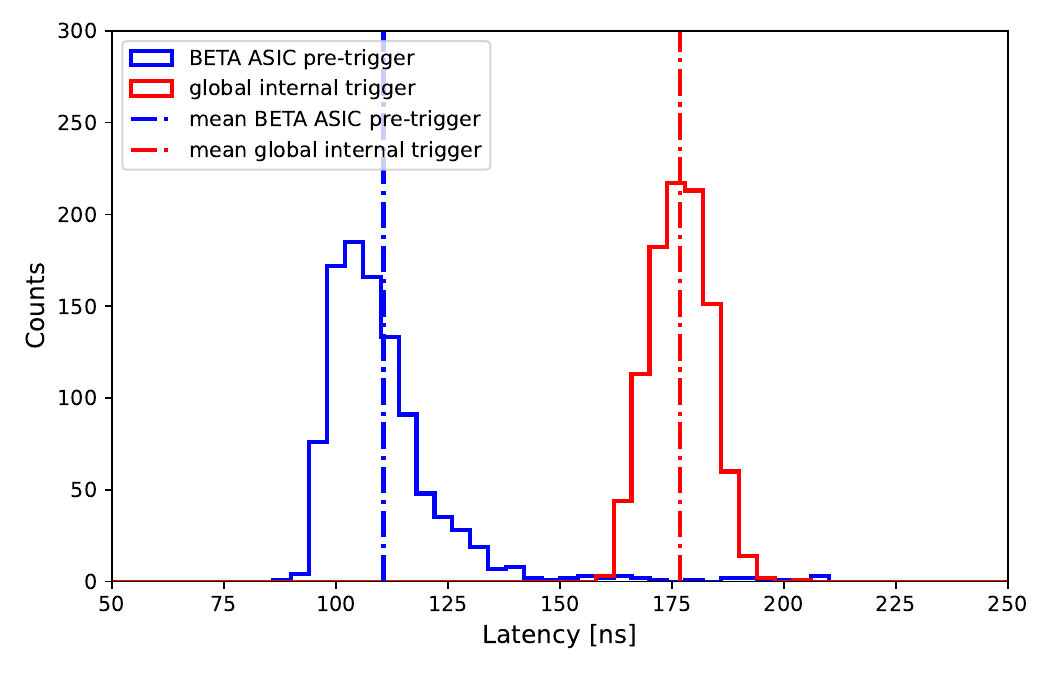}
\caption{\label{fig:Latency_Int6} Distribution of the latency of a single BETA ASIC pre-trigger and of the global internal trigger for a LED intensity setting of $6$, corresponding to $\sim10$ photoelectrons recorded by the SiPMs. $t = 0$ corresponds to the rising edge of the Agilent 81160A signal, used to drive the LED.}
\end{figure}

\begin{figure}[htbp]
\centering
\includegraphics[width=0.7\columnwidth]{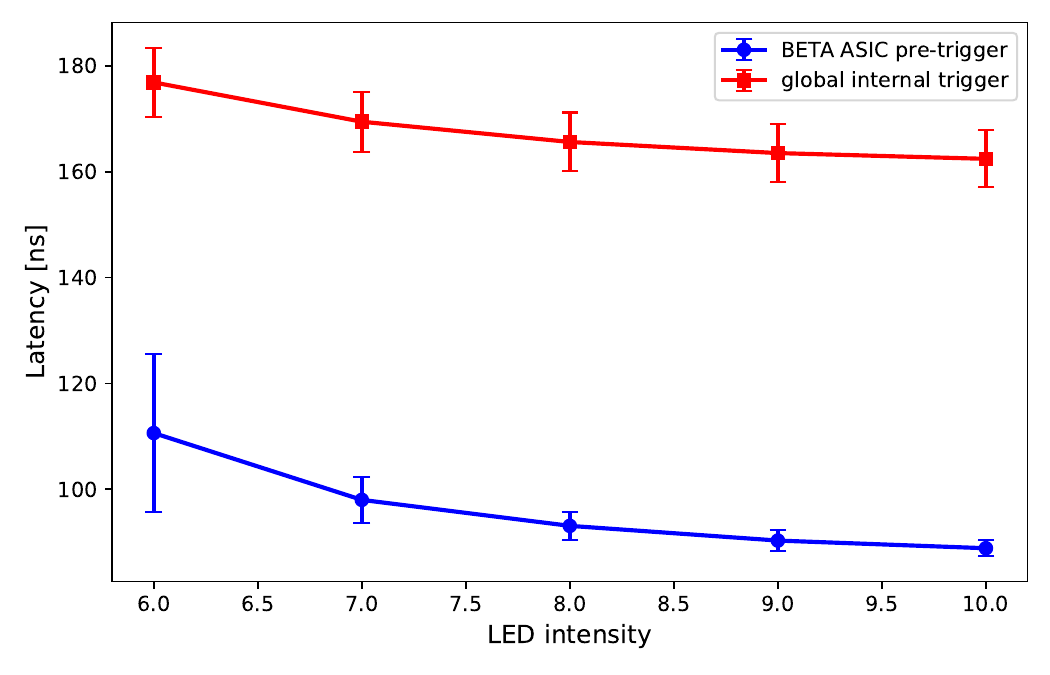}
\caption{\label{fig:Latency} Latency of one BETA ASIC pre-trigger and the global internal trigger as a function of the LED intensity, set by the potentiometer of the Picoquant PDL 800-D driver.}
\end{figure}

Latency measurements were performed at various LED intensities to investigate the timing performance as a function of the mean recorded signal in photoelectrons. Figure~\ref{fig:Latency_Int6} shows the latency of the pre-trigger of one BETA ASIC trigger involved in the internal trigger generation, and of the global internal trigger, both relative to the Agilent 81160A pulse function generator. Measurements were performed for an LED Intensity $= 6$, corresponding to a light level of about $10$ phe recorded in the SiPMs. The dependence of the trigger latency on the LED intensity is shown in Figure~\ref{fig:Latency}.

The BETA ASIC trigger timing response is dominated by the preamplifier rise time and its spread is determined by the time walk effects associated with different signal amplitudes, i.e. varying numbers of photoelectrons. Both effects decrease with increasing LED intensity, reaching a plateau for the highest illumination levels. Concerning the global internal trigger, its latency includes the additional time required for the FPGA logic computation. Its distribution is further smeared by the asynchronous pulsing of the LED with respect to the FPGA clock. Consequently, at the highest LED intensities, the global internal trigger spread is limited to $20/\sqrt{12} \approx
5.8\,\mathrm{ns}$\footnote{$20\,\mathrm{ns}$ is the FPGA clock period.}.

The results in Figure~\ref{fig:Latency_Int6} and Figure~\ref{fig:Latency} were not corrected for additional delays introduced in the experimental setup. Specifically, the PDL 800-D Driver introduces a delay of $35 \pm 5\,\mathrm{ns}$ when pulsing the LED with an external trigger,\footnote{as reported in the datasheet provided by the manufacturer.} and the cabling adds approximately $20\,\mathrm{ns}$. Therefore, the total latency for internal trigger generation in our system is $\lesssim 120\,\mathrm{ns}$, well within HERD's operational requirements. A fast signal is essential to minimize light losses in the Intensified scientific CMOS (IsCMOS) cameras employed in the HERD calorimeter. If the trigger delay remains below $200\,\mathrm{ns}$, more than $85\%$ of the photons are kept \cite{HERD_trigger, HERD_CALO}.

\section{Summary and conclusions}
\label{sec:conclusions}
In this work, we present the development and performance characterization of a scalable BETA ASIC-based readout system for the Scintillating Fiber Tracker (FIT), validated in a dedicated optical test setup. The front-end electronics and firmware are developed to manage communication and data acquisition across multiple ASIC devices. A global internal trigger, based on combinational logic of pre-triggers from individual BETA ASICs, was developed for the identification of ionizing particles in the FIT detector.

The experimental results demonstrate that the readout system coupled to the S13552-10 SiPM arrays provides a wide dynamic range, from single-photoelectron levels up to $8000$ photoelectrons,  with non-linearities below $3\%$. This range is suitable for the detection of cosmic-ray nuclei up to iron. The trigger thresholds were calibrated to single photoelectron levels, to ensure a reliable and efficient trigger for the identification of ionizing particles, while keeping the fake trigger rate from coincident dark-count events at a manageable level. The overall system latency for generating the internal trigger was measured to be $\lesssim 120\,\mathrm{ns}$.

Between $2024$ and $2025$, multiple BETA ASICs were configured and tested in the optical setup. The system was further tested in CERN-SPS test beam campaigns, where the readout electronics and firmware were successfully operated with a prototype miniFIT detector. Future work will focus on scaling the system to a larger number of channels, further validation with the miniFIT prototype in test beam and cosmic-ray campaigns, and qualification of the SiPM arrays for space applications.

\acknowledgments
This work was supported by grants PID2020-116075GB-C21, PID2020-116075GB-C22, PID2023-146071NB-C21, PID2023-146071NB-C22, PRE2021-097518, and PREP2023-001838, funded by MICIU/AEI /10.13039/501100011033, FEDER, UE and FSE+.\\
Additional funding was provided by the European Union – NextGenerationEU and the Department of Research and Universities of the Government of Catalonia, within the framework of the R\& D\& I Complementary Plans of the Government of Spain (“Recovery, Transformation and Resilience Plan (C17.I1) – Funded by the European Union – NextGenerationEU”).\\
This work was supported by the financial contributions from the Swiss National Foundation (SNSF-PZ00P2\textunderscore 193523 and SNSF-TMSGI2\textunderscore 226294).
\vspace{1em}

\noindent
\includegraphics[height=1.cm]{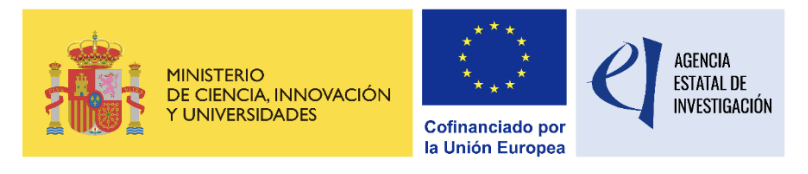}

\vspace{0.5em}

\noindent
\includegraphics[height=1.cm]{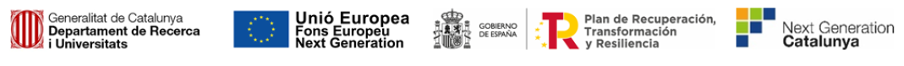}


\newpage

\appendix
\section{Configuration and optimization of the BETA ASIC registers}
\label{sec:BETA_optimization}
The BETA ASIC allows configuration of several operational parameters, including gain settings, bias currents, DC offsets, shaping and sampling times, individual channel trigger thresholds, and automatic path selection thresholds. All parameters are configured via an I2C interface.

For accurate calibration and stable operation, several parameters must be tuned relative to the manufacturer's default configuration. This section describes the main optimization procedures employed in the experimental setup for the optical characterization of the S13552-10 SiPM arrays.

\subsection{Bandgap reference voltage: \texttt{D\textunderscore Vref\textunderscore BGR}}
To ensure stable performance under varying temperature and environmental conditions, the BETA ASIC employs a main bandgap reference voltage. An operating value of $500\,\mathrm{mV}$ is recommended, with a tolerance of $\pm 15\,\mathrm{mV}$. The reference voltage can be adjusted through the $8$-bit register \texttt{D\textunderscore Vref\textunderscore BGR}. The actual voltage value is verified by a direct measurement with a multimeter at the dedicated test point on the FIB. If the measured voltage deviates from the recommended range for operation, the \texttt{D\textunderscore Vref\textunderscore BGR} register value is tuned individually for each BETA ASIC until compliance is achieved. The optimal register values for all tested BETA-16 and BETA-64 ASICs can be found in Figure~\ref{fig:ConfigSummary}.

\subsection{ADC ramp slope: \texttt{D\textunderscore SLOPE}}
\begin{figure}[htbp]
\centering
\includegraphics[width=0.7\columnwidth]{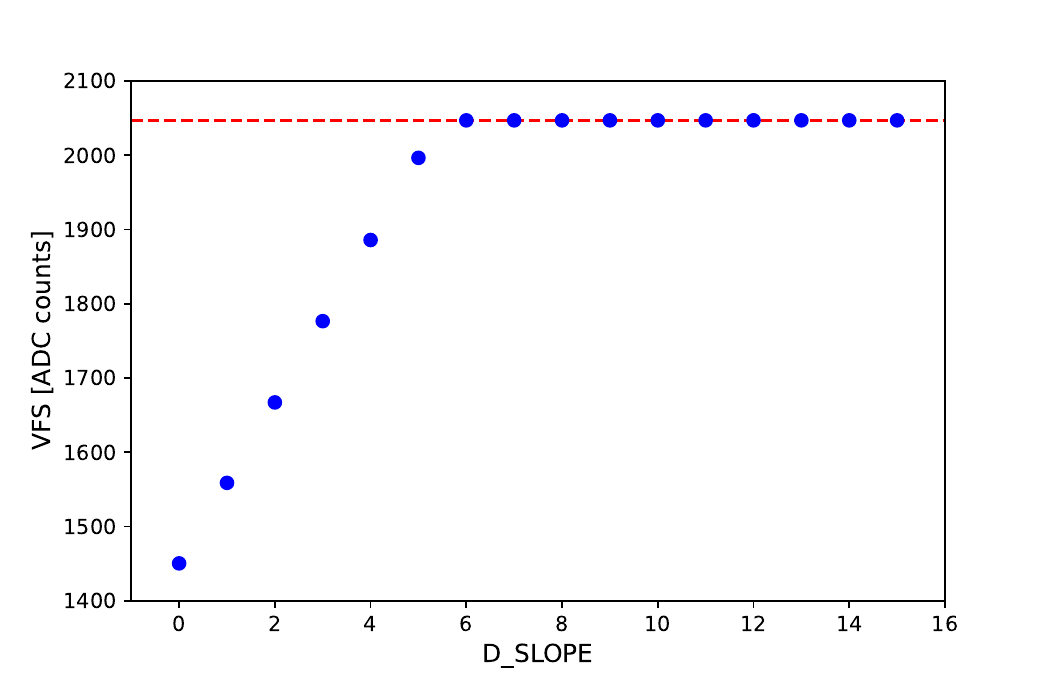}
\caption{\label{fig:DSLOPE} Full-scale voltage, VFS, measured as a function of the \texttt{D\textunderscore SLOPE} register value, for one ASIC of the experimental setup. For the specific ASIC shown in the plot, the optimal configuration is \texttt{D\textunderscore SLOPE} $= 5$.}
\end{figure}

To fully exploit the BETA ASIC dynamic range, the ramp slope of its ADC is adjusted via the $4$-bits register \texttt{D\textunderscore SLOPE}. A full-scale voltage value, VFS, is recorded at the end of each conversion of the analog signal, and used to calibrate the device. An example of the VFS dependence as a function of the \texttt{D\textunderscore SLOPE} register value is shown in Figure~\ref{fig:DSLOPE}. The optimal value for the \texttt{D\textunderscore SLOPE} register corresponds to a measured VFS as close as possible to $2000$, just below the 11 bit ADC saturation at $2047$. A summary of the optimal \texttt{D\textunderscore SLOPE} register values obtained for all tested BETA-16 and BETA-64 ASICs is provided in Figure~\ref{fig:ConfigSummary}.

\subsection{Sampling time: \texttt{D\textunderscore WAIT\textunderscore INTERNAL}}
\label{sec:SamplingTime}
In the BETA ASIC, the peak of the shaper output is captured using a flip-around track-and-hold (T\&H) circuit. The delay between the rising edges of the trigger signal (internal or external) and the T\&H is controlled by the configuration register \texttt{D\textunderscore WAIT\textunderscore INTERNAL}. After this time delay, the T\&H is set to hold, maintaining the output signal constant, and the ADC conversion begins simultaneously.

\begin{figure}[ht]
	\centering
	\subfloat[Mean signal as a function of the internal delay.]{
	\includegraphics[width=0.65\columnwidth]{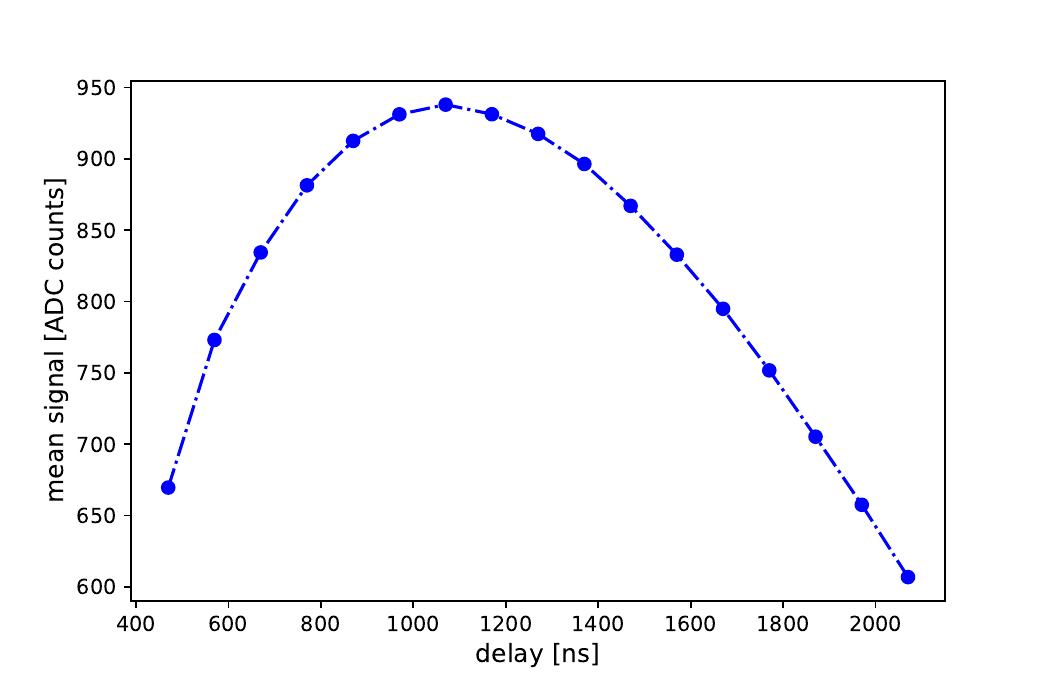}
	\label{fig:MeanVsIntDelay}}
	\vspace{-0.05cm}
	\subfloat[Signal distributions obtained for different internal delay settings.]{
	\includegraphics[width=0.65\columnwidth]{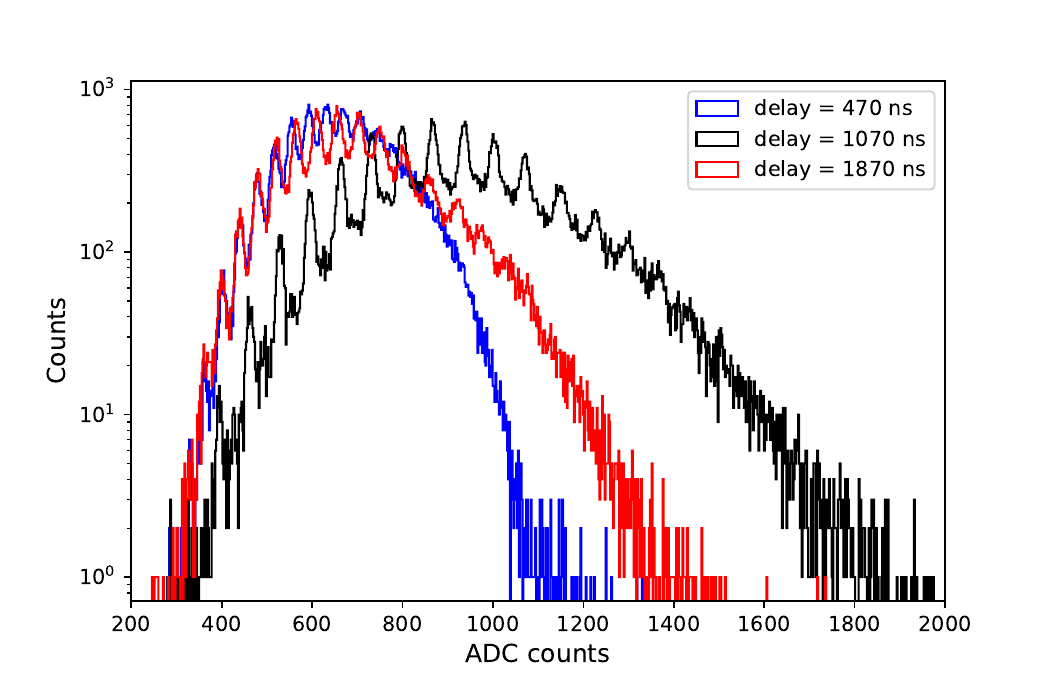}
	\label{fig:DistributionIntDelay}}
	\caption{\label{fig:IntDelay} Optimization of the internal delay. Best performance is achieved when the shaper output is sampled at its peak value.}
\end{figure}

The optimal configuration corresponds to sampling the shaper output at its peak, which maximizes both signal amplitude and photopeak resolution. The time delay is optimized by scanning the \texttt{D\textunderscore WAIT\textunderscore INTERNAL} register for a fixed LED light intensity. Figure~\ref{fig:MeanVsIntDelay} shows the mean signal value as a function of the internal delay. The analog shaper output signal reaches its maximum approximately $1\,\mathrm{\mu s}$ after the LED is pulsed. Figure \ref{fig:DistributionIntDelay} illustrates the measured signal distributions for a few values of the delay setting: when signals are sampled too early or too late with respect to the peak position, photopeak separation deteriorates.

\subsection{Preamplifier and low-pass filter offsets}
The DC offsets of the preamplifier (for both HG and LG paths) and of the low-pass filter are controlled by the $8$-bit registers \texttt{Voffset\textunderscore CH\textunderscore Preamp} and \texttt{Voffset\textunderscore CH\textunderscore LPF}, respectively. Proper adjustment of these offsets is crucial for reliable triggering and optimal use of the ADC dynamic range.

\subsubsection{Preamplifier offset: \texttt{D\textunderscore SELFTRIGV\textunderscore CHx}}
The preamplifier offset is optimized to allow stable trigger generation at the level of a few photoelectrons. The optimization procedure run as follows:
\begin{itemize}
\item Trigger thresholds (configuration register \texttt{D\textunderscore SELFTRIGV\textunderscore CHx}) were set to their minimum level for all channels. To ensure an accurate and precise measurement, channels exhibiting excessive noise were disabled. These channels were identified by analyzing the spread of the signal distribution measured in dark conditions.
\item The HG path was selected and the gain was set to the maximum value, HG00.
\item The \texttt{Voffset\textunderscore CH\textunderscore Preamp} register was scanned, while recording the ASIC pre-trigger counts in dark conditions.
\end{itemize}

\begin{figure}[htbp]
\centering
\includegraphics[width=0.7\columnwidth]{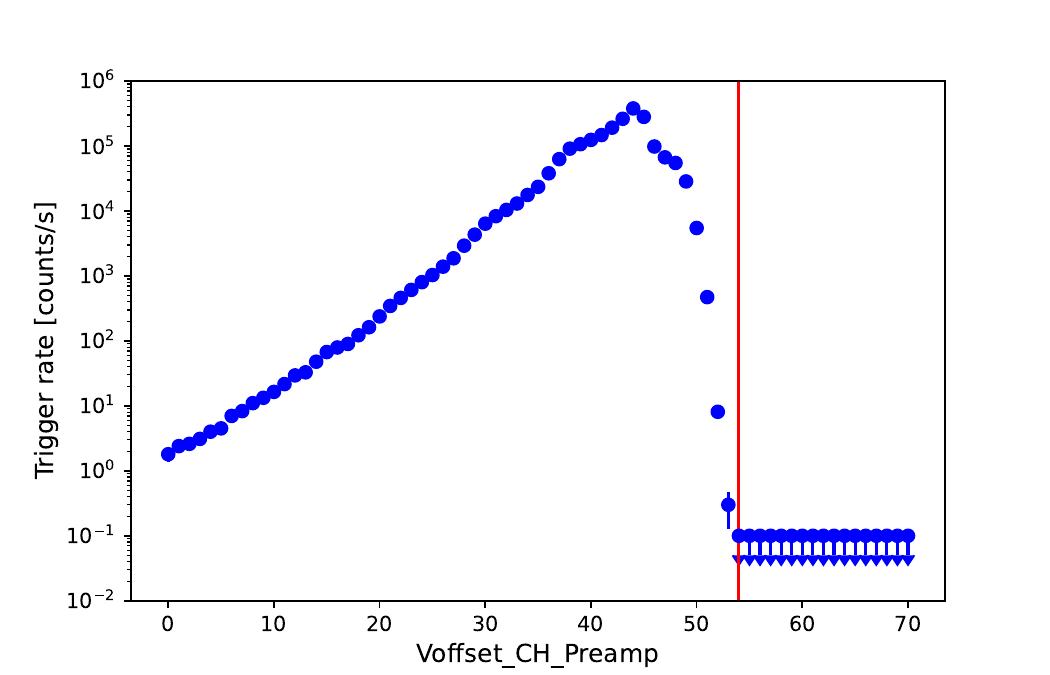}
\caption{\label{fig:OffsetPreamp} Recorded trigger rates as a function of the \texttt{Voffset\textunderscore CH\textunderscore Preamp} register value.}
\end{figure}

The trigger rate evolution as a function of the \texttt{Voffset\textunderscore CH\textunderscore Preamp} register value is shown in Figure \ref{fig:OffsetPreamp}. As the register value increases, the trigger rate initially rises because the DC offset of the preamplifier moves closer to the voltage corresponding to the threshold level set, allowing triggers to be generated starting from lower photoelectron levels. The trigger rate reaches a maximum until the baseline is crossed, after which the trigger rate abruptly drops to zero. The preamplifier offset is set to the register value indicated by the vertical line in the plot; this configuration allows trigger thresholds to be set starting from the baseline and incremented in integer photoelectron steps, by adjusting the \texttt{D\textunderscore SELFTRIGV\textunderscore CHx} as illustrated in Section~\ref{sec:thr_calibration}. The optimal register values for all tested BETA-16 and BETA-64 ASICs are summarized in Figure~\ref{fig:ConfigSummary}.

\subsubsection{Low-Pass Filter offset: \texttt{Voffset\textunderscore CH\textunderscore LPF}}
\begin{figure}[htbp]
	\centering
	\subfloat[ADC spectra obtained in dark conditions for different values of the \texttt{Voffset\textunderscore CH\textunderscore LPF} register in the BETA-16 ASIC.]{
	\includegraphics[width=0.65\columnwidth]{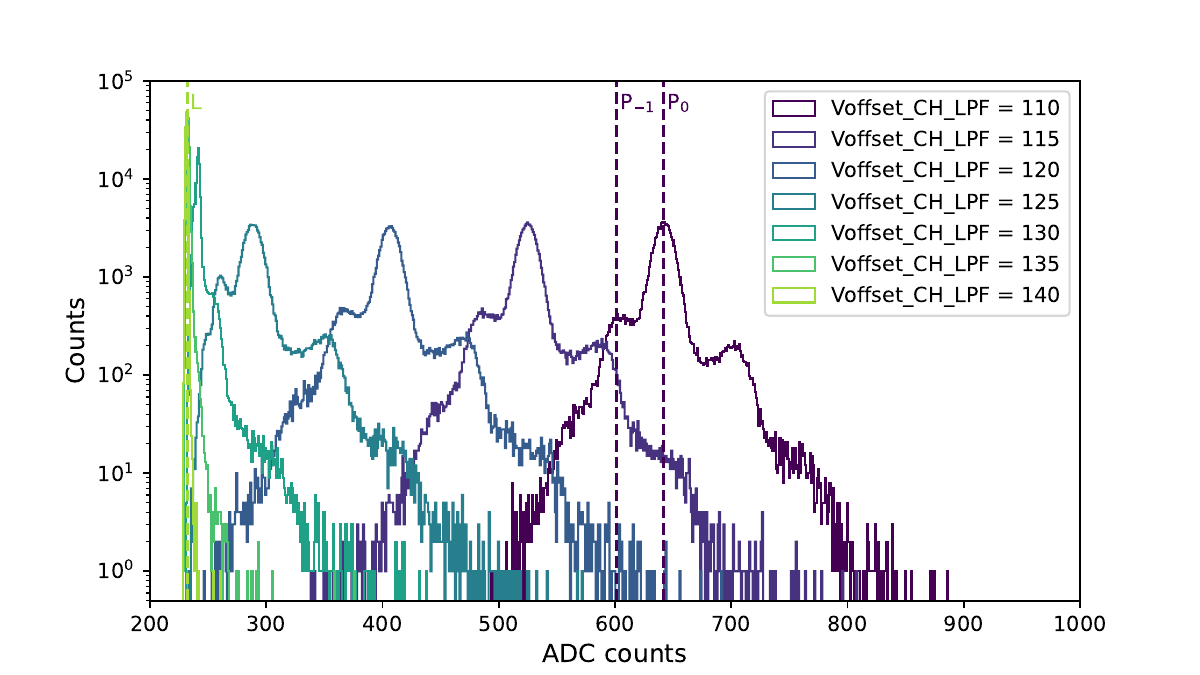}
	\label{fig:LPF_histo}}
	\vspace{-0.05cm}
	\subfloat[Pedestal peak position as a function of the \texttt{Voffset\textunderscore CH\textunderscore LPF} register value. The optimal values for the pedestal and the \texttt{Voffset\textunderscore CH\textunderscore LPF} register are shown by the dotted lines.]{
	\includegraphics[width=0.65\columnwidth]{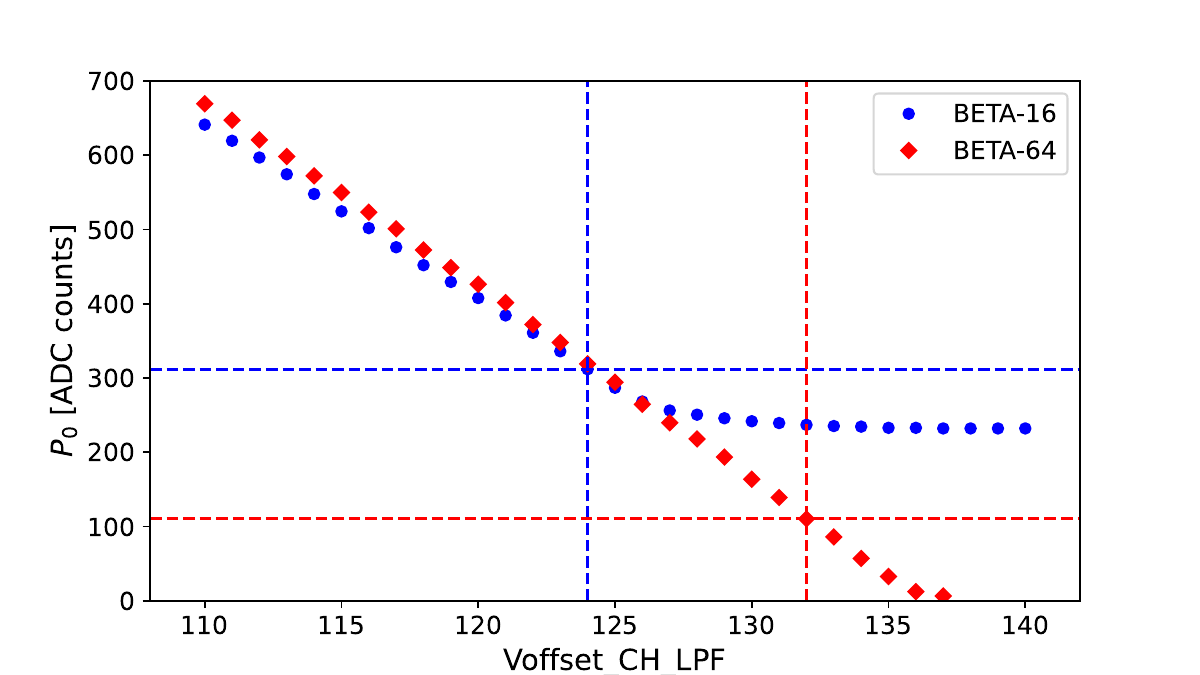}
	\label{fig:LPF_optimization}}
	\caption{\label{fig:OffsetLPF} Optimization of the low-pass filter osset, \texttt{Voffset\textunderscore CH\textunderscore LPF}.}
\end{figure}

After setting the preamplifier offset, the low-pass filter offset is tuned to maximize the usable ADC range. The baseline level of the low-pass filter is optimized by scanning the \texttt{Voffset\textunderscore CH\textunderscore LPF} register in dark conditions. The distribution of the measured ADC spectra as a function of the \texttt{Voffset\textunderscore CH\textunderscore LPF} register value is shown in Figure~\ref{fig:OffsetLPF}. As the register values increase, the measured distribution\footnote{See Section~\ref{sec:calibration} for details on the shape and features of the measured distribution.} shifts toward lower ADC values, eventually shrinking and asymptotically approaching a lowest recordable ADC count (see Figure~\ref{fig:LPF_histo}). As outlined in the plot, we define as $L$ the lowest recordable ADC count, $P_0$ the pedestal peak position, $P_{-1}$ the undershoot peak position, and $D = P_0 - P_{-1}$. The optimal pedestal position $P_{\mathrm{opt}}$ is then defined as:
\begin{equation}
P_{\mathrm{opt}} = L + 2 \cdot D\, .
\end{equation}
This choice maximizes the usable ADC range while preserving a good linearity response. The optimal \texttt{Voffset\textunderscore CH\textunderscore LPF} corresponds to the pedestal position closest to the computed $P_{\mathrm{opt}}$ value, as shown in Figure~\ref{fig:LPF_optimization}. Due to differences in the internal biasing circuitry, the minimum achievable pedestal differs between the two ASIC versions. For the BETA-16 ASIC, the lowest recordable ADC count is limited to approximately $L \simeq 200$, whereas for the BETA-64 ASIC the extended DC bias range allows operation over the full ADC dynamic range. The optimal register values for all tested BETA-16 and BETA-64 ASICs are summarized in Figure~\ref{fig:ConfigSummary}.

\subsection{Path selection comparator: \texttt{D\textunderscore PATHSELV\textunderscore CHx}}
When the auto-path mode is enabled, a discriminator selects between the HG and LG paths, based on a threshold level defined by the $8$-bit register \texttt{D\textunderscore PATHSELV\textunderscore CHx}. If the HG signal exceeds this threshold, the readout automatically switches to the LG path, thus extending the dynamic range. The \texttt{D\textunderscore PATHSELV\textunderscore CHx} register can be configured individually for each channel of the BETA ASIC, allowing precise tuning and calibration for optimal channel-to-channel uniformity.

\begin{figure}[htbp]
\centering
\includegraphics[width=0.95\columnwidth]{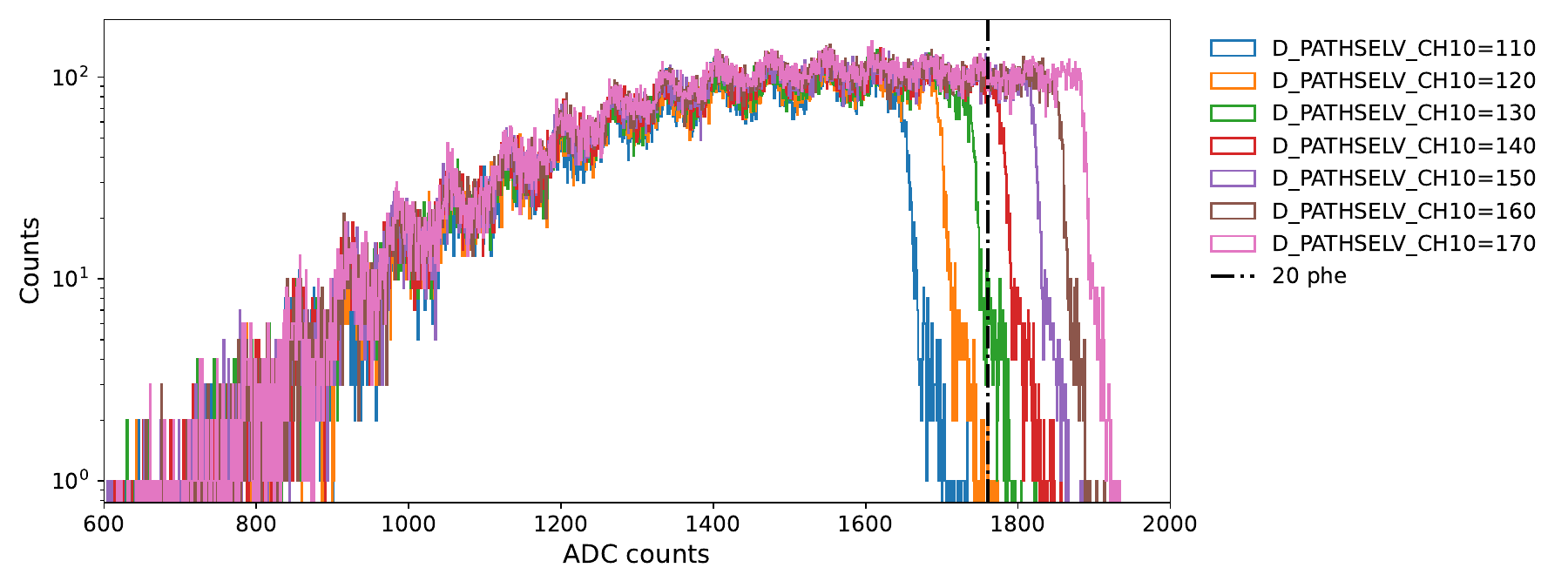}
\caption{\label{fig:PathSelector} HG signal distribution as a function of the \texttt{D\textunderscore PATHSELV\textunderscore CHx} register value. Increasing the threshold comparator level shifts the HG to LG transition toward higher ADC counts.}
\end{figure}

The threshold is optimized by illuminating the SiPMs with a fixed LED intensity while scanning the \texttt{D\textunderscore PATHSELV\textunderscore CHx} register. The LED intensity was chosen so that, with forced HG read-out, the photoelectron spectrum reaches saturation. In contrast, with auto-path mode enabled, the resulting ADC spectra are shown in Figure \ref{fig:PathSelector}: as the \texttt{D\textunderscore PATHSELV\textunderscore CHx} register value increases, the transition from HG to LG shifts at higher ADC values. The transition edge is smeared by electronic noise and random coincidence of dark-count events with the LED pulses.

\begin{figure}[htbp]
    \centering
    \subfloat[\texttt{D\textunderscore Vref\textunderscore BGR}.]{
        \includegraphics[width=0.48\columnwidth]{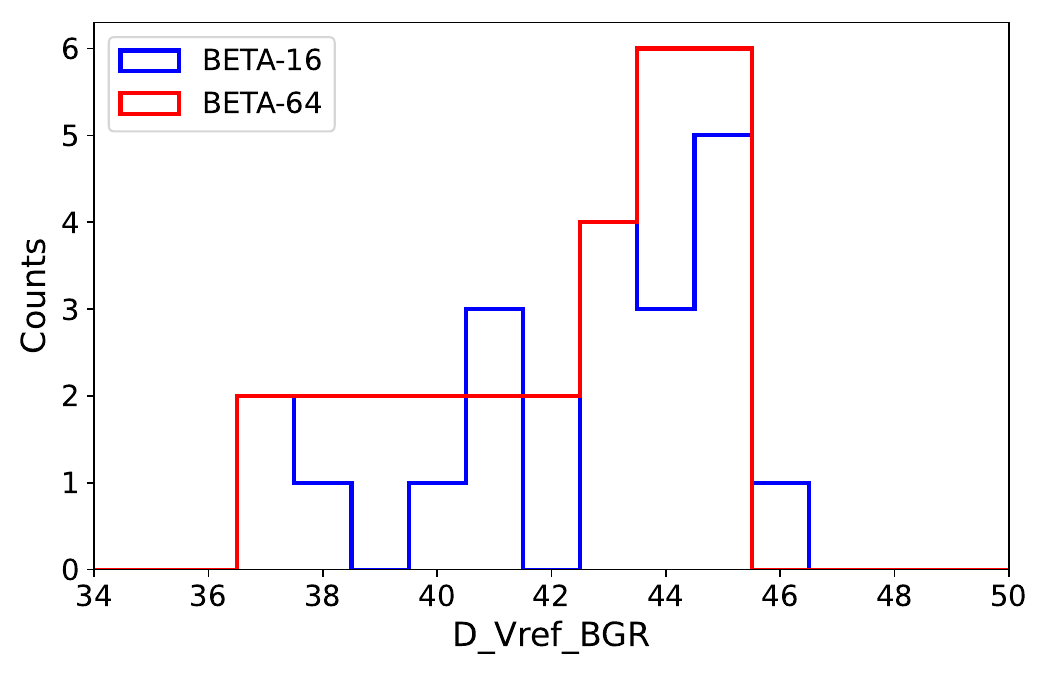}
    }
    \subfloat[\texttt{D\textunderscore SLOPE}.]{
        \includegraphics[width=0.48\columnwidth]{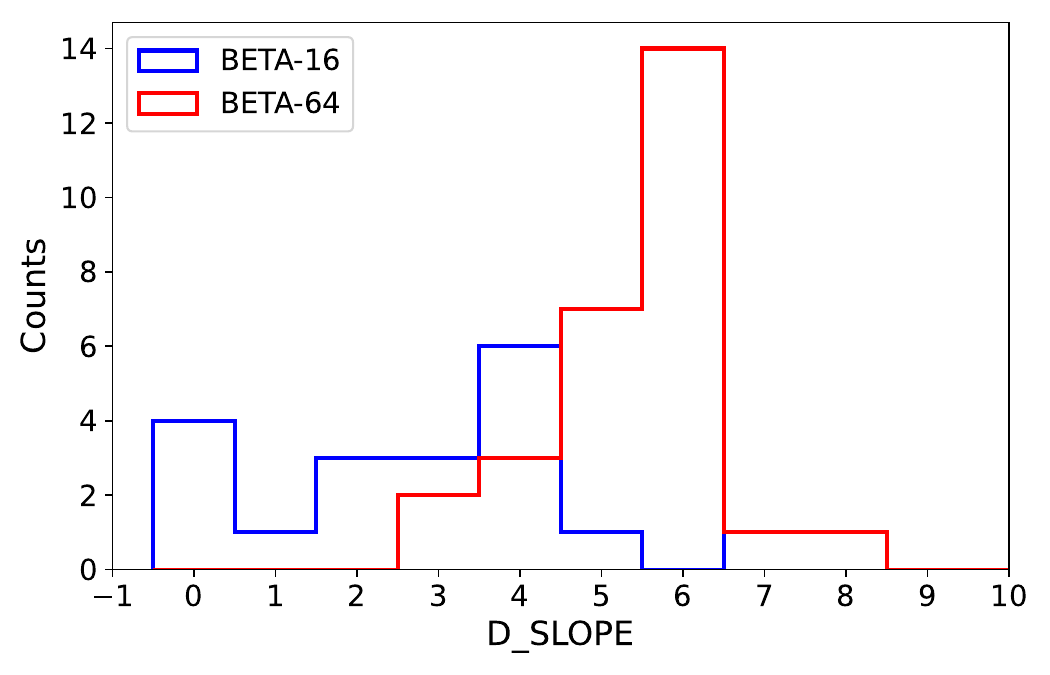}
    }
    
    \vspace{-0.05cm}
    \centering
    \subfloat[\texttt{Voffset\textunderscore CH\textunderscore Preamp}.]{
        \includegraphics[width=0.48\columnwidth]{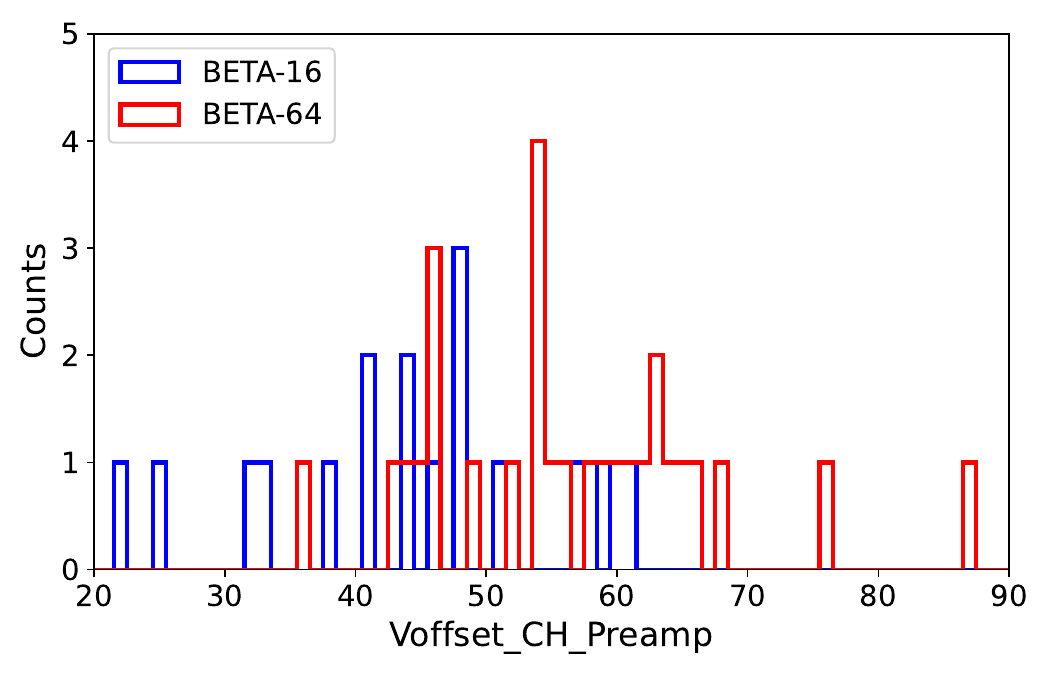}
    }
    \subfloat[\texttt{Voffset\textunderscore CH\textunderscore LPF}.]{
        \includegraphics[width=0.48\columnwidth]{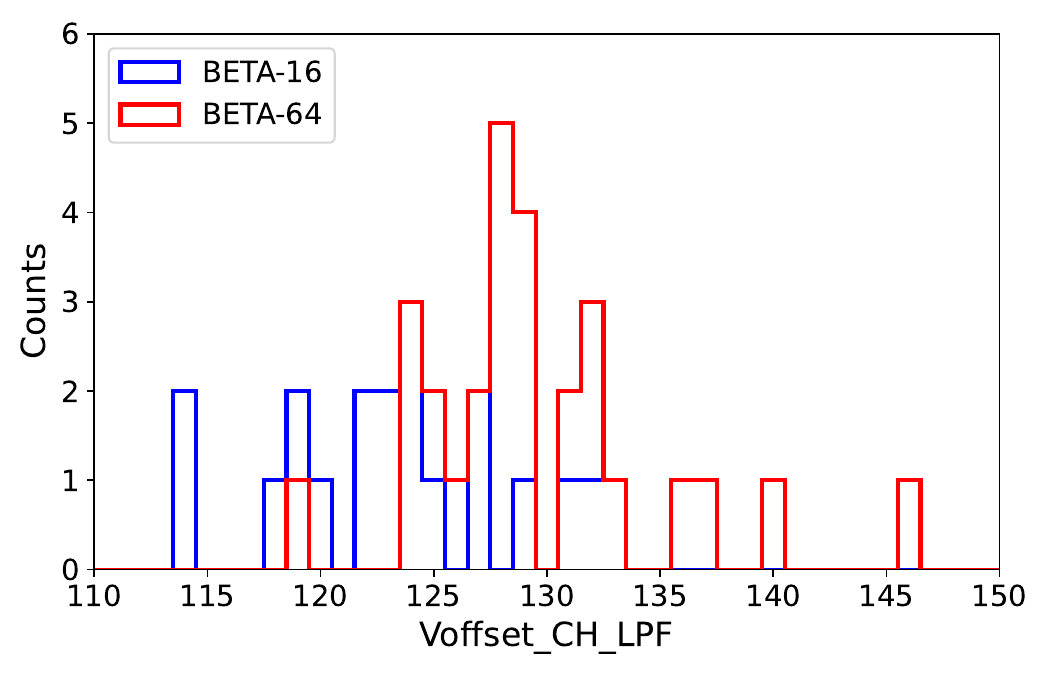}
    }

    \vspace{-0.05cm}
    \centering
    \subfloat[\texttt{D\textunderscore PATHSELV\textunderscore CHx}.]{
        \includegraphics[width=0.48\columnwidth]{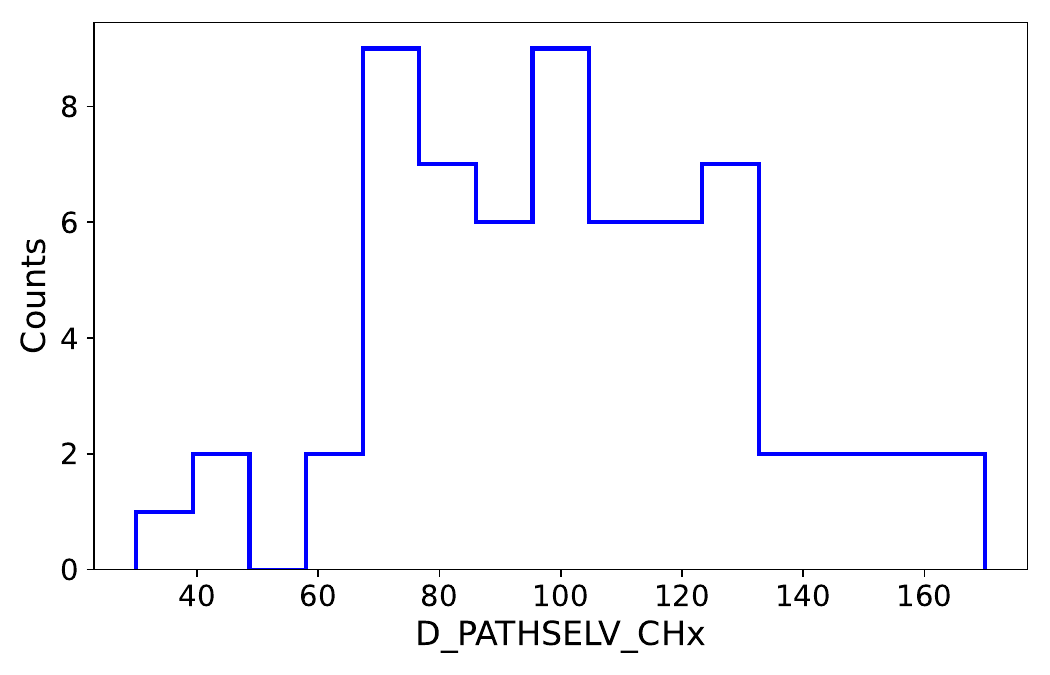}
    }
    \caption{
        \label{fig:ConfigSummary}
        Summary of the optimized values of the BETA ASIC configuration registers determined from the calibration measurements.
    }

\end{figure}

It is convenient to express the path selection threshold in either ADC or photoelectron units. Moreover, the path selector threshold is optimized such that the HG-to-LG transition occurs as close as possible to the HG saturation point without begin affected by it. For the CERN test beam campaigns, the thresholds are conventionally set to correspond to $20\,\mathrm{photoelectrons}$ at HG00 for all BETA-16 ASICs. The resulting values are summarized in Figure~\ref{fig:ConfigSummary}. For the BETA-64 ASICs, this register was not configured, since the corresponding test beam campaigns were performed exclusively with protons, for which automatic path selection was not required.

\section{Statistical error propagation in threshold calibration}
\label{sec:error_thr}
The threshold values corresponding to integer numbers of photoelectrons are determined from the zero-crossing points of the second derivative of the trigger count rates. Denoting by $(x_1, y_1)$ the last point before the zero crossing and by $(x_2, y_2)$ the first point after it, the straight line connecting these two points intersects the $x$-axis at $(x_0, 0)$, where:
\begin{equation}
x_0 = x_1 - \frac{y_1 \cdot (x_2 - x_1)}{y_2 - y_1}\, .
\end{equation}

The uncertainty on $x_0$ is obtained by standard error propagation, as:
\begin{equation}
\begin{split}
\sigma_{x_0} &= \sqrt{\left( \frac{\partial x_0}{\partial y_1} \cdot \sigma_{y_1} \right)^2 + \left( \frac{\partial x_0}{\partial y_2} \cdot \sigma_{y_2} \right)^2} \\
&= \sqrt{\left( y_2 \cdot \frac{x_2 - x_1}{(y_2 - y_1)^2} \cdot \sigma_{y_1} \right)^2 + \left( y_1 \cdot \frac{x_2 - x_1}{(y_2 - y_1)^2} \cdot \sigma_{y_2} \right)^2} \\
&= \frac{\lvert x_2 - x_1 \rvert}{(y_2 - y_1)^2} \cdot \sqrt{y_2^2 \cdot \sigma_{y_1}^2 + y_1^2 \cdot \sigma_{y_2}^2} \, .
\end{split}
\end{equation}

In Figure~\ref{fig:Thr_staircase}, the $y$-axis values correspond to the second derivative of the trigger rate normalized to the rate itself, i.e. $y_i = \frac{r_i^{\prime \prime}}{r_i}$, with $r_i^{\prime \prime} = \frac{d^2 r}{dx^2}$. The associated uncertainty $\sigma_{y_i}$ is therefore given by:
\begin{equation}
\sigma_{y_i} = \lvert y_i \rvert \cdot \sqrt{\left( \frac{\sigma_{r_i^{\prime \prime}}}{r_i^{\prime \prime}} \right)^2 + \left( \frac{\sigma_{r_i}}{r_i} \right)^2}\, .
\end{equation}
The second derivative of a function can be approximated numerically at any point and for a small step size using the second-order central difference method:
\begin{equation}
\label{eq:secondorder_diff}
f^{\prime \prime }(x)\approx \frac{f(x+h)-2f(x)+f(x-h)}{h^{2}} \, .
\end{equation}
Assuming a unit step size, $h \equiv 1$, Equation~\ref{eq:secondorder_diff} can be written as:
\begin{equation}
r_i^{\prime \prime } \approx r_{i+1} -2r_i + r_{i-1} \, ;
\end{equation}
from which the propagated uncertainty follows as:
\begin{equation}
\sigma_{r_i^{\prime \prime }} = \sqrt{ \sigma_{r_{i+1}}^2 + 4 \sigma_{r_i}^2 + \sigma_{r_{i-1}}^2 } \, .
\end{equation}
Finally, since the trigger counts $N$ recorded over a time interval $t$ follow Poisson statistics, the uncertainty on the rate is:
\begin{equation}
\sigma_{r_i} = \frac{\sqrt{N}}{t} \, .
\end{equation}

\end{document}